\documentclass[fleqn,usenatbib]{mnras}

\usepackage{mathptmx}
\usepackage[T1]{fontenc}
\usepackage{ae,aecompl}
\usepackage{graphicx}	
\usepackage{amsmath}	
\usepackage{amssymb}	
\usepackage{verbatim} 
\usepackage[english]{babel}
\usepackage[dvipsnames]{xcolor}
\usepackage{natbib}
\usepackage{txfonts}
\usepackage{multirow}
\usepackage{pifont}
\usepackage{longtable} 
\usepackage[export]{adjustbox} 
\usepackage{booktabs} 

\title[Photometric study of the three intermediate age open clusters]{Photometric and Kinematic study of the three intermediate age open clusters NGC 381, NGC 2360, and Berkeley 68}

\author[J. Maurya et al.]{
Jayanand Maurya,$^{1,2}$\thanks{E-mail: jayanand@aries.res.in}
Y. C. Joshi$^{1}$
\\
$^{1}$Aryabhatta Research Institute of Observational Sciences, Nainital-263002, India
\\
$^{2}$School of Studies in Physics and Astrophysics, Pandit Ravishankar Shukla University, Chattisgarh 492 010, India
}

\date{Accepted XXX. Received YYY; in original form ZZZ}

\pubyear{2020}

\begin{document}

\maketitle

\begin{abstract}
We present UBVR$_{c}$I$_{c}$ photometric study of three intermediate age open star clusters NGC 381, NGC 2360, and Berkeley 68 (Be 68). We examine the cluster membership of stars using recently released \textit{Gaia} DR2 proper motions and obtain a total of 116, 332, and 264 member stars in these three clusters. The mean reddening of $E(B-V)$ = 0.36$\pm$0.04, 0.08$\pm$0.03, and 0.52$\pm$0.04 mag are found in the direction of these clusters where we observe an anomalous reddening towards NGC 381. We fitted the solar metallicity isochrones to determine age and distance of the clusters which are found to be log(Age) = 8.65$\pm$0.05, 8.95$\pm$0.05, and 9.25$\pm$0.05 yr with the respective distance of 957$\pm$152, 982$\pm$132, and 2554$\pm$387 pc for the clusters NGC 381, NGC 2360, and Be 68. A two-stage power law in the mass function (MF) slope is observed in the cluster NGC 381, however, we observe only a single MF slope in the clusters NGC 2360 and Be68. To study a possible spatial variation in the slope of MF we estimate slopes separately in the inner and the outer regions of these clusters and notice a steeper slope in outer region. The dynamic study of these clusters reveals deficiency of low-mass stars in their inner regions suggesting the mass segregation process in all these clusters. The relaxation times of 48.5, 78.9, and 87.6 Myr are obtained for the clusters NGC 381, NGC 2360, and Be 68, respectively which are well below to their respective ages. This suggests that all the clusters are dynamically relaxed.
\end{abstract}

\begin{keywords}
open clusters: individual: NGC 381, NGC 2360, Be 68-techniques: photometric - stars: formation - stars: luminosity function, mass function, mass-segregation
\end{keywords}       

\section{Introduction}\label{introduction}
Open clusters (OCs) are very important in the studies of stellar evolution as member stars of a cluster are born from the same molecular cloud, therefore have approximately same age, distance, and chemical composition but different stellar mass. By comparing the colour-magnitude diagram (CMD) and two-colour diagram (TCD) of OCs with the theoretical evolutionary models, one can obtain their age, distance, chemical composition and interstellar extinction in the direction of cluster. Along with the knowledge of stellar population distribution and cluster MF, these properties of OCs offer constrain on the stellar evolution models and enable us to understand the star formation processes in the Galaxy \citep{2003ARA&A..41...57L}. The systematic observations of large number of OCs in different locations and environments in the Galaxy can further be used to probe the Galactic structure \citep{1998MNRAS.296.1045C, 2003AJ....125.1397C, 2005MNRAS.362.1259J, 2007MNRAS.378..768J, 2006A&A...445..545P, 2016A&A...593A.116J}.
%
\begin{table*} 
\caption{The values of parameters listed in the \citet{2018A&A...618A..93C} and WEBDA for the clusters in the present study. The values of RA, DEC, l, b, Dist. are from \citet{2018A&A...618A..93C} and remaining parameters are from WEBDA.}
\label{webda}
  \begin{tabular}{c c c c c c c c c c }  
  \hline  
  Cluster & RA&  DEC& l& b& Dia.& E(B-V)& Log(age) &  Dist.\\
          & (deg.)& (deg.)&  (deg.)& (deg.)& ($\arcmin$)& (mag)& (yr)& (pc)\\
  \hline
  NGC 381& 17.094& +61.586& 124.95& -1.22& 6& 0.40& 8.505& 1024\\
  NGC 2360& 109.443& -15.631& 229.80& -1.41& 13& 0.11& 8.75& 970\\
  Berkeley 68& 71.053& +42.134& 162.04& -2.40& 12& 0.671& 8.39& 2437\\
  \hline
  \end{tabular}
\end{table*}

\begin{figure*}  
\vspace{-0.3cm}
\hbox{ 
\hspace{0.0 cm}  
  \includegraphics[width=5.8 cm, height=6.0 cm]{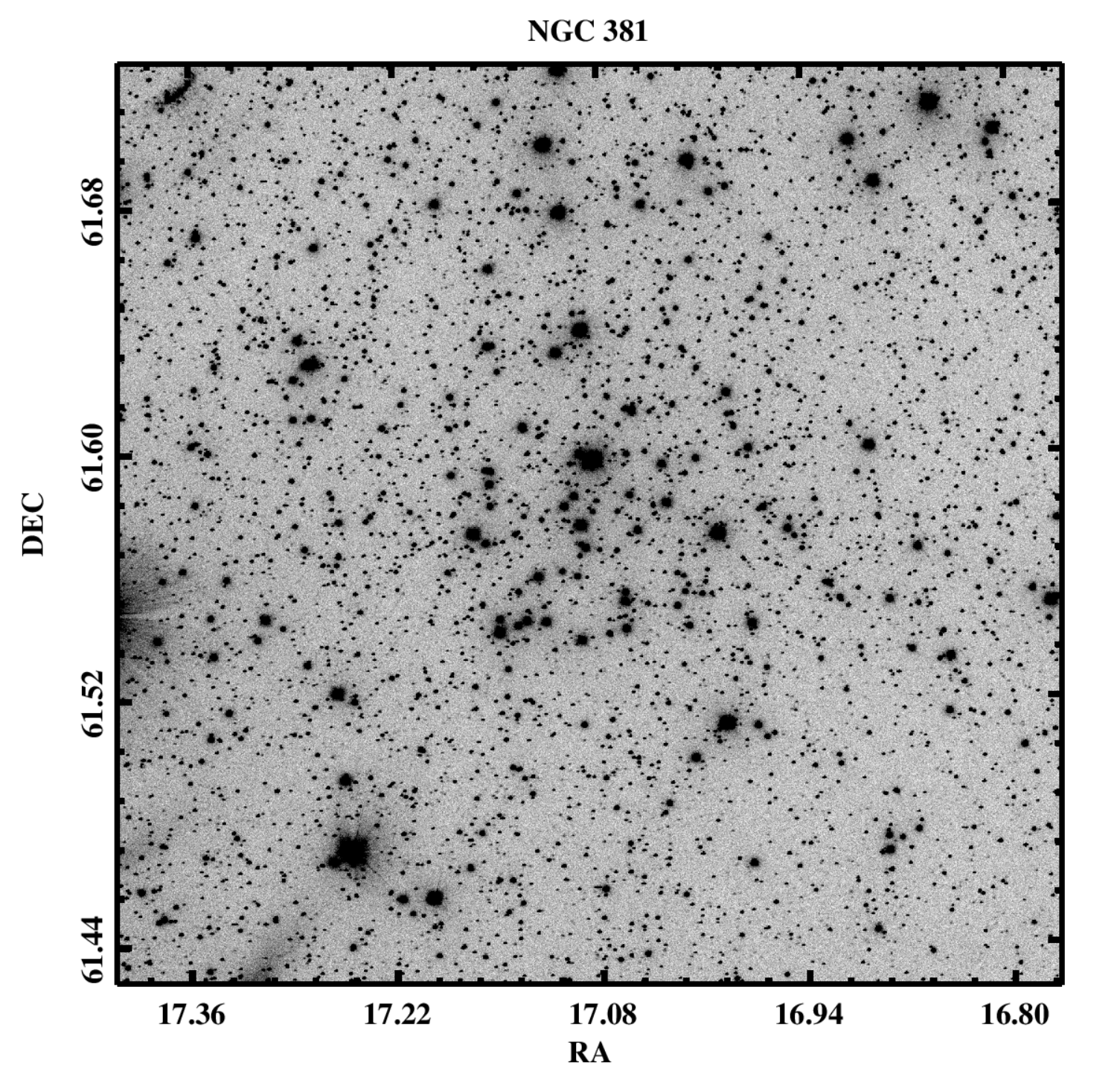}
  \includegraphics[width=5.8 cm, height=6.0 cm]{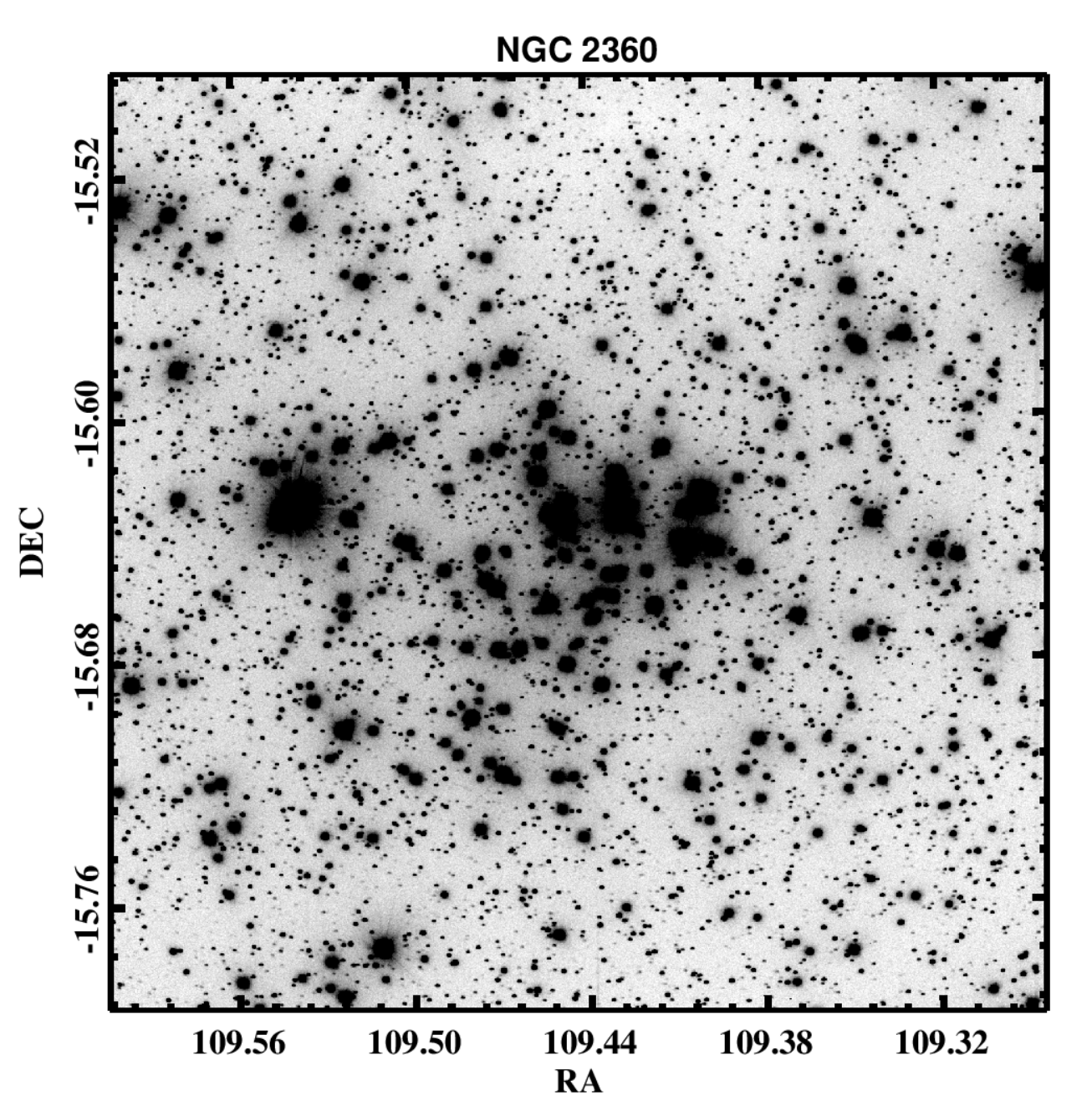}      
  \includegraphics[width=5.8 cm, height=6.0 cm]{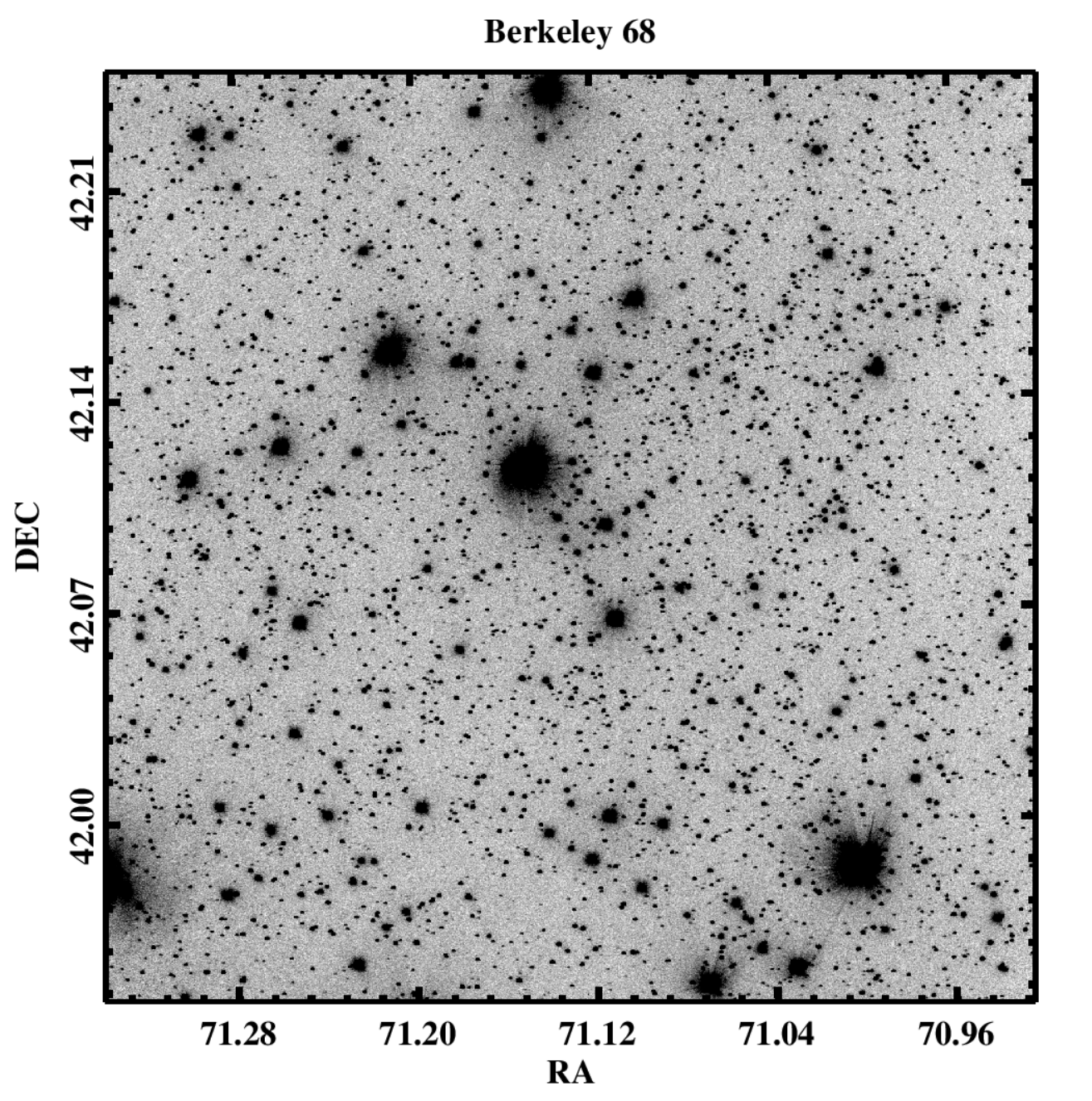}  
  }
\caption{Finding charts of the three clusters studied in this paper. The image sizes are ~$18^{'}$ $\times$ $18^{'}$.}
  \label{fchart}
\end{figure*}
%

Since most of the OCs are likely to be affected by the field star contamination, the knowledge of membership of the cluster stars is absolutely necessary to investigate the cluster properties \citep{2008MNRAS.386.1625C, 2013A&A...558A..53K, 2018A&A...618A..93C}. There are many methods to determine the membership of cluster stars using photometric, proper motions, radial velocities, polarimetric data and their combinations \citep{2012MNRAS.419.2379J, 2013MNRAS.430.1334M, 2014RAA....14..159G}. Recently, with the availability of the \textit{Gaia} DR2 catalog \citep{2018A&A...616A...1G} having unprecedented astrometric precision, the membership determination based on kinematic method (proper motions or radial velocities) becomes a reliable tool to identify the cluster members \citep[e.g.,][]{2018A&A...618A..93C, 2019A&A...624A.126C, 2018A&A...618A..59C, 2019A&A...627A..35C, 2019MNRAS.482.1471B, 2020MNRAS.492.3602J}. The precise knowledge of cluster membership helps us to study the distribution of stellar mass, also known as MF, which has a fundamental importance in the study of dynamic evolution of star clusters. In the recent years, several authors have determined the present day MF for a number of OCs \citep[e.g.,][] {2012AJ....143...41H,2013MNRAS.434.3236K,2017MNRAS.464.1738D,2018Ap&SS.363..143B,2020MNRAS.492.3602J}. It is however still not clear if the IMF is universal in time and space or depends upon different star forming conditions.

The dynamic evolution processes such as stellar encounters and interactions with the Galactic tidal field cause loss of a significant fraction of the cluster stars to the field regions. The tidal fields sometimes cause a star cluster to have longer dynamical relaxation time and a lower degree of mass segregation by reducing star cluster expansion ability. Tidal stripping also plays a role in reducing observation of mass segregation by preferentially removing low mass stars from the outskirts of the tidally perturbed cluster \citep{2018MNRAS.479.3708W, 2019ApJ...877...37R}. Further studies are required to draw a firm conclusion whether mass segregation is a result of dynamic evolution or an imprint of star formation process itself \citep{2016MNRAS.459L.119P}. The study of mass segregation in OCs can be carried out by studying mass distribution and variation in the slope of the MF with the radial distances \citep{2013AJ....145...46L}. The intermediate age OCs are spread in most regions of the Galactic disk and have a wide range in their ages \citep{2016A&A...593A.116J}, therefore, they are very important in the understanding of stellar evolution and dynamical evolution of the stars, particularly having intermediate to low masses. 

In this paper, we present a comprehensive study of the three intermediate age OCs, namely, NGC 381, NGC 2360, and Be 68. Some basic parameters of these clusters are summarized in Table~\ref{webda}. This research is the continuation of our ongoing efforts to investigate some poorly studied OCs in the Galaxy \citep{2012MNRAS.419.2379J, 2014MNRAS.437..804J, 2020MNRAS.492.3602J}. The CCD photometric study of NGC 381 has been done previously by \citet{1994ApJS...90...31P} and \citet{2002AJ....123..905A} in UBV and UBVI filters, respectively. The physical parameters of the cluster has also been derived by \citet{2019A&A...623A.108B}. The significant disagreement in the cluster ages derived in the mentioned studies as well as no previous study of radius, dynamic evolution and MF of this cluster makes it very interesting target for the present study. The photometric study of NGC 2360 has been carried out by \citet{2015NewA...34..195O}. However, further study of spatial structure, MF, and dynamic evolution is required to understand the star formation and dynamic evolution scenario in this cluster. NGC 2360 is also included recently in the list of clusters with an extended Main Sequence Turn Off  \citep{2018ApJ...863L..33M,2018ApJ...869..139C,2019AJ....158...35S,2020MNRAS.491.2129D}. Be 68 is listed as an intermediate age OC in the WEBDA\footnote{https://webda.physics.muni.cz} but no photometric study of this cluster has been reported so for. Therefore, we carry out photometric analysis of these three intermediate age OCs in the present study. The finding charts of these three clusters are shown in Figure~\ref{fchart}.

This paper is organized as follows. We have described observation and data analysis in Section~\ref{observ}. The structural parameters are studied in the Section~\ref{struct}. The derivation of cluster parameters like reddening, age, distance have been carried out in Section~\ref{param}. The Section~\ref{dynamic} is devoted to the dynamic evolution of the clusters. Finally, we discuss results of the present study in Section~\ref{discussion} and summarized our work in Section~\ref{summary}.
\section{Observations and Data Analysis}\label{observ}
\begin{table} 
\caption{The observation log for the three selected clusters.}
  \label{log}
  \begin{tabular}{c c c c c c}  
  \hline  
  Cluster &  Date& Filter&  Exposure time (sec) \\
  &&  &  ($\times$ no. of exposures)\\
  \hline
  NGC 381 &  21 Oct. 2017 &  U&  400$\times$2, 500 \\
  &  &  B&  300$\times$3, 30$\times$3  \\
  &  &  V&  200$\times$3, 20$\times$3  \\
  &  &  R&  100$\times$3, 10$\times$3 \\
  &  &  I&  60, 80$\times$2, 10$\times$2 \\ \\
 NGC 2360 &  13 Jan. 2018 &  U&  260, 300$\times$2 \\ 
  &  &  B&  200$\times$3, 20, 30 \\
  &  &  V&  150$\times$3, 20$\times$3 \\
  &  &  R&  80$\times$3, 10$\times$3 \\
  &  &  I&  40, 60$\times$2, 8$\times$2, 10 \\ \\ 
Berkeley 68 &  21 Oct. 2017 &  U&  400$\times$2, 500 \\ 
  &  &  B&  300$\times$2, 400 \\
  &  &  V&  200$\times$3, 20$\times$3 \\
  &  &  R&  100$\times$2, 10$\times$2 \\
  &  &  I&  60$\times$3, 8$\times$3 \\ \hline
  \end{tabular}
\end{table}

\subsection{ Observations and calibration}
The observations in Johnson UBV and Cousins RI filters of three selected clusters were obtained using 1.3-m Devasthal Fast Optical Telescope (DFOT) at Devasthal, India. The telescope is equipped with a 2k$\times$2k CCD camera having a field of view of $\sim$ $18^{\prime}$ $\times$ $18^{\prime}$. The gain of the CCD is 2 e$^{-}$/ADU and read out noise is 6.5 e$^{-}$. We made master bias by averaging numerous bias frames taken on each observing night. We took multiple twilight sky flat fields with small offset on the sky between successive flats in each filter to avoid any star in the master flat. We also observed Landolt's standard fields at different airmasses in each band to obtain a reliable estimate of the atmospheric extinction coefficients. The observational log is given in Table~\ref{log}. Raw photometric data was processed using IRAF\footnote{IRAF is distributed by the National Optical Astronomy Observatory, which is operated by the Association of Universities for Research in Astronomy under cooperative agreement with the National Science Foundation.}. Instrumental magnitudes of the stars were determined by carrying out aperture photometry in the frames using DAOPHOT II \citep{1987PASP...99..191S}. Calibration of the instrumental magnitudes to the standard system was done using the procedure outlined by \citet{1992ASPC...25..297S}. For the conversion, the following transformation equations were used:\\

\noindent u = U + A$_{0}$ + A$_{1}$(U-B) + A$_{2}$X\\
b = B + B$_{0}$ + B$_{1}$(B-V) + B$_{2}$X\\
v = V + C$_{0}$ + C$_{1}$(B-V) + C$_{2}$X\\
r = R + D$_{0}$ + D$_{1}$(R-I) + D$_{2}$X\\
i = I + E$_{0}$ + E$_{1}$(V-I) + E$_{2}$X\\

\noindent where u, b, v, r, and i denote the aperture instrumental magnitudes and U, B, V, R, and I represent the standard magnitudes. The airmass is denoted by X. The values obtained for these coefficients on the two photometric nights are given in Table~\ref{coeff}. Using the transformation equations, we converted all the instrumental magnitudes obtained for 5 different filters to the respective standard magnitudes of the stars. The mean photometric error in each magnitude bin are summarized in Table~\ref{mag_err}. For all the three clusters, we retrieved stars down to 22 mag in U, B, and V bands but 20-21 mag in $R$ and $I$ passbands. However, the photometric error is higher towards fainter stars which is well expected.
\begin{table} 
\caption{The values of standardization coefficients obtained on two different observing nights.}
\label{coeff}
  \begin{tabular}{c c c c c c}  
  \hline  
  Date & Filter& Zero& Color& Extinction \\

 & & point& coefficient& coefficient\\

\hline

 21 Oct. 2017 &  U&4.895$\pm$0.015&0.024$\pm$0.020&0.545$\pm$0.023\\
  & B&3.193$\pm$0.011&-0.127$\pm$0.013&0.289$\pm$0.011\\
  & V&2.478$\pm$0.008&0.028$\pm$0.010&0.194$\pm$0.010\\
  & R&2.002$\pm$0.012&0.215$\pm$0.026&0.143$\pm$0.010\\
  & I&2.566$\pm$0.013&-0.001$\pm$0.013&0.109$\pm$0.012\\ \\ 
  13 Jan. 2018 & U&4.873$\pm$0.007&-0.068$\pm$0.007&0.452$\pm$0.008\\
  & B&3.231$\pm$0.006&-0.170$\pm$0.005&0.198$\pm$0.006\\
  & V&2.464$\pm$0.008&0.067$\pm$0.006&0.135$\pm$0.009\\
  & R&2.079$\pm$0.007&0.102$\pm$0.009&0.098$\pm$0.007\\
  & I&2.617$\pm$0.007&-0.032$\pm$0.004&0.064$\pm$0.008\\ \hline
  \end{tabular}
\end{table}


\hspace{-1 cm}
\begin{table} \fontsize{6.8}{7.0}\selectfont
\caption{The average photometric error ($\sigma$) and number of stars having error estimation, given in brackets, in each magnitude bin of UBVRI bands for the three clusters.}
\label{mag_err}
  \begin{tabular}{l l l l l l l l l l l}  
  \hline  
 Mag.&\multicolumn{5}{c}{NGC 381} \\
    \cmidrule(lr){2-6}
range&\hspace{-0.2 cm} $\sigma_{U}$&\hspace{-0.2 cm} $\sigma_{B}$&\hspace{-0.2 cm} $\sigma_{V}$&\hspace{-0.2 cm} $\sigma_{R}$&\hspace{-0.2 cm} $\sigma_{I}$ \\ 
\hline
  \hspace{-0.2 cm} 10-11&\hspace{-0.2 cm} -&\hspace{-0.2 cm} 0.002 (1)&\hspace{-0.2 cm} 0.017 (1)&\hspace{-0.2 cm} 0.012 (2)&\hspace{-0.2 cm} 0.009 (6) \\
  \hspace{-0.2 cm} 11-12&\hspace{-0.2 cm} 0.003 (4)&\hspace{-0.2 cm} 0.006 (5)&\hspace{-0.2 cm} 0.007 (11)&\hspace{-0.2 cm} 0.003 (17)&\hspace{-0.2 cm} 0.003 (27) \\
  \hspace{-0.2 cm} 12-13&\hspace{-0.2 cm} 0.004(17)&\hspace{-0.2 cm} 0.029 (15)&\hspace{-0.2 cm} 0.016 (28)&\hspace{-0.2 cm} 0.003 (44)&\hspace{-0.2 cm} 0.020 (66) \\
  \hspace{-0.2 cm} 13-14&\hspace{-0.2 cm} 0.004 (21)&\hspace{-0.2 cm} 0.002 (34)&\hspace{-0.2 cm} 0.003 (68)&\hspace{-0.2 cm} 0.004 (89)&\hspace{-0.2 cm} 0.004 (142) \\
  \hspace{-0.2 cm} 14-15&\hspace{-0.2 cm} 0.004 (56)&\hspace{-0.2 cm} 0.002 (59)&\hspace{-0.2 cm} 0.018 (99)&\hspace{-0.2 cm} 0.005 (187)&\hspace{-0.2 cm} 0.005 (252) \\
  \hspace{-0.2 cm} 15-16&\hspace{-0.2 cm} 0.005 (75)&\hspace{-0.2 cm} 0.003 (109)&\hspace{-0.2 cm} 0.005 (226)&\hspace{-0.2 cm} 0.006 (314)&\hspace{-0.2 cm} 0.008 (485)\\
  \hspace{-0.2 cm} 16-17&\hspace{-0.2 cm} 0.007 (155)&\hspace{-0.2 cm} 0.005 (212)&\hspace{-0.2 cm} 0.006 (351)&\hspace{-0.2 cm} 0.008 (567)&\hspace{-0.2 cm} 0.009 (883)\\
  \hspace{-0.2 cm} 17-18&\hspace{-0.2 cm} 0.013 (278)&\hspace{-0.2 cm} 0.008 (346)&\hspace{-0.2 cm} 0.010 (638)&\hspace{-0.2 cm} 0.012 (989)&\hspace{-0.2 cm} 0.016 (1291)\\
  \hspace{-0.2 cm} 18-19&\hspace{-0.2 cm} 0.025 (469)&\hspace{-0.2 cm} 0.014 (601)&\hspace{-0.2 cm} 0.017 (1059)&\hspace{-0.2 cm} 0.021 (1355)&\hspace{-0.2 cm} 0.032 (912)\\
  \hspace{-0.2 cm} 19-20&\hspace{-0.2 cm} 0.060 (770)&\hspace{-0.2 cm} 0.025 (962)&\hspace{-0.2 cm} 0.029 (1279)&\hspace{-0.2 cm} 0.035 (550)&\hspace{-0.2 cm} 0.063 (60)\\
  \hspace{-0.2 cm} 20-21&\hspace{-0.2 cm} 0.121 (326)&\hspace{-0.2 cm} 0.052 (1237)&\hspace{-0.2 cm} 0.053 (397)&\hspace{-0.2 cm} 0.063 (8)&\hspace{-0.2 cm} -\\
  \hspace{-0.2 cm} 21-22&\hspace{-0.2 cm} 0.369 (12)&\hspace{-0.2 cm} 0.106 (569)&\hspace{-0.2 cm} 0.094 (4)&\hspace{-0.2 cm} -&\hspace{-0.2 cm} -\\
  \hline 
  &\multicolumn{5}{c}{NGC 2360} \\
  \cmidrule (lr){2-6}
  \hspace{-0.2 cm} 10-11&\hspace{-0.2 cm} -&\hspace{-0.2 cm} -&\hspace{-0.2 cm} 0.002 (7)&\hspace{-0.2 cm} 0.003 (23)&\hspace{-0.2 cm} 0.040 (24)\\
  \hspace{-0.2 cm} 11-12&\hspace{-0.2 cm} 0.004 (9)&\hspace{-0.2 cm} 0.014 (18)&\hspace{-0.2 cm} 0.003 (43)&\hspace{-0.2 cm} 0.004 (48)&\hspace{-0.2 cm} 0.043 (60)\\
  \hspace{-0.2 cm} 12-13&\hspace{-0.2 cm} 0.005 (64)&\hspace{-0.2 cm} 0.019 (73)&\hspace{-0.2 cm} 0.004 (84)&\hspace{-0.2 cm} 0.005 (90)&\hspace{-0.2 cm} 0.051 (118)\\
  \hspace{-0.2 cm} 13-14&\hspace{-0.2 cm} 0.007 (94)&\hspace{-0.2 cm} 0.021 (88)&\hspace{-0.2 cm} 0.005 (104)&\hspace{-0.2 cm} 0.006 (140)&\hspace{-0.2 cm} 0.048 (166)\\
  \hspace{-0.2 cm} 14-15&\hspace{-0.2 cm} 0.006 (110)&\hspace{-0.2 cm} 0.022 (116)&\hspace{-0.2 cm} 0.007 (153)&\hspace{-0.2 cm} 0.006 (190)&\hspace{-0.2 cm} 0.060 (288)\\
  \hspace{-0.2 cm} 15-16&\hspace{-0.2 cm} 0.008 (119)&\hspace{-0.2 cm} 0.026 (144)&\hspace{-0.2 cm} 0.008 (267)&\hspace{-0.2 cm} 0.009 (363)&\hspace{-0.2 cm} 0.056 (458)\\
  \hspace{-0.2 cm} 16-17&\hspace{-0.2 cm} 0.011 (206)&\hspace{-0.2 cm} 0.029 (287)&\hspace{-0.2 cm} 0.012 (371)&\hspace{-0.2 cm} 0.012 (511)&\hspace{-0.2 cm} 0.061 (721)\\
  \hspace{-0.2 cm} 17-18&\hspace{-0.2 cm} 0.021 (274)&\hspace{-0.2 cm} 0.039 (314)&\hspace{-0.2 cm} 0.019 (549)&\hspace{-0.2 cm} 0.020 (754)&\hspace{-0.2 cm} 0.068 (895)\\
  \hspace{-0.2 cm} 18-19&\hspace{-0.2 cm} 0.047 (362)&\hspace{-0.2 cm} 0.057 (559)&\hspace{-0.2 cm} 0.033 (755)&\hspace{-0.2 cm} 0.036 (838)&\hspace{-0.2 cm} 0.082 (516)\\
  \hspace{-0.2 cm} 19-20&\hspace{-0.2 cm} 0.084 (469)&\hspace{-0.2 cm} 0.065 (651)&\hspace{-0.2 cm} 0.058 (757)&\hspace{-0.2 cm} 0.057 (320)&\hspace{-0.2 cm} 0.099 (35)\\
  \hspace{-0.2 cm} 20-21&\hspace{-0.2 cm} 0.147 (96)&\hspace{-0.2 cm} 0.073 (713)&\hspace{-0.2 cm} 0.052 (206)&\hspace{-0.2 cm} 0.047 (12)&\hspace{-0.2 cm} 0.116 (1)\\
  \hspace{-0.2 cm} 21-22&\hspace{-0.2 cm} 0.425 (3)&\hspace{-0.2 cm} 0.128 (333)&\hspace{-0.2 cm} 0.113 (14)&\hspace{-0.2 cm} 0.120 (1)&\hspace{-0.2 cm} -\\
  \hline  
  &\multicolumn{5}{c}{Berkeley 68} \\
  \cmidrule (lr){2-6}
  \hspace{-0.2 cm} 10-11&\hspace{-0.2 cm} 0.008 (1)&\hspace{-0.2 cm} 0.004 (1)&\hspace{-0.2 cm} 0.018 (3)&\hspace{-0.2 cm} 0.004 (4)&\hspace{-0.2 cm} 0.003 (8)\\
  \hspace{-0.2 cm} 11-12&\hspace{-0.2 cm} 0.014 (3)&\hspace{-0.2 cm} 0.004 (3)&\hspace{-0.2 cm} 0.014 (5)&\hspace{-0.2 cm} 0.004 (10)&\hspace{-0.2 cm} 0.006 (12)\\
  \hspace{-0.2 cm} 12-13&\hspace{-0.2 cm} 0.004 (7)&\hspace{-0.2 cm} 0.006 (6)&\hspace{-0.2 cm} 0.003 (18)&\hspace{-0.2 cm} 0.007 (27)&\hspace{-0.2 cm} 0.003 (43)\\
  \hspace{-0.2 cm} 13-14&\hspace{-0.2 cm} 0.030 (14)&\hspace{-0.2 cm} 0.003 (24)&\hspace{-0.2 cm} 0.004 (30)&\hspace{-0.2 cm} 0.006 (67)&\hspace{-0.2 cm} 0.003 (90)\\
  \hspace{-0.2 cm} 14-15&\hspace{-0.2 cm} 0.007 (29)&\hspace{-0.2 cm} 0.004 (33)&\hspace{-0.2 cm} 0.005 (85)&\hspace{-0.2 cm} 0.008 (109)&\hspace{-0.2 cm} 0.003 (157)\\
  \hspace{-0.2 cm} 15-16&\hspace{-0.2 cm} 0.008 (56)&\hspace{-0.2 cm} 0.006 (70)&\hspace{-0.2 cm} 0.006 (143)&\hspace{-0.2 cm} 0.008 (236)&\hspace{-0.2 cm} 0.005 (376)\\
  \hspace{-0.2 cm} 16-17&\hspace{-0.2 cm} 0.011 (104)&\hspace{-0.2 cm} 0.004 (154)&\hspace{-0.2 cm} 0.006 (294)&\hspace{-0.2 cm} 0.010 (464)&\hspace{-0.2 cm} 0.008 (682)\\
  \hspace{-0.2 cm} 17-18&\hspace{-0.2 cm} 0.023 (275)&\hspace{-0.2 cm} 0.006 (301)&\hspace{-0.2 cm} 0.007 (525)&\hspace{-0.2 cm} 0.014 (748)&\hspace{-0.2 cm} 0.016 (872)\\
  \hspace{-0.2 cm} 18-19&\hspace{-0.2 cm} 0.036 (376)&\hspace{-0.2 cm} 0.011 (481)&\hspace{-0.2 cm} 0.014 (730)&\hspace{-0.2 cm} 0.024 (885)&\hspace{-0.2 cm} 0.034 (845)\\
  \hspace{-0.2 cm} 19-20&\hspace{-0.2 cm} 0.078 (585)&\hspace{-0.2 cm} 0.022 (660)&\hspace{-0.2 cm} 0.028 (859)&\hspace{-0.2 cm} 0.042 (616)&\hspace{-0.2 cm} 0.062 (112)\\
  \hspace{-0.2 cm} 20-21&\hspace{-0.2 cm} 0.125 (327)&\hspace{-0.2 cm} 0.055 (732)&\hspace{-0.2 cm} 0.052 (506)&\hspace{-0.2 cm} 0.071 (28)&\hspace{-0.2 cm} -\\
  \hspace{-0.2 cm} 21-22&\hspace{-0.2 cm} 0.283 (15)&\hspace{-0.2 cm} 0.125 (702)&\hspace{-0.2 cm} 0.111 (17)&\hspace{-0.2 cm} -&\hspace{-0.2 cm} -\\
  \hline
  \end{tabular}
\end{table}

\subsection{Completeness of the data}
%
\begin{figure}
  \includegraphics[height=12.0cm, width=8.0cm]{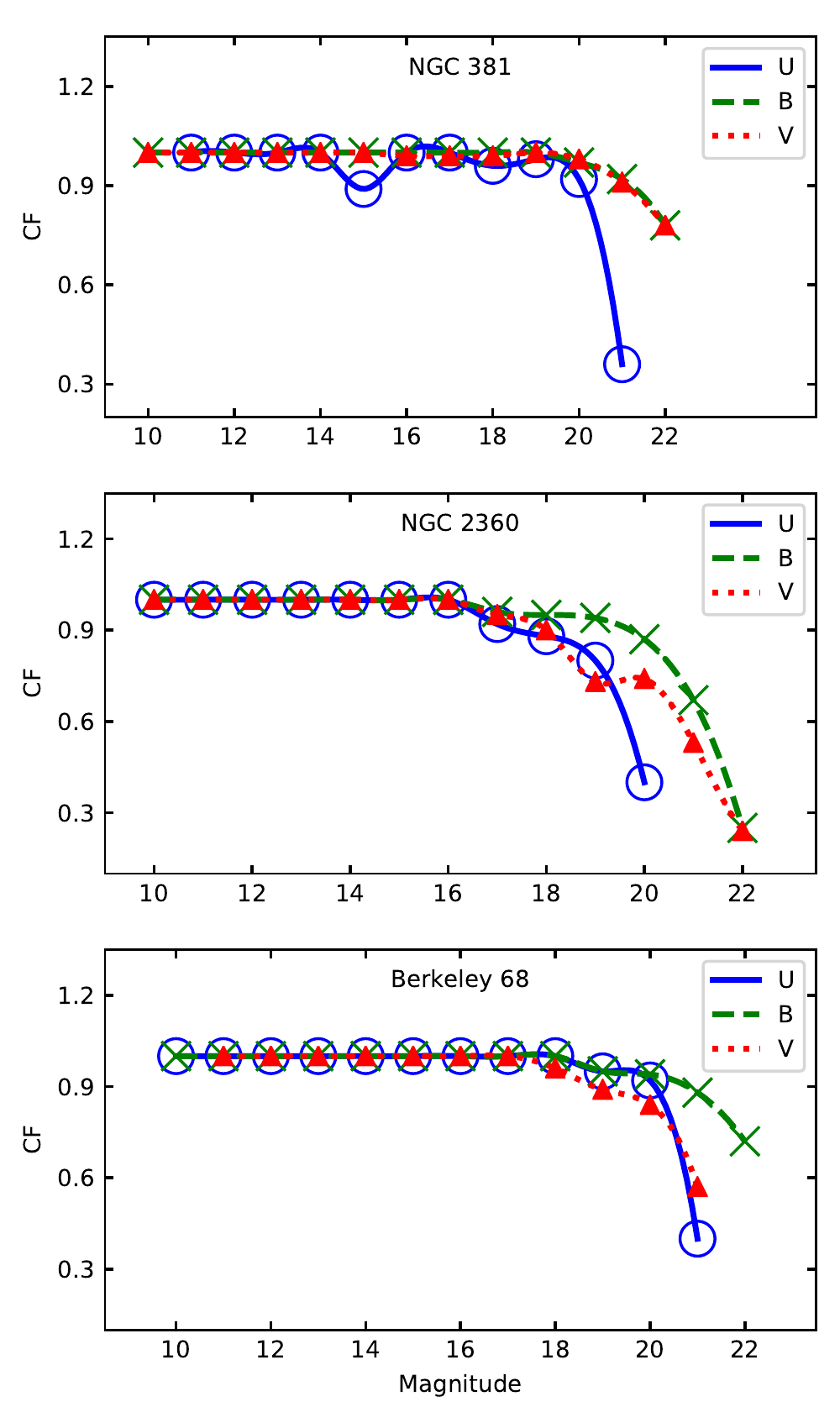} 
\caption{Plots of completeness factor versus magnitude for the open clusters NGC 381, NGC 2360, and Be 68.} 
\label{f_cf}
\end{figure}
%
It may not always be possible to detect every star in the CCD images, particularly for fainter stars. This data incompleteness may arise due to various reasons, e.g., saturation of bright stars, over-crowding of the stars, lack of detection of faint stars due to telescope detection limit, poor observing conditions, etc. The determination of completeness factor (CF) of the data is required in order to derive the luminosity function (LF) and MF of the cluster as well as to estimate the radial density profile (RDP) of the cluster \citep[e.g.,][]{2014MNRAS.437..804J}. We used ADDSTAR routine in DAOPHOT to determine CF in each magnitude bin which is defined as the ratio of number of stars recovered to the number of artificially added stars in corresponding magnitude bin. In this method random artificial stars were added with different, but known, magnitudes and positions to the original frames. Only 10-15 percent of actually detected stars were added in each magnitude bin so that crowding characteristics of the original frame remain unchanged. We added the artificial stars to all bands in a way that they have similar geometrical locations. The images with artificially added stars were re-reduced using the same way as that of original images. A plot of CF versus corresponding magnitude is shown in Figure~\ref{f_cf}. Our photometric data is complete down to 20, 18, and 19 mag in V band for the clusters NGC 381, NGC 2360, and Be 68, respectively.
\subsection{Comparison with previous photometries}
\begin{figure}
   \centering
  \includegraphics[width=8.0 cm, height=5.3 cm]{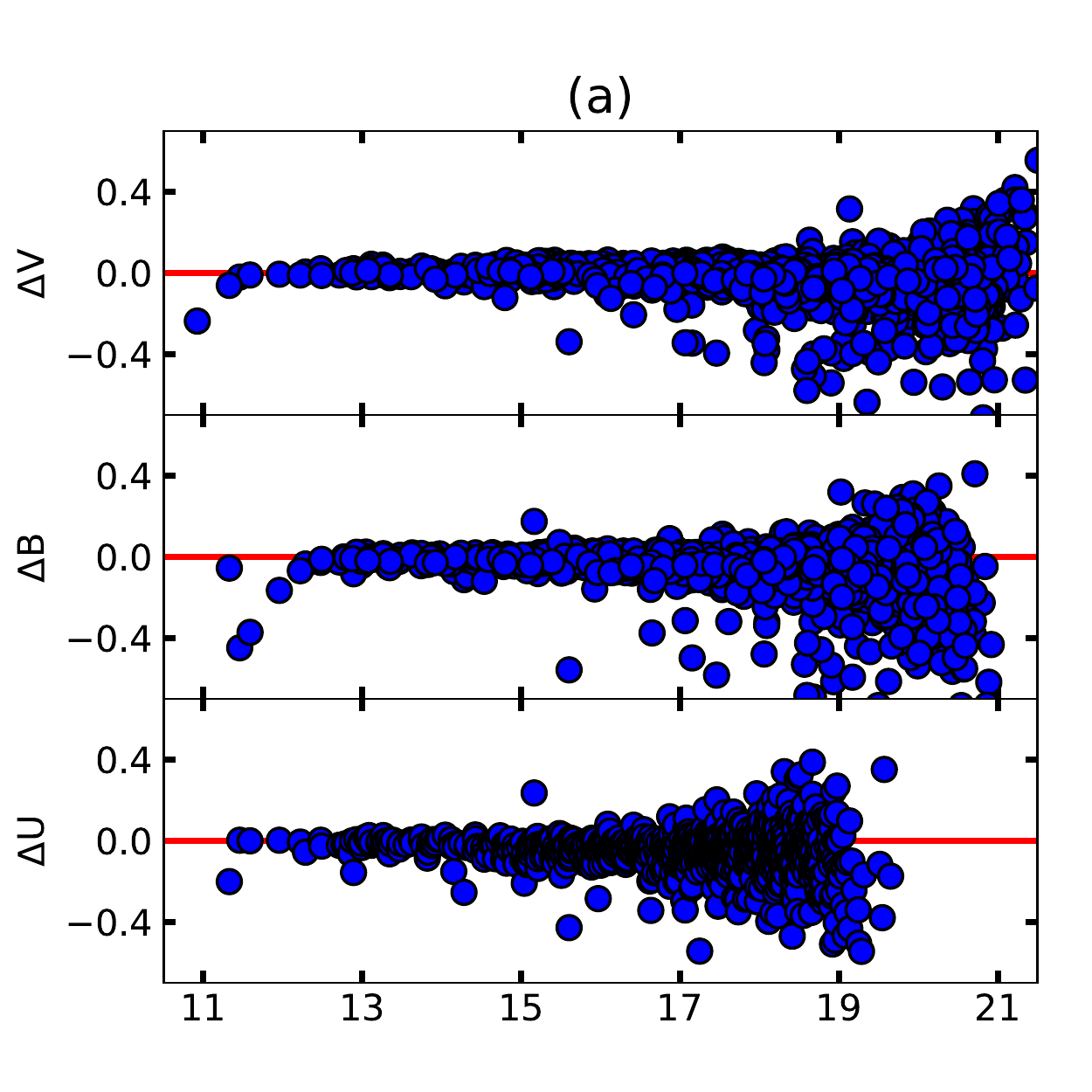} 
  \includegraphics[width=8.0 cm, height=5.3 cm]{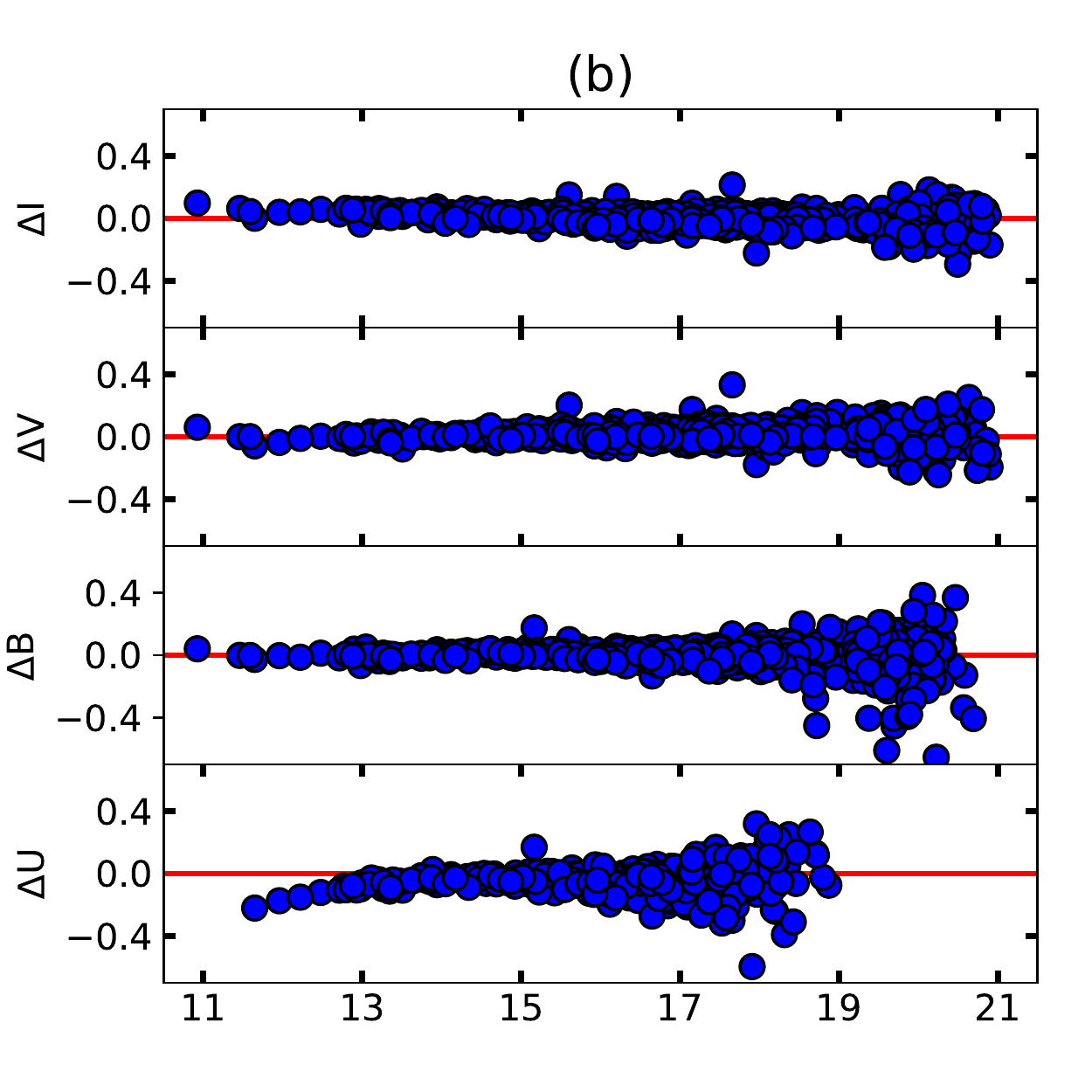}
  \includegraphics[width=8.0 cm, height=5.3 cm]{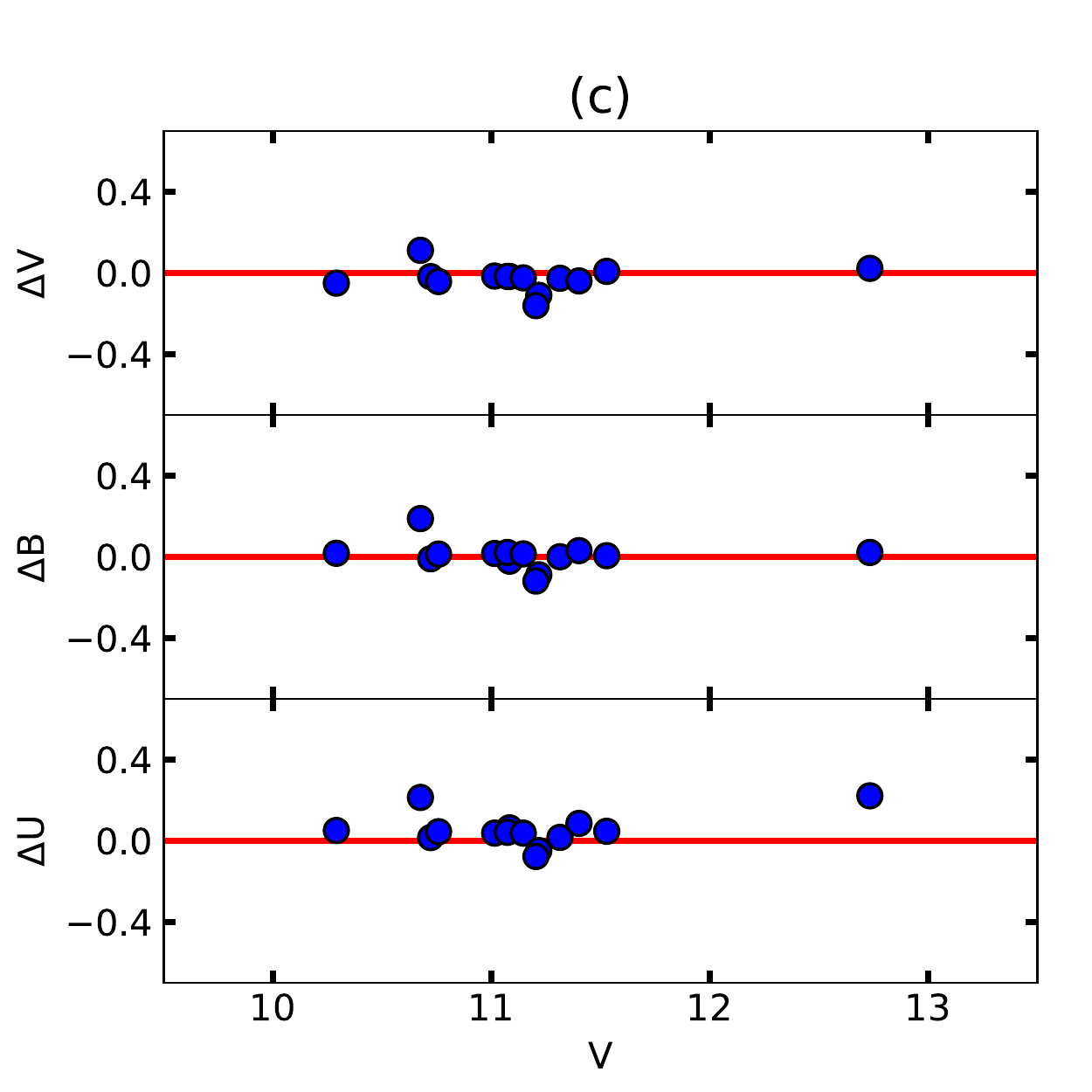} 
\caption{Comparison of present photometric data with data of previous photometric studies as a function of $V$ magnitudes (a) \citet{1994ApJS...90...31P} for NGC 381, (b) \citet{2002AJ....123..905A} for NGC 381, and (c) \citet{2008AN....329..609C} for NGC 2360 (see text for the detail).} 
 \label{comp}
\end{figure}

Photometric studies of NGC 381 have been carried out by \citet{1994ApJS...90...31P} and \citet{2002AJ....123..905A} in UBV and UBVI filters, respectively. Photoelectric data for the cluster NGC 2360 is provided by of \citet{2008AN....329..609C} in UBV bands. We compared present photometric data with the previous studies and plots of V band magnitude versus differences between our data and measurements presented in previous studies in each bands are shown in Figure~\ref{comp}. We found that our data is in good agreement with the previous photometries except there is some scattering in magnitude difference of bright stars of \citet{1994ApJS...90...31P} and a trend in U band towards brighter end of \citet{2002AJ....123..905A}. However, we could not find photometric data for the clusters NGC 2360 and Be 68 to make any comparison. For the user community, we provide photometric catalogue of all the three clusters studied in the present paper and will be available on CDS.
%
%
\subsection{Archived Data}
We used archived near-Infrared data in the $J$, $H$, and $K_s$ magnitudes from the Two Micron All-Sky Survey (2MASS) archive \citep{2006AJ....131.1163S}. The data has limiting magnitude of 15.8, 15.1, and 14.3 mag in J, H, and K$_{s}$ bands, respectively, with a signal-to-noise ratio (S/N) greater than 10. We also used data from the recently released \textit{Gaia} archive DR2 data for the proper motion (PM) studies \citep{2018A&A...616A...1G}. The \textit{Gaia} DR2 provides trigonometric parallaxes with mean parallax error up to 0.04 mas for sources having G $\leq$ 15 mag and around 0.7 mas for sources having G = 20 mag. The DR2 provides proper motions of more than 1.3 billions sources with uncertainties up to 0.06 mas yr$^{-1}$ for the sources having G $<$ 15 mag, 0.2 mas yr$^{-1}$ for G = 17 mag, and 1.2 mas yr$^{-1}$ for sources upto G = 20 mag \citep{2018A&A...616A...1G}. In the following analysis, we use the \textit{Gaia} proper motions and parallaxes to identify the cluster member stars in order to determine various cluster parameters.
\section{Stuctural parameters of the clusters}\label{struct}
\subsection{Spatial structure: cluster size and richness}\label{RDP}
\begin{figure}
   \centering
  \includegraphics[width=9.0 cm,height=14.0 cm]{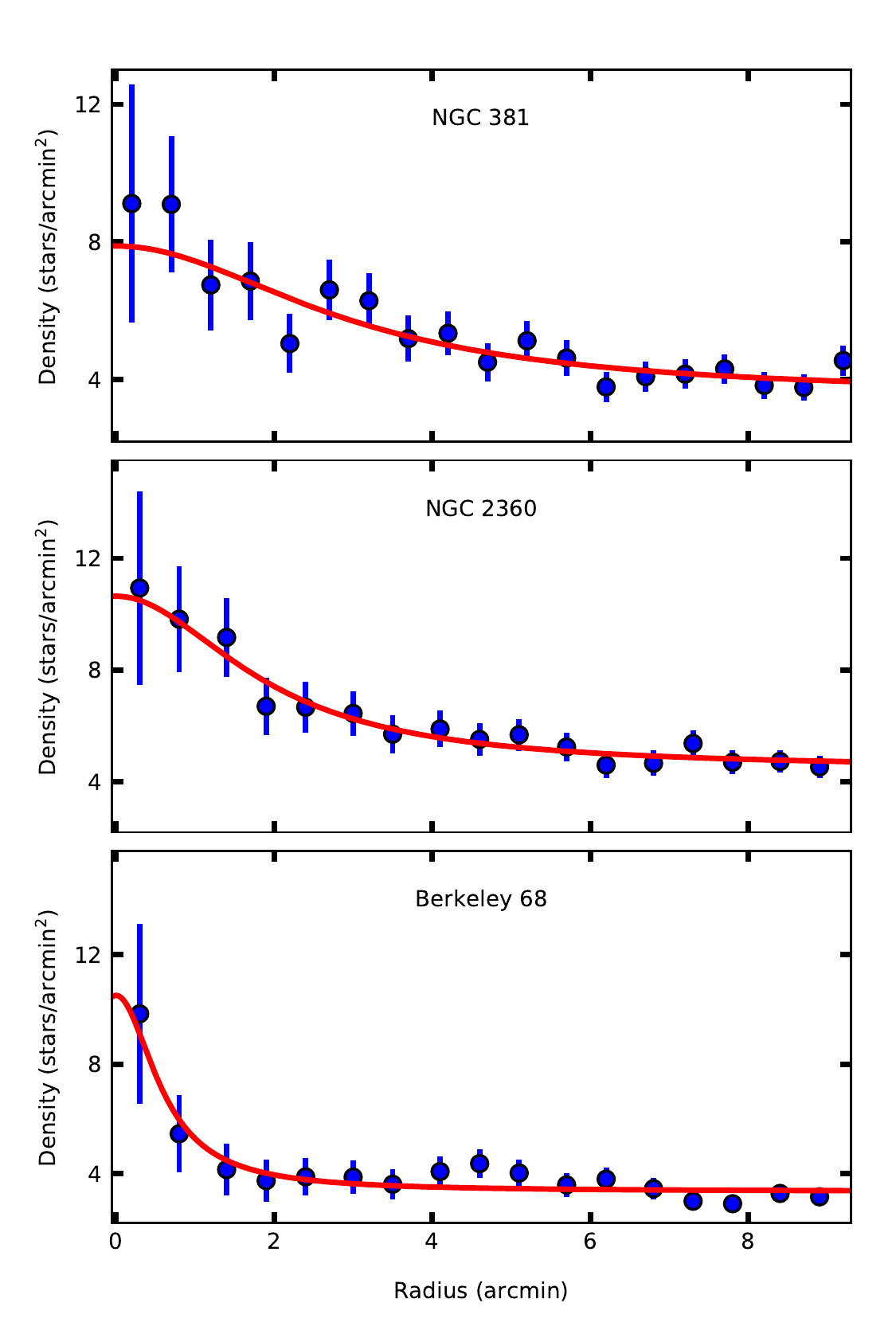}
   \vspace{-0.5cm}
 \caption{Radial density profile for the clusters NGC 381, NGC 2360, and Be 68. The red solid curve represents the best fit of stellar density profile given by \citet{1962AJ.....67..471K}.}
 \label{rdp}
\end{figure}

The knowledge of cluster size is important to study the dynamic evolution of the clusters although an irregular shape of the cluster makes it difficult to determine their precise cluster center and radial extent. To determine cluster center in the present study, we considered all the stars down to the limiting magnitudes where CFs were found to be better than 90$\%$ in these clusters. The cluster center is defined as the pixel coordinate at which stellar density is maximum. We derived the stellar density profile (RDP) by estimating stellar density in concentric annular rings of 0$^{\prime}$.5 width centred at the cluster center. The method used here is described in \citet{2012MNRAS.419.2379J}. We found center of the clusters at the pixel coordinates (1040, 1070), (1100, 1002), and (1052, 1054) which correspond to celestial coordinates (01:08:19.57, +61:35:18.24), (07:17:43.48, -15:38:39.85), and (04:44:29.97, +42:05:55.81) for the clusters NGC 381, NGC 2360, and Be 68, respectively. These derived cluster centers are close to the center derived by \citet{2018A&A...618A..93C} as (01:08:22.56, +61:35:09.6), (07:17:46.32, -15:37:51.6)  for NGC 381 and NGC 2360, respectively but differ significantly in DEC for Be 68 as center reported by \citet{2018A&A...618A..93C} is (04:44:12.72, +42:08:02.4). The determined cluster centres are however close to the center coordinates (01:08:25.92, 61:34:44.40) and (04:44:27.60, 42:06:00.00) reported by \citet{2018AJ....155...91Y} for the clusters NGC 381 and Be 68, respectively obtained using PPMXL data. The stellar density distributions for all the three clusters in the V band are shown in Figure~\ref{rdp}. The spatial structure and radial density of the OCs were derived by fitting RDP given by \citet{1962AJ.....67..471K} and \citet{1992AcA....42...29K} as following: 
$$
\rho(r) = \rho_{f} + \frac{\rho_{0}}{1 + \left(\frac{r}{r_{c}}\right)^{2}}
$$
Here $\rho_{f}$ is the field density and r$_{c}$ is the core radius of the cluster defined as radial distance from the center where the stellar density, $\rho$(r), becomes half of its central value, $\rho_{0}$. We found core radii to be 3$^{\prime}$.0$\pm$0$^{\prime}$.9, 1$^{\prime}$.9$\pm$0$^{\prime}$.4, and 0$^{\prime}$.6$\pm$0$^{\prime}$.3 for the clusters NGC 381, NGC 2360, and Be 68, respectively. A $\chi^{2}$ best fit to the RDP for all the three clusters are shown in Figure~\ref{rdp} where solid curves represent the King profile. We considered cluster boundary as a point in radial direction where $\rho$(r) is 1\,$\sigma$ above $\rho_{f}$. This can be defined as 
\textbf{$$
\rho_{lim} = \rho_{f} + \sigma_{f}
$$}
\noindent where $\sigma_{f}$ is uncertainty in the $\rho_{f}$ value. The radius, r$_{cluster}$, of the cluster can be estimated as
\textbf{$$
r_{cluster} = r_{c} \sqrt{\frac{\rho_{0}}{\sigma_{f}} - 1}
$$}
\noindent Using the best fit in radial density profile of the clusters, we found values of the cluster radii to be 10$^{\prime}$.4$\pm$0$^{\prime}$.1, 12$^{\prime}$.1$\pm$0$^{\prime}$.1, and 4$^{\prime}$.7$\pm$0$^{\prime}$.1 for the clusters NGC 381, NGC 2360 and Be 68, respectively. 

The density contrast parameter, $\delta_{c}$, was determined to estimate the compactness of the clusters. One can determine the density contrast parameter using the following formula:
$$
\delta_{c} = 1 + \frac{\rho_{0}}{\rho_{f}}
$$
Using the stellar density in the cluster and field regions obtained through the radial density profiles, we estimated the values of $\delta_{c}$ as 2.2, 2.4, and 3.14 for the clusters NGC 381, NGC 2360, Be 68, respectively. The resulting values of $\delta_{c}$ are less than the values defined for the compact clusters derived by \citet{2009MNRAS.397.1915B} (7 $\leq \delta_{c} \leq$ 23). This suggests that these three clusters can be assigned as sparse clusters. The derived values of structural parameters are summarized in Table~\ref{struc_par} for these clusters. The open clusters are of irregular shape and may have weak contrast between members and field stars which makes precise determination of cluster center and radial extent very difficult \citep{2002A&A...383..153N,2013A&A...558A..53K}. The cluster radius determined through RDP method only provides an approximate value of the cluster extent, particularly for the sparse clusters, as it not only depends on the limiting magnitude of the observed cluster \citep{2014MNRAS.437..804J} but also on the cut-off point in the profile where cluster density merges with the field density. Therefore, the radii estimated for these clusters are just minimum limit of radial extent of the clusters and stars belonging to these clusters might be found beyond the radii derived by us.
\begin{table}
  \centering
  \caption{The derived values of structural parameters of the clusters}
  \label{struc_par}
  \begin{tabular}{c c c c c c c c c}  
  \hline  
  Cluster &  \multicolumn{2}{c}{Central coordinates} & r$_{c}$    & r$_{cluster}$ & R$_{t}$ & $\delta_{c}$\\
          &  RA (J2000)      & DEC (J2000)           & ($\prime$) &   ($\prime$)  & (pc)    &             \\ \hline
   NGC 381 &01:08:19.6&+61:35:18.2&3.0& 10.4& 6.49&2.2\\
   NGC 2360 &07:17:43.5&-15:38:39.8&1.9& 12.1& 9.46&2.4\\
   Be 68 &04:44:30.0&+42:05:55.8&0.6& 4.7& 9.04&3.1\\ \hline
  \end{tabular}
\end{table}

%
\subsection{Proper motion membership}\label{Membership}

%
\begin{figure}
\centering
\includegraphics[width=8.0cm, height=7.2cm]{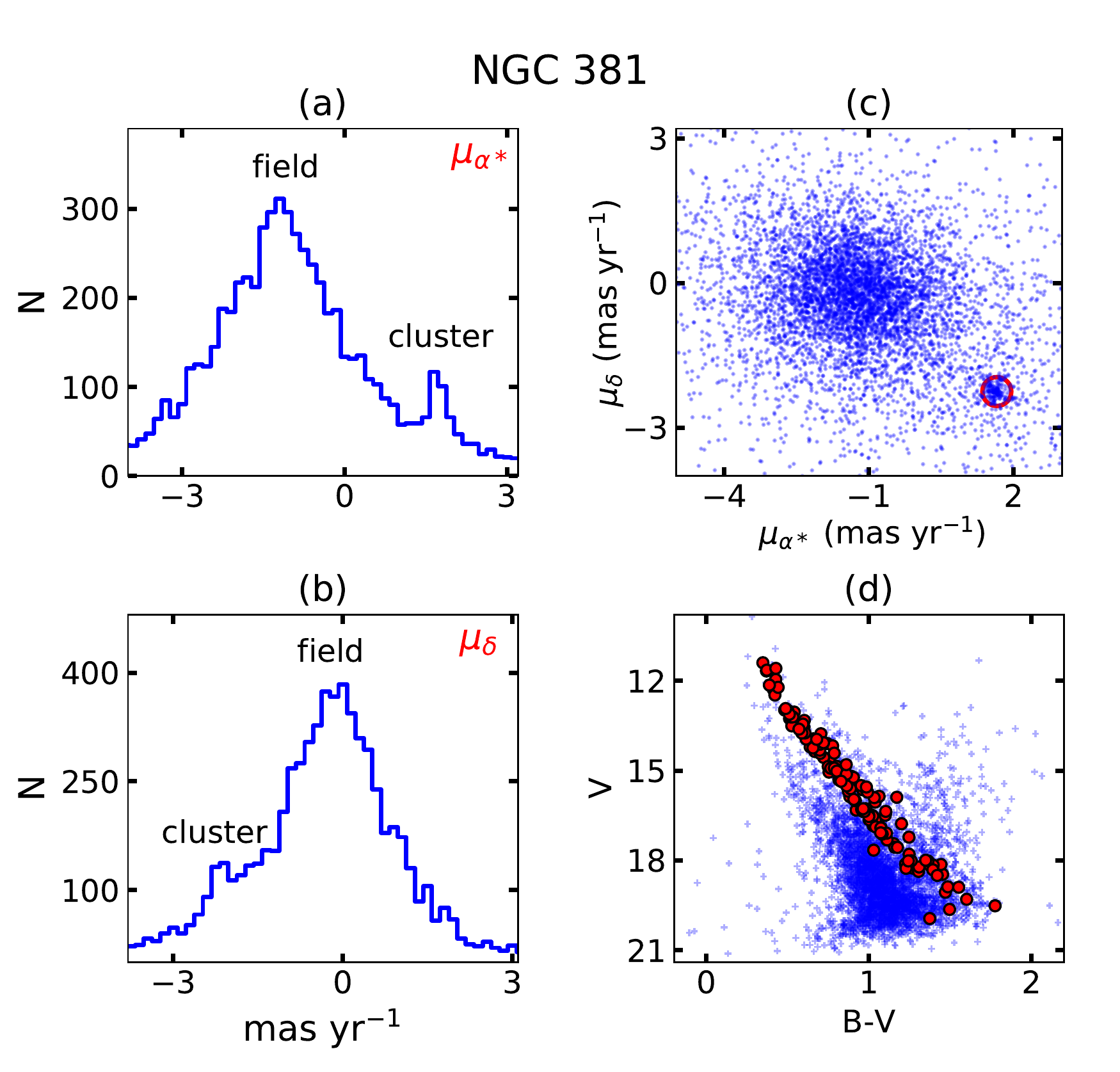}
\includegraphics[width=8.0cm, height=7.2cm]{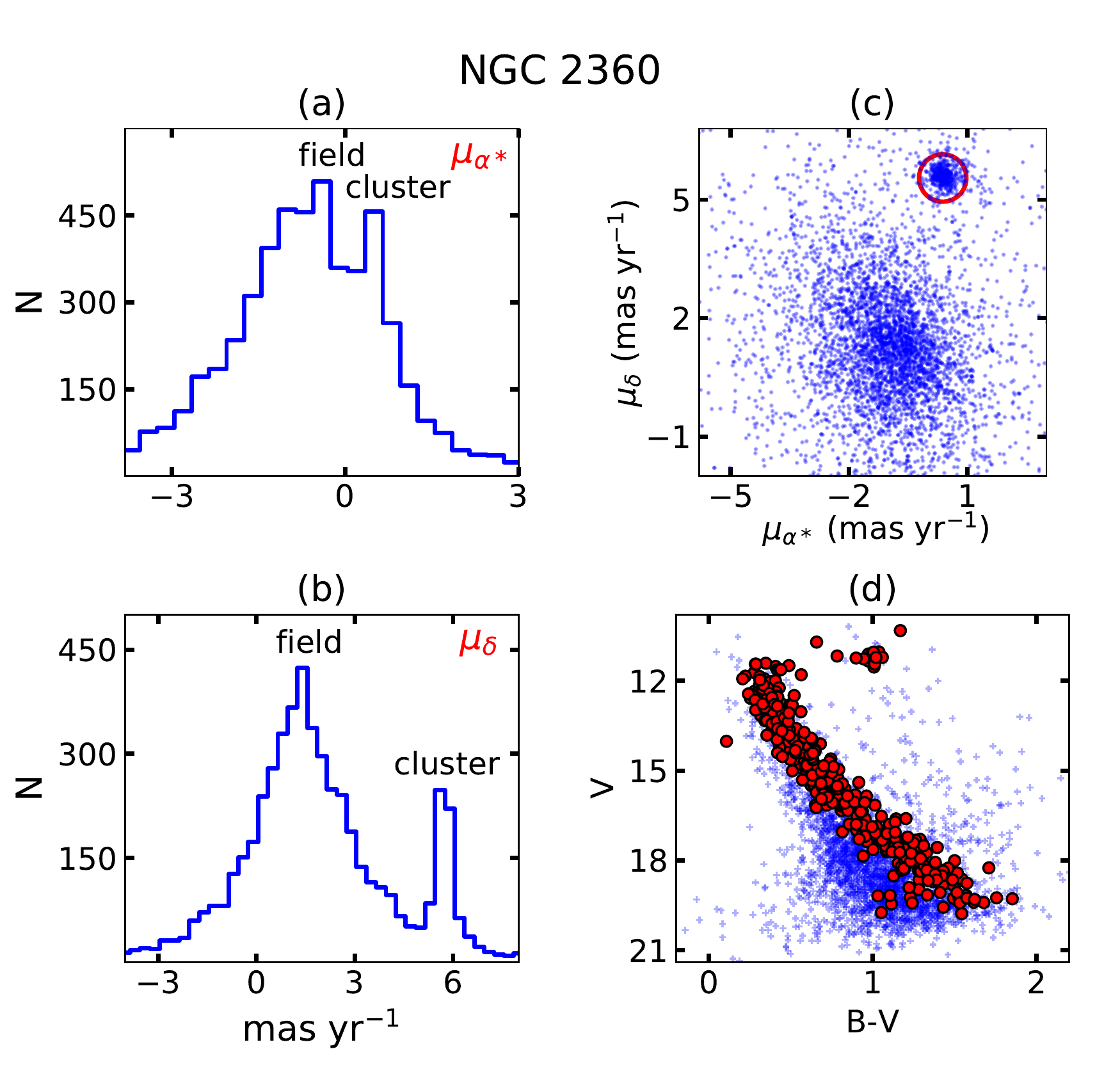}
\includegraphics[width=8.0cm, height=7.2cm]{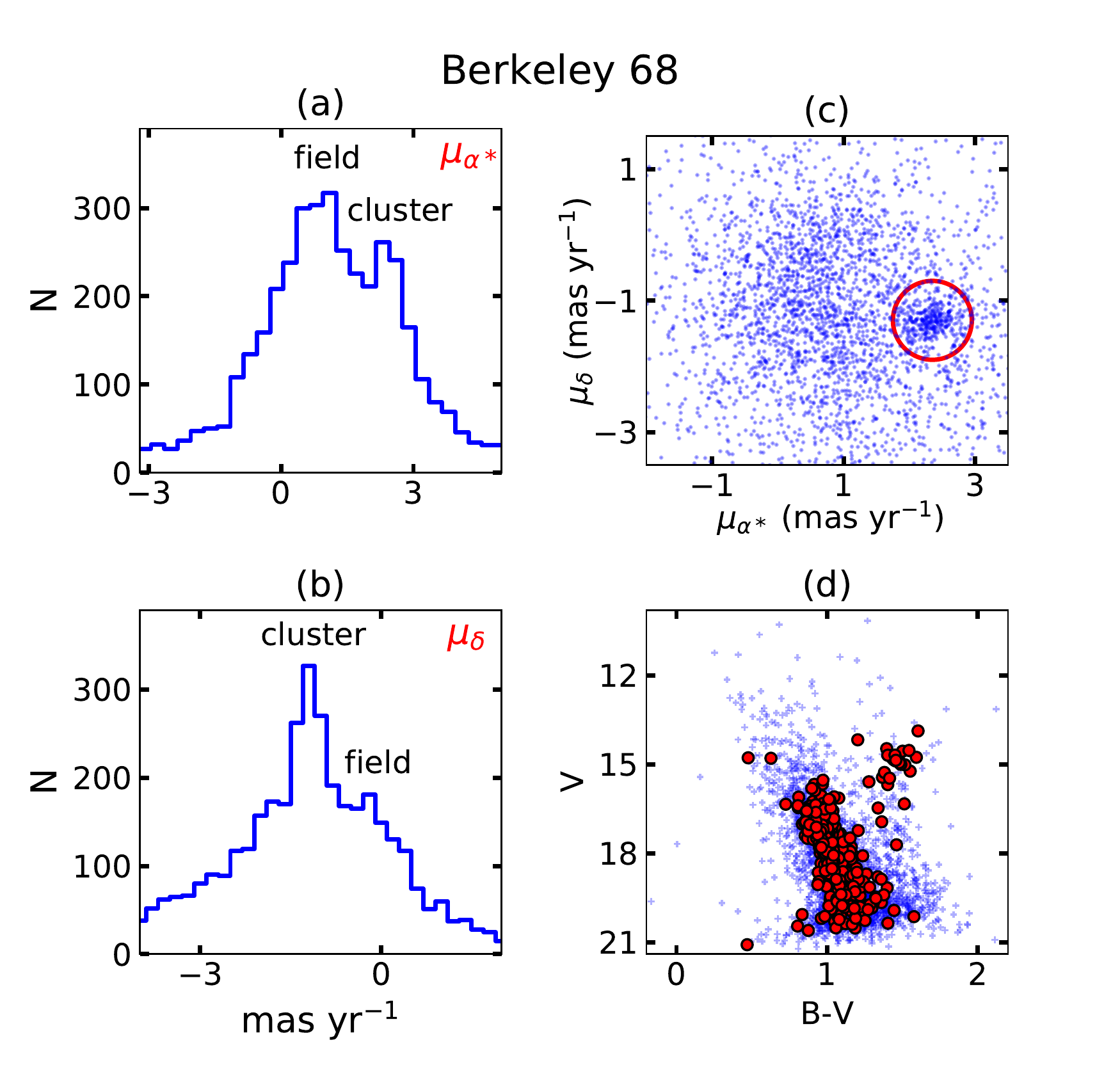}
\caption{Plots for membership determination using VPD constructed from proper motion. The plots for clusters mentioned in the title contains histogram of (a) $\mu_{\alpha*}$ and (b)  $\mu_{\delta}$ as well as (c) VPD and (d) CMD. The histograms have two peak corresponding to cluster and field stars. The region encircled by red circle represents member stars in the clusters. The member stars and field stars are shown by red points and blue '+' signs in CMDs. }
\label{vpd}
\end{figure}

The kinematic study of a cluster is important in determining membership of the cluster stars as determination of the membership is essential in the accurate estimation of cluster parameters \citep[e.g.,][]{2011AJ....142...71C, 2015A&A...584A..59S}. The major difficulty in determination of the cluster parameters using isochrone fitting is contamination by the field stars. The proper motions based vector point diagram is very effective in separating field stars from the member stars of a cluster \citep{2013MNRAS.430.3350Y,2019A&A...623A..22S}. With the availability of the \textit{Gaia} DR2 catalog with unprecedented astrometric precision, kinematic study of the OCs has become more reliable \citep{2018A&A...618A..93C, 2018A&A...618A..59C, 2019A&A...627A..35C, 2019ApJS..245...32L, 2020MNRAS.492.3602J}. We therefore used data from the \textit{Gaia} DR2 archive for proper motion studies of the OCs. We extracted only those stars from the \textit{Gaia} DR2 catalog which were also present in our photometric data taken for the $18^{\prime}$ $\times$ $18^{\prime}$ region around the cluster center. Proper motions in the RA-DEC plane are plotted as Vector-point diagram (VPD) for the stars of three selected clusters in Figure~\ref{vpd}. It is evident from the plot that there are two separate bunches of points in each of these VPDs. After plotting CMD for stars belonging to these two different regions in a cluster, it is apparent that these two regions correspond to field stars and probable cluster members. In the right panels of Figure~\ref{vpd}, we show these probable members by small region within the red circle in the VPDs which are the same stars those lie exactly on the Main-sequence and shown by filled circles in the CMDs of the same figure. To quantify membership of stars we assigned membership probabilities to stars in the clusters which were estimated using a statistical method based on PMs of stars as described in previous studies \citep{1971A&A....14..226S,2019MNRAS.487.3505M, 2020MNRAS.492.3602J}. 
In this method i$^{th}$ star is assigned with membership probability calculated as
$$P_{\mu}(i) = \frac{n_{c}~.~\phi_c^{\nu}(i)}{n_{c}~.~\phi_c^{\nu}(i) + n_f~.~\phi_f^{\nu}(i)}$$
\noindent where $n_{c}$ and $n_{f}$ are the normalized number of stars for cluster and field i.e. $n_c + n_f = 1$. The $\phi_c^{\nu}$ and $\phi_f^{\nu}$ denote the frequency distribution functions for the cluster and field stars. The $\phi_c^{\nu}$ for the i$^{th}$ star is calculated as below:

~~~~~$\phi_c^{\nu}(i) =\frac{1}{2\pi\sqrt{{(\sigma_{\alpha* c}^2 + \epsilon_{\alpha* i}^2 )} {(\sigma_{\delta c}^2 + \epsilon_{\delta i}^2 )}}}$

~~~~~~~~~~~~~~$ exp\{{-\frac{1}{2}[\frac{(\mu_{\alpha* i} - \mu_{\alpha* c})^2}{\sigma_{\alpha* c}^2 + \epsilon_{\alpha* i}^2 } + \frac{(\mu_{\delta i} - \mu_{\delta c})^2}{\sigma_{\delta c}^2 + \epsilon_{\delta i}^2}] }\}$ 

\noindent where $\mu_{\alpha*i}$ and $\mu_{\delta i}$ are the proper motions of $i^{th}$ star. The errors in the proper motions of $i^{th}$ star are denoted by $\epsilon_{\alpha* i}$ and $\epsilon_{\delta i}$. The $\mu_{\alpha* c}$ and $\mu_{\delta c}$ are proper motion center with dispersions $\sigma_{\alpha* c}$ and $\sigma_{\delta c}$ for the cluster. The frequency function $\phi_f^{\nu}(i)$ for the i$^{th}$ is calculated using following formula:

$\phi_f^{\nu}(i) =\frac{1}{2\pi\sqrt{(1-\gamma^2)}\sqrt{{(\sigma_{\alpha* f}^2 + \epsilon_{\alpha* i}^2 )} {(\sigma_{\delta f}^2 + \epsilon_{\delta i}^2 )}}} exp\{{-\frac{1}{2(1-\gamma^2)}}$

$[\frac{(\mu_{\alpha*i} - \mu_{\alpha* f})^2}{\sigma_{\alpha* f}^2 + \epsilon_{\alpha* i}^2} -\frac{2\gamma(\mu_{\alpha*i} - \mu_{\alpha* f})(\mu_{\delta i} - \mu_{\delta f})} {\sqrt{(\sigma_{\alpha* f}^2 + \epsilon_{\alpha* i}^2 ) (\sigma_{\delta f}^2 + \epsilon_{\delta i}^2 )}} + \frac{(\mu_{\delta i} - \mu_{\delta f})^2}{\sigma_{\delta f}^2 + \epsilon_{\delta i}^2}]\}$ 

\noindent where $\mu_{\alpha* f}$ and  $\mu_{\delta f}$ are the proper motion center for the field star with dispersions $\sigma_{\alpha* f}$ and $\sigma_{\delta f}$, respectively. The correlation coefficient $\gamma$ in the above formula is defined as
$$\gamma = \frac{(\mu_{\alpha*i} - \mu_{\alpha* f})(\mu_{\delta i} - \mu_{\delta f})}{\sigma_{\alpha* f}\sigma_{\delta f}}$$
We chose stars having membership probability above 60$\%$ and parallax within 2$\sigma$ of mean parallax of the stars in the cluster region of the VPD as member stars for the clusters NGC 381 and NGC 2360. However, for the cluster Be 68, the stars having membership probability above 60$\%$ and parallax within 1$\sigma$ of mean parallax were chosen as member stars. This is due to the fact that the cluster Be 68 lies relatively at larger distance (see, Table~\ref{webda}) hence have smaller parallax values. We thus identified a total of 116, 332, and 264 stars to be member stars in the clusters NGC 381, NGC 2360, and Be 68, respectively. The membership study of these clusters has also been done by \citet{2018A&A...618A..93C} in which they identified 148, 696, and 247 member stars in the clusters NGC 381, NGC 2360, and Be 68, respectively. These numbers are larger than our estimated numbers for NGC 381 and NGC 2360. If we take 50$\%$ as probability cut-off similar to the probability cut-off used by \citet{2018A&A...618A..93C} then number of members would be 166 and 466 for NGC 381 and NGC 2360, respectively. If we apply additional 3$\sigma$ parallax cut-off then there would be 132 stars in NGC 381 and 375 stars in NGC 2360 having probability greater than 50$\%$. \citet{2018A&A...618A..93C} searched members for NGC 381 and NGC 2360 in wider regions of $\sim$869 and $\sim$4471 arcmin$^{2}$, respectively in comparison to $\sim$324 arcmin$^{2}$ region in our study. Thus the greater number of member stars found in case of NGC 381 and NGC 2360 can be attributed to larger regions chosen for these clusters by \citet{2018A&A...618A..93C}. Using our defined cluster members, we estimated the mean proper motion values for the member stars as well as field stars which are given in Table~\ref{pm}. The PM values are consistent within the errors with those by \citet{2018A&A...618A..93C}. The standard deviation in the mean proper motion of member stars are relatively small which can be attributed to the precise \textit{Gaia} astrommetry.

\begin{table}
  \caption{The mean proper motions obtained for the three clusters in mas yr$^{-1}$. The values mentioned in brackets are from \citet{2018A&A...618A..93C}}
  \label{pm}
  \hspace{-0.5 cm}
  \hbox{ 
  \begin{tabular}{c c c c c c c c c c c}
  \hline
   \hspace{-0.3 cm} &\hspace{-0.3 cm} NGC 381&\hspace{-0.3 cm} NGC 2360&\hspace{-0.3 cm} Be 68\\ 
   \hline
 \hspace{-0.3 cm} \underline{Member stars:}\\
  \hspace{-0.3 cm} $\bar{\mu}_{\alpha*}$&\hspace{-0.3 cm} 1.62$\pm$0.10 (1.60)&\hspace{-0.3 cm} 0.39$\pm$0.18 (0.38)&\hspace{-0.3 cm} 2.29$\pm$0.37 (2.31)\\
 \hspace{-0.3 cm} $\bar{\mu}_{\delta}$ &\hspace{-0.3 cm} -2.25$\pm$0.13 (-2.26)&\hspace{-0.3 cm} 5.59$\pm$0.19 (5.60)&\hspace{-0.3 cm} -1.29$\pm$0.24 (-1.31)\\
 \hspace{-0.3 cm} $\bar{\mu}$ &\hspace{-0.3 cm} 2.78$\pm$0.12&\hspace{-0.3 cm} 5.61$\pm$0.19&\hspace{-0.3 cm} 2.65$\pm$0.34\\
 \hspace{-0.3 cm} \underline{Field stars:}\\
 \hspace{-0.3 cm} $\bar{\mu}_{\alpha*}$ &\hspace{-0.3 cm} -0.75 $\pm$4.10&\hspace{-0.3 cm} -1.03 $\pm$ 2.59&\hspace{-0.3 cm} 1.26$\pm$4.90\\
 \hspace{-0.3 cm} $\bar{\mu}_{\delta}$ &\hspace{-0.3 cm} -0.53 $\pm$ 2.42&\hspace{-0.3 cm} 1.19 $\pm$ 3.30&\hspace{-0.3 cm} -2.15$\pm$4.90\\ 
 \hline
  \end{tabular}
}
\end{table}


\begin{figure}
\includegraphics[width=8.0cm, height=13.0cm]{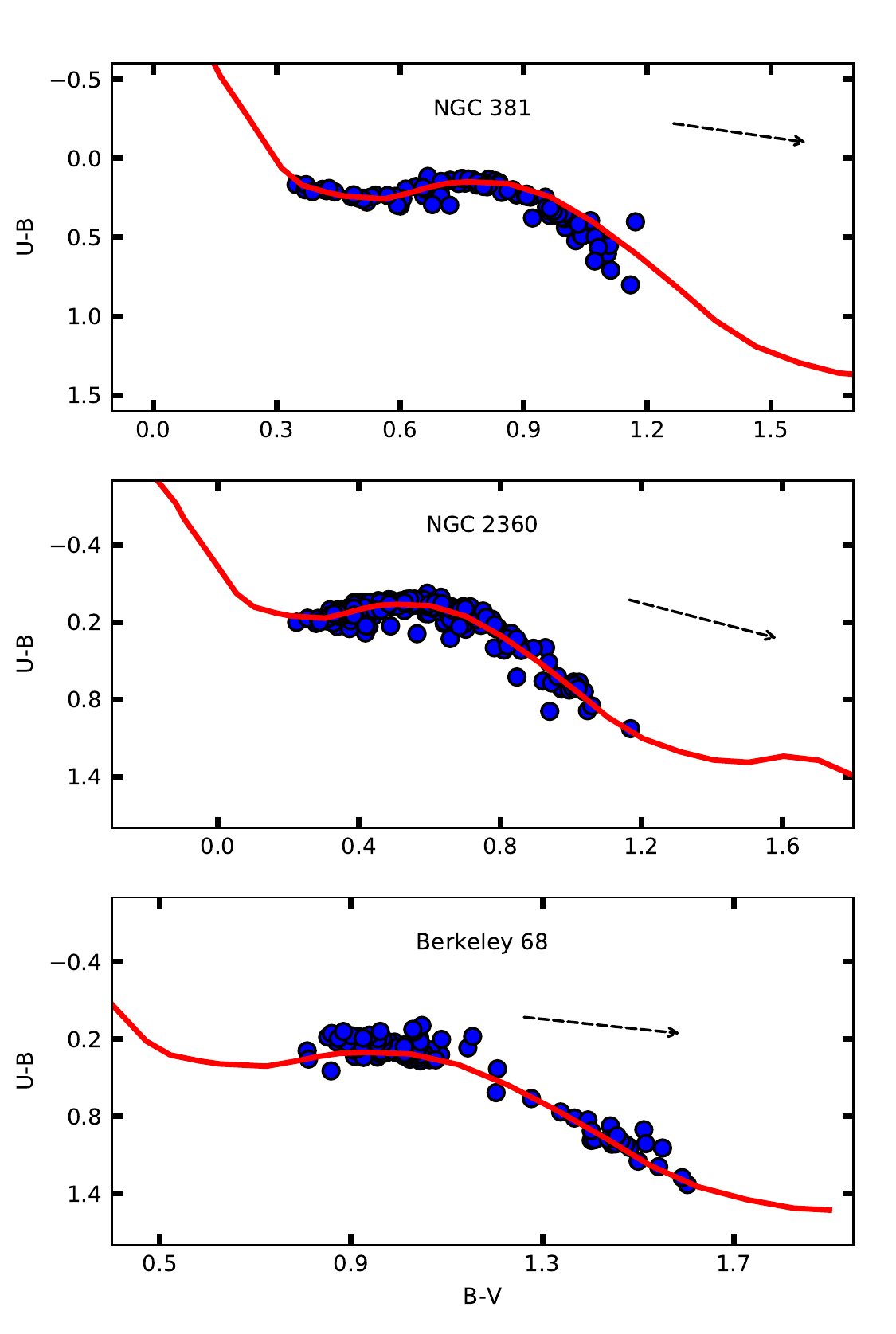}
\caption{The (U-B) versus (B-V) two colour diagrams for the clusters NGC 381, NGC 2360, and Be 68. The solid red lines present best fit of ZAMS isochrone given by \citet{Schmidt-Kaler}. The dashed lines represent slope and direction of the reddening vector.}
\label{reddening}
\end{figure}

\begin{figure*}
\includegraphics[width=16.0 cm, height=12.0 cm]{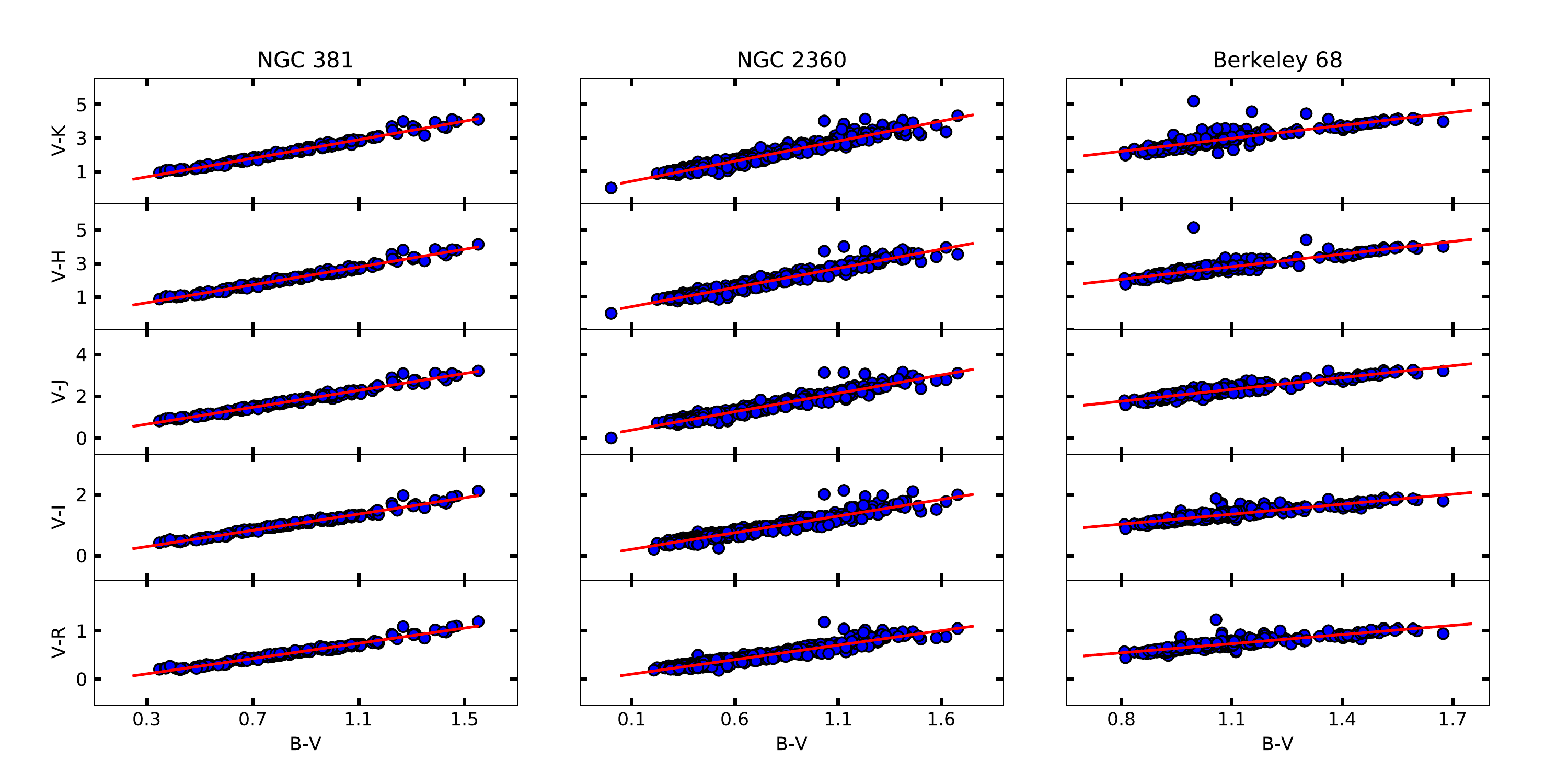}
\caption{The ($\lambda$-V)/(B-V) two-colour diagrams for stars in the regions of clusters NGC 381, NGC 2360 and Be 68. The red solid lines present best fit slopes.}
\label{tcd}
\end{figure*}
%
\section{CLUSTER PARAMETERS}\label{param}

\subsection{Reddening law and two-colour diagrams}\label{rv}
It is necessary to determine the reddening law to estimate interstellar extinction in the direction of the observed clusters. The value of the ratio of the total-to-selective extinction R$_{V}$ can be different for the cluster stars and the foreground stars if the size distribution of the dust is different in the cluster region and that in the interstellar medium along the same line of sight \citep{1990ARA&A..28...37M}. The normal reddening law, R$_{V}$ = $\frac{A_{v}}{E(B-V)}$ is not applicable for lines of sight that pass through stellar dust, gas and dense molecular clouds \citep{1978ApJ...223..168S}. The clusters associated with gas and dust or behind the dusty Galactic spiral arms may give a different value of R$_{V}$. 

The slope of the TCDs of the form ($\lambda$-V)/(B-V), where $\lambda$ is any broad band filter, distinguishes normal extinction due to diffused interstellar medium from the extinction caused by the abnormal dust grains \citep{1990A&A...227..213C}. For the present study we plotted TCDs from (V-R), (V-I), (V-J), (V-H), and (V-K) plotted against (B-V) as shown in Figure~\ref{tcd}. The near-Infrared J, H, and K magnitudes are taken from the 2MASS archive \citep{2006AJ....131.1163S}. The K$_{s}$ magnitude of 2MASS are converted into K magnitude using the expression given by \citet{2001AJ....121.2851C}. The derived slope from these TCDs are given in Table~\ref{rv_slope}. The slope for the cluster NGC 381 is higher in comparison of that obtained for the diffuse ISM, however, it is broadly similar to the normal value \citep[e.g.,][]{2020MNRAS.492.3602J} in case of NGC 2360 and Be 68. We thus conclude that there is anomalous reddening law towards the cluster NGC 381 while no such anomaly is present in the clusters NGC 2360 and Be 68.

\begin{table} 
  \centering
  \caption{The slopes of the ($\lambda$-V)/(B-V) diagrams for the member stars in the direction of the clusters. The normal values of the slopes in the same diagrams are given in brackets.}
  \label{rv_slope}
  \hbox{
\hspace{-0.4 cm}  
  \begin{tabular}{c c c c c c c c}  
  \hline  
 \hspace{-0.3 cm} & \hspace{-0.3 cm} $\frac{(R-V)}{(B-V)}$ & \hspace{-0.3 cm} $\frac{(I-V)}{(B-V)}$ & \hspace{-0.3 cm} $\frac{(J-V)}{(B-V)}$ & \hspace{-0.3 cm} $\frac{(H-V)}{(B-V)}$ & \hspace{-0.3 cm} $\frac{(K-V)}{(B-V)}$ \\ \hline
 \hspace{-0.3 cm} NGC 381& \hspace{-0.3 cm} 0.79$\pm$0.01 & \hspace{-0.3 cm}  1.33$\pm$0.02& \hspace{-0.3 cm} 2.02$\pm$0.03& \hspace{-0.3 cm} 2.65$\pm$0.04& \hspace{-0.3 cm} 2.75$\pm$0.04\\
 \hspace{-0.3 cm}  NGC 2360& \hspace{-0.3 cm} 0.60$\pm$0.01& \hspace{-0.3 cm} 1.08$\pm$0.02& \hspace{-0.3 cm} 1.75$\pm$0.03& \hspace{-0.3 cm} 2.28$\pm$0.03& \hspace{-0.3 cm} 2.39$\pm$0.04\\
  \hspace{-0.3 cm}  Be 68 & \hspace{-0.3 cm} 0.63$\pm$0.02 & \hspace{-0.3 cm} 1.09$\pm$0.03 & \hspace{-0.3 cm} 1.89$\pm$0.04 & \hspace{-0.3 cm} 2.51$\pm$0.10 & \hspace{-0.3 cm} 2.58$\pm$0.11 \\
  \hspace{-0.3 cm} Normal & (0.55) & (1.10) & (1.96) & (2.42) & (2.60) \\ \hline
  \end{tabular}
  }
\end{table}


We derived the total-to-selective extinction R$_{cluster}$ in the direction of the clusters using the approximate relation given by \citet{1981A&AS...45..451N} as following:
$$
R_{cluster} = \frac{m_{c}}{m_{n}} \times R_{normal}
$$      
where, R$_{normal}$ = 3.1, m$_{n}$ is the slope of TCDs for the normal ISM and m$_{c}$ is slope of the linear fit for the member stars in the cluster regions. Using above relation, we found the average value of R$_{cluster}$ to be 3.6, 3.0, and 3.2 in the direction of the clusters NGC 381, NGC 2360 and Be 68, respectively. The high value of R$_{cluster}$, like that of NGC 381, is generally attributed to the presence of larger dust grain size within the cluster or towards the line of sight of the selected region \citep{1989ApJ...345..245C, 2013ApJ...764..172P}.
\subsection{Reddening determination: \textbf{(U-B) vs (B-V)}}\label{ebv}

A precise knowledge of reddening is essential for estimating age and distance of the clusters. Here, we used (U-B)/(B-V) colour-colour variation for the reddening estimation in the selected clusters. In Figure~\ref{reddening}, we show two-colour diagram drawn for the member stars of all the three selected clusters. To avoid the large uncertainties in some of the stars, we did not consider stars having photometric error larger than 0.04, 0.03 and 0.02 mag in $U$, $B$ and $V$ bands, respectively. We fit intrinsic zero-age main sequence (ZAMS) isochrones given by \citet{Schmidt-Kaler} to observational TCD of cluster by shifting E(B-V) and E(U-B) along different values of the reddening vector $\dfrac{E(U-B)}{E(B-V)}$. In the case of NGC 381, a best fit was achieved for $\dfrac{E(U-B)}{E(B-V)}$ = 0.44$\pm$0.03 and corresponding value of reddening E(B-V) was found to be 0.36$\pm$0.04 mag in the direction of NGC 381. The values of the reddening vector $X = \dfrac{E(U-B)}{E(B-V)}$ are estimated as 0.71$\pm$0.03 and 0.60$\pm$0.03 for NGC 2360 and Be 68, respectively. We found the value of E(B-V) to be 0.08$\pm$0.03 and 0.52$\pm$0.04 mag for NGC 2360 and Be 68, respectively. The TCDs with best fit isochrones are shown in Figure~\ref{reddening}. Our estimated value of colour excess E(B-V) = 0.36$\pm$0.04 for NGC 381 is in agreement with the value E(B-V)=0.36 obtained by \citet{1994ApJS...90...31P} but slightly lower than 0.40$\pm$0.10 estimated by \citet{2002AJ....123..905A}. The derived reddening E(B-V) = 0.08$\pm$0.03 for NGC 2360 is in agreement with value E(B-V) = 0.09$\pm$0.06 obtained in the previous CCD photometric study done by \citet{2015NewA...34..195O} while significantly smaller than the value E(B-V) = 0.16 obtained by \citet{2018ApJ...869..139C}. There is no previous photometric study of the cluster Be 68 to compare.

The value of reddening vector $X$ for the clusters NGC 2360 and Be 68 are found to be close to the normal value of 0.72 \citep{1988Ap&SS.143..317G, 1989AJ.....98.2300T} but we yield a slope of 0.46 for NGC 381 that significantly low than the normal value. This implies that the dust grain size is larger than average size in the direction of the cluster NGC 381. This finding is in perfect agreement of our conclusion from the infra-red colour-colour diagrams discussed in the previous section where we found a larger value of total-to-selective extinction $R$ suggesting an anomalous reddening in the direction of this cluster due to larger dust grains. We do not see any such abnormal behaviour in the case of the clusters NGC 2360 and Be 68 from both optical as well as near-IR colour-colour diagrams.

\begin{figure*}
  \centering
  \hbox{
  \includegraphics[width=17.0cm, height=8.0cm]{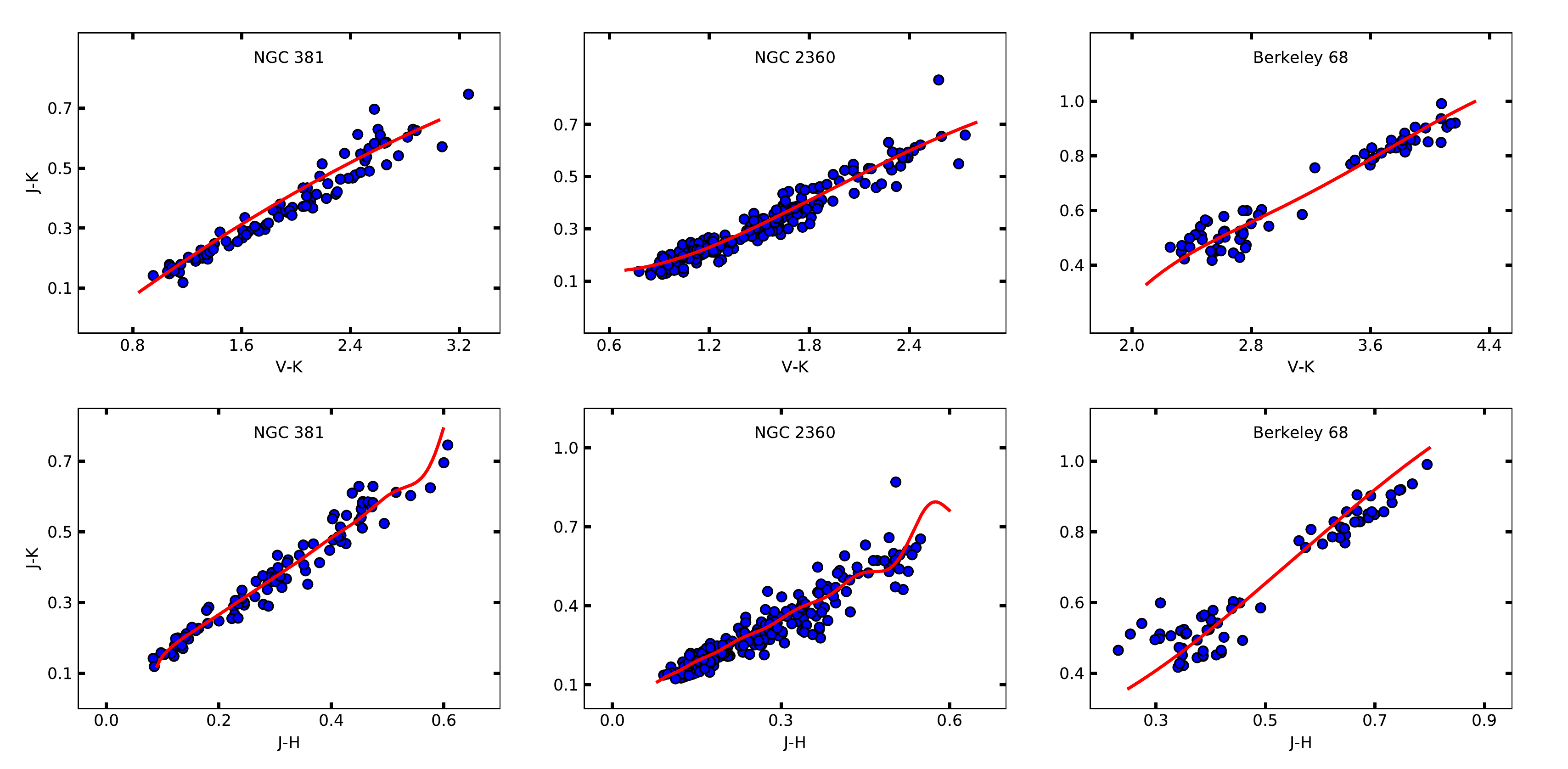}   
  }
 \caption{The plots of (V-K)/(J-K) (upper panel) and (J-H)/(J-K) (lower panel) for the clusters NGC 381, NGC 2360 and Be 68. The red solid lines represent best fit isochrone of \citet{2017ApJ...835...77M}}\label{2mass}
\end{figure*}
\subsection{Interstellar extinction in the near-IR}
\begin{table}
\centering
\caption{Parameters estimated from the extinction study in the near-IR.}
\label{ir}
\begin{tabular}{lccc}
\hline
 & NGC 381& NGC 2360& Be 68\\ 
 \hline
 E(V-K) & 1.04$\pm$0.02& 0.43$\pm$0.02& 1.60$\pm$0.02\\
 E(J-K) & 0.16$\pm$0.02& 0.07$\pm$0.02& 0.27$\pm$0.02\\
$\frac{E(J-K)}{E(V-K)}$& 0.15& 0.16& 0.17\\ \\
 E(J-H) & 0.11$\pm$0.02& 0.030$\pm$0.006& 0.16$\pm$0.02\\
 E(J-K) & 0.17$\pm$0.02& 0.054$\pm$0.006& 0.26$\pm$0.02\\
$\frac{E(J-H)}{E(J-K)}$& 0.65    & 0.55& 0.62\\ 
\hline 
\end{tabular}
\end{table}

The use of near-IR photometry has been very helpful in the study of interstellar extinction \citep{1984IAUS..105..353T,1988MNRAS.232..661T}. We used the V-JHK magnitudes of member stars to determine interstellar extinction in the near-IR towards the cluster region. The IR colours are preferred against the optical colours because they are less affected by blanketing of metallic lines, thus are less sensitive to metallicity than the optical colours \citep{1996A&A...313..873A, 2005AJ....129..656C}. The advantage of combining optical and near-IR photometry
is that it gives a very long range of the wavelength baselines and colours.
\subsubsection{(V-K) vs (J-K) diagram}
We fitted the isochrones given by \citet{2017ApJ...835...77M} on the (V-K)/(J-K) diagrams and best fit are shown in Figure~\ref{2mass}. We used only stars with photometric error less than 0.02 mag in V magnitude and less than 0.06 mag in J, H, and K to plot near-IR colour-colour diagrams as shown in Fig~\ref{2mass}. This is to avoid large uncertainty in the resulting near-IR reddening values. The best fit values of E(V-K), E(J-K) and ratio $\frac{E(J-K)}{E(V-K)}$ are given in Table~\ref{ir} for these clusters. We found the ratio $\frac{E(J-K)}{E(V-K)}$ to be 0.15$\pm$0.03, 0.16$\pm$0.03, and 0.17$\pm$0.03 for the clusters NGC 381, NGC 2360, and Be 68, respectively which are in broad agreement with the value of 0.19 given by \citet{1989ApJ...345..245C}.

\subsubsection{(J-H) vs (J-K) diagram}
The reddening, E(B-V), value can also be determined from the near-IR (J-H)/(J-K) colour-colour plot using the following relations:
$$E(J-H) = 0.273 \times E(B-V)$$
$$E(J-K) = 0.50 \times E(B-V)$$
We fitted the isochrone provided by \citet{2017ApJ...835...77M} to the (J-H)/(J-K) colour-colour diagram which are shown in Figure~\ref{2mass}. The best fit values of E(J-H) and E(J-K) are given in Table~\ref{ir} for these clusters. It can be seen from the table that the obtained values of E(J-K) from the both plots (V-K) versus (J-K) and (J-H) versus (J-K) are in close agreement. A best fit was achieved for the ratio $\frac{E(J-H)}{E(J-K)}$ to be $\sim$ 0.65 and 0.62 for clusters NGC 381 and Be 68, respectively which are slightly higher than the normal interstellar extinction ratio of 0.55 given by \citet{1989ApJ...345..245C}. We derived the ratio $\frac{E(J-H)}{E(J-K)}$ = 0.55 for NGC 2360 which is equal to the normal interstellar extinction ratio. The reddening E(B-V) derived using above relations are found to be 0.37 mag, 0.11 mag, and 0.55 mag for NGC 381, NGC 2360 and Be 68, respectively. To calculate extinction A$_{v}$, we used E(B-V) estimated using near-IR data and total-to-selective extinction R$_{cluster}$ derived in Sect.~\ref{rv}. The A$_{v}$ was found to be 1.33$\pm$0.11, 0.33$\pm$0.09, and 1.76$\pm$0.10 for the clusters NGC 381, NGC 2360 and Be 68, respectively. The A$_{v}$ values calculated by us are in broad agreement with the values 1.06$\pm$0.37, 0.32$\pm$0.22 and 1.69$\pm$0.29 mag derived in \citet{2019A&A...628A..94A} for the clusters NGC 381, NGC 2360, and Be 68, respectively.  We also observed that the reddening E(B-V) derived from the near-IR photometry are in agreement with the the values derived from the optical photometry for these clusters.
\subsection{Distance and Age determination}
\subsubsection{Age and distance through CMD}
%
\begin{figure*}
\includegraphics[width=16.0cm, height=9.0cm]{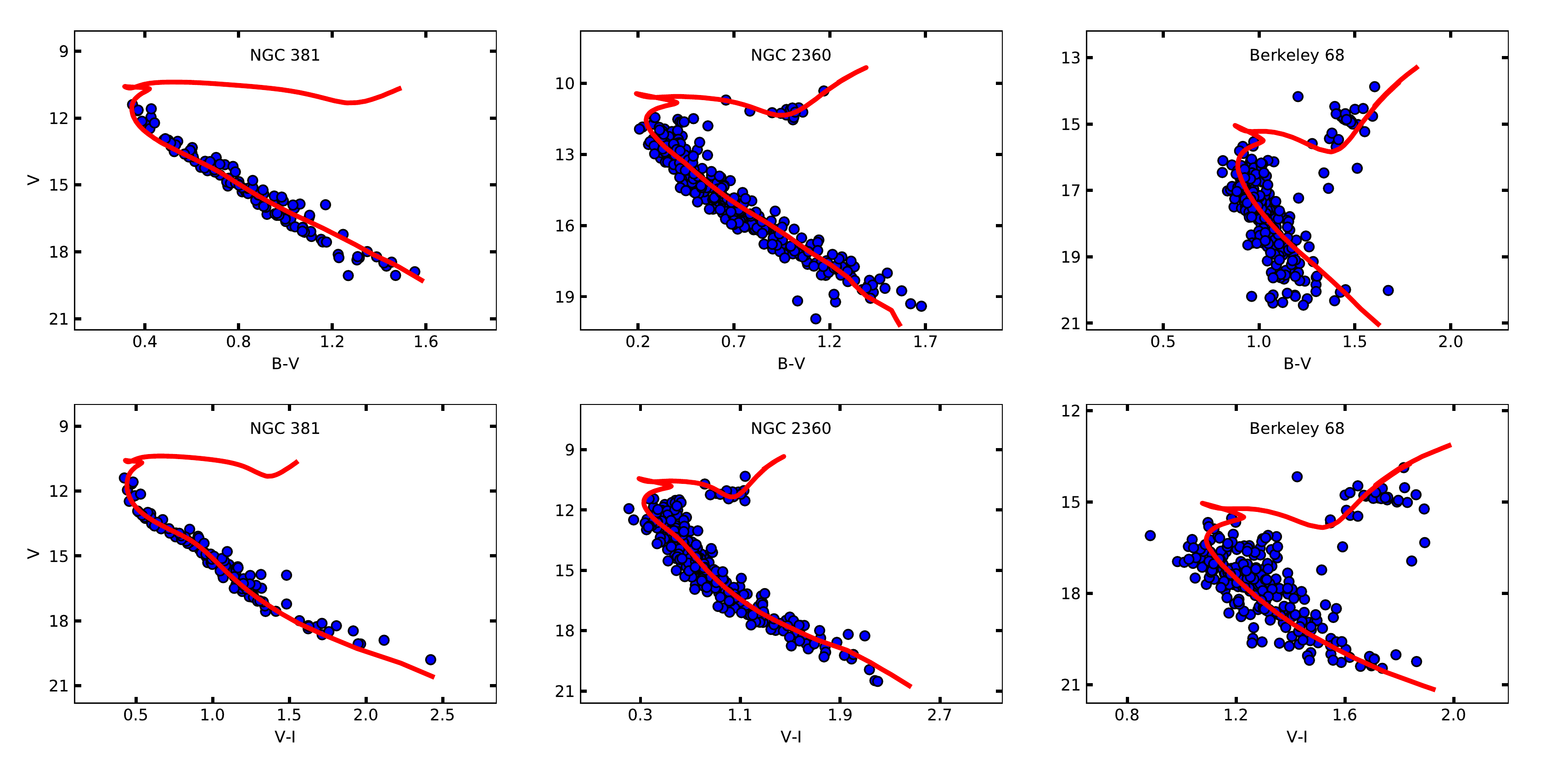}
\caption{The plots of (B-V)/V (upper panel) and (V-I)/V (lower panel) CMDs for clusters NGC 381, NGC 2360 and Be 68. The red solid lines represent best fit isochrone of \citet{2017ApJ...835...77M}. The best fits were achieved for log(Age)=8.65, 8.95, and 9.25 yr for the clusters NGC 381, NGC 2360, and Be 68, respectively.}
\label{col_mag}
\end{figure*}

The distance and age of the Galactic OCs are important parameters in tracing the Galactic structure and chemical evolution of the Galaxy using OCs as tracers \citep{1993A&A...267...75F, 2016A&A...593A.116J}. The CMD of the cluster play important role in identification of the member stars as well as determination of the cluster parameters like reddening, metallicity, age, and distance modulus. We used simultaneous (B-V)/V and (V-I)/V CMDs, as shown in Figures~\ref{col_mag}, to estimate the age and distance of the clusters. The (B-V)/V and (V-I)/V CMDs of the clusters NGC 2360 and Be 68 are showing presence of red giant stars. The shape and presence of the different features in the CMDs depend on the age and metallicity of that particular cluster. We used classical method of reddening determination through isochrone fitting on (U-B)/(B-V) TCD as already discussed in the Sect.~\ref{ebv}. The age and distance of the cluster have been derived by fitting theoretical isochrones of solar metallicity Z = 0.0152 given by \citet{2017ApJ...835...77M} on CMD of the clusters. The reddening, E(B-V), values determined earlier are used to achieve the best fit while shifting the isochrone of certain age for varying distance-modulus. The reddening, E(V-I), derived from (V-I)/V CMD were found to be 0.46, 0.18, and 0.65 mag for NGC 381, NGC 2360 and Be 68 clusters, respectively. We achieved best fit on both (B-V)/V and (V-I)/V CMD for log(Age) = 8.65$\pm$0.05, 8.95$\pm$0.05, 9.25$\pm$0.05 years and apparent distance-modulus (V$_{0}$ - M$_{v}$) = 11.2$\pm$0.2 , 10.2$\pm$0.2, and 13.7$\pm$0.2 mag for NGC 381, NGC 2360 and Be 68, respectively. The red lines in Figures~\ref{col_mag} show to best fit isochrones in the CMDs.
\begin{figure*}
\includegraphics[width=18.3 cm, height=10.0cm]{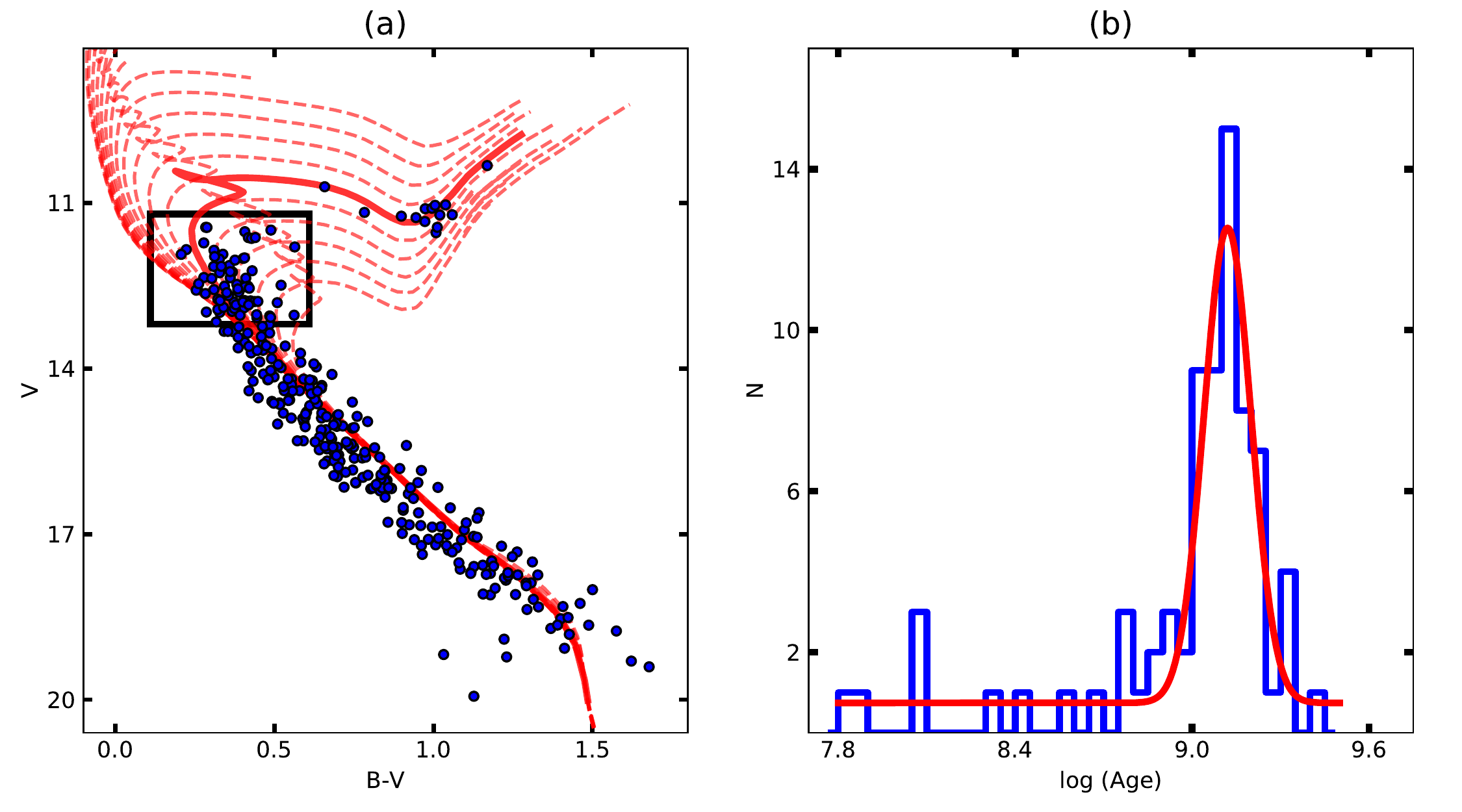}\\
\caption{(a) The plot of (B-V)/V for the cluster NGC 2360 along with rectangular eMSTO region. The red dashed lines correspond to isochrones of \citet{2017ApJ...835...77M} with log(Age) interval of 0.1 yr while red solid line presents best fit \citet{2017ApJ...835...77M} isochrone of log(Age)=8.95 yr. The minimum and the maximum log(Age) values of the isochrones are 7.80 and 9.40 yr, respectively. (b) Fitted Gaussian profile on the histograms of number of stars (N) versus age-bin (log(Age)) calculated from the rectangular region of panel (a).}
\label{emsto}
\end{figure*}

We found extended main sequence turn-off (eMSTO) in the cluster NGC 2360. Previously, eMSTO were known to be present in the OCs of Magellanic Clouds \citep{2007MNRAS.379..151M} but now eMSTO are found in Galactic OCs too \citep{2018ApJ...863L..33M,2018ApJ...869..139C,2019AJ....158...35S}. In order to quantify the apparent spread in the age due to eMSTO in the cluster NGC 2360, we fitted a grid of \citet{2017ApJ...835...77M} isochrones having the same metallicity, reddening, and distance-modulas but separated by 0.01 year in log(Age) on the rectangular part of the CMD of the cluster as shown in Figure~\ref{emsto}(a). The selected rectangular region is the region near turn-off where the spread in the CMD due to apparent age variation is clearly visible. We fitted the Gaussian profile on the histogram of the distribution of ages found from the grid of the isochrones as shown in Figure~\ref{emsto}(b). The isochrones in Figure~\ref{emsto} have log(Age) interval of 0.1 yr. We found FWHM of Gaussian profile to be 0.19 which corresponds to the apparent age spread of $\sim$595 Myr. Our estimated apparent spread in the age is higher than 210$\pm$35 and 370 Myr obtained by \citet{2018ApJ...869..139C} and \citet{2020MNRAS.491.2129D}, respectively. The apparent age spread in a cluster can be due to photometric uncertainty, binary, metallicity, stellar rotation, and multi-populations \citep{2018ApJ...869..139C, 2019MNRAS.490.2414P, 2020MNRAS.491.2129D}. The detailed study of eMSTO for NGC 2360 can be found in some of the recent studies \citep{2018ApJ...863L..33M,2018ApJ...869..139C,2019AJ....158...35S,2020MNRAS.491.2129D}

The distances from the apparent distance modulus were derived using the reddening law R$_{cluster}$ in the Sect.~\ref{rv}. The distance to these clusters are found to be 957$\pm$152, 982$\pm$132, and 2554$\pm$387 pc for the clusters NGC 381, NGC 2360, Be 68, respectively. The derived distances are in agreement with the distances of 1024, 970, and 2437 for the clusters NGC 381, NGC 2360, and Be 68 obtained by \citet{2018A&A...618A..93C}. The present estimated age of the cluster NGC 381 is 447$\pm$52 Myr which is in close agreement with the age $\sim$500 Myr obtained by \citet{2019A&A...623A.108B} and slightly higher than $\sim$320 Myr reported by \citet{2002AJ....123..905A} but is significantly less than 1.1 Gyr reported by \citet{1994ApJS...90...31P}. The distance-modulus for NGC 381 in the present study is found to be 9.90$\pm$0.34 mag which is lower than the value 10.13, 10.3$\pm$0.3, and $\sim$10.52 mag obtained by \citet{1994ApJS...90...31P}, \citet{2002AJ....123..905A}, and \citet{2019A&A...623A.108B}, respectively. We estimated distance modulus for NGC 2360 to be 9.96$\pm$0.29 mag which is in agreement, within error, with the calculated values 10.25$\pm$0.02 and 10.23 mag by \citet{2015NewA...34..195O} and \citet{2018ApJ...869..139C}, respectively. We found log(Age/yr) to be 8.95$\pm$0.05 for the cluster NGC 2360 which is in agreement with 9.05$\pm$0.05 and 9.01 reported by \citet{2015NewA...34..195O} and \citet{2018ApJ...869..139C}, respectively. We estimated the values of age and distance-modulus of the cluster Be 68 for the first time in the present study so no comparison can be made for this cluster.

\subsubsection{Distance estimation using parallax}
\begin{table*}
  \centering
  \caption{The derived values of physical parameters of the clusters}
  \label{clust_par}
  \begin{tabular}{c c c c c c c c c c}  
  \hline  
   \hspace{-0.2 cm} Cluster & \hspace{-0.3 cm} r$_{c}$ & r & \hspace{-0.3 cm} $\bar{\mu}$ &R$_{cluster}$ & \hspace{-0.3 cm} E(B-V) & \hspace{-0.3 cm} log(Age) & \hspace{-0.3 cm} D$_{isochrone}$& D$_{parallax}$\\
    & \hspace{-0.3 cm} (pc) & (pc)& \hspace{-0.3 cm} (mas/yr)& &(mag) & \hspace{-0.3 cm} (Myr) & \hspace{-0.3 cm} (pc) & (pc)\\
  \hline
  NGC 381 & 0.8 & 2.9& 2.78$\pm$0.12& 3.6& 0.36$\pm$0.04& 8.65$\pm$0.05& 957$\pm$152& 1131$\pm$48\\
  NGC 2360 & 0.5& 3.4& 5.61$\pm$0.19& 3.0& 0.08$\pm$0.03& 8.95$\pm$0.05& 982$\pm$132& 1078$\pm$ 50\\
  Berkeley 68 & 0.4& 3.5& 2.62$\pm$0.33& 3.2& 0.52$\pm$0.04& 9.25$\pm$0.05& 2554$\pm$387& 3135$\pm$757\\ 
  \hline
  \end{tabular}
\end{table*}

\begin{table*}
  \centering
  \caption{Derived LF and MF from the clusters in the present study.}
  \label{lf}
  \begin{tabular}{c c c c c c c c c c c c c c}
  \hline
   \hspace{0.3 cm} V \hspace{0.3 cm} & \multicolumn{3}{c}{NGC 381} \hspace{0.3 cm} & \multicolumn{3}{c}{NGC 2360} \hspace{0.3 cm} & \multicolumn{3}{c}{Be 68} \\
   \cmidrule(lr){2-4}\cmidrule(lr){5-7}\cmidrule(lr){8-10}
   Range & \hspace{0.3 cm} Mass Range& \hspace{0.3 cm}$\bar{m}$& \hspace{0.3 cm} N& \hspace{0.3 cm} Mass Range& \hspace{0.3 cm} $\bar{m}$& \hspace{0.3 cm} N& \hspace{0.3 cm} Mass Range& \hspace{0.3 cm} $\bar{m}$& \hspace{0.3 cm} N \\
 (mag)& \hspace{0.3 cm} (M$_{\odot}$) & \hspace{0.3 cm} (M$_{\odot}$) & \hspace{0.3 cm} & \hspace{0.3 cm} (M$_{\odot}$) & \hspace{0.3 cm} (M$_{\odot}$) & \hspace{0.3 cm} & \hspace{0.3 cm} (M$_{\odot}$) & \hspace{0.3 cm} (M$_{\odot}$) \\
    \hline
    10-11& \hspace{0.3 cm} -& \hspace{0.3 cm}-& \hspace{0.3 cm} -& \hspace{0.3 cm} 2.21-2.12& \hspace{0.3 cm} 2.20& \hspace{0.3 cm} 2& \hspace{0.3 cm} -& \hspace{0.3 cm} -& \hspace{0.3 cm}- \\ 
    11-12& \hspace{0.3 cm} 2.72-2.35& \hspace{0.3 cm} 2.49& \hspace{0.3 cm} 5& \hspace{0.3 cm} 2.12-1.80& \hspace{0.3 cm}2.03& \hspace{0.3 cm} 27& \hspace{0.3 cm} -& \hspace{0.3 cm} -& \hspace{0.3 cm}- \\    
    12-13& \hspace{0.3 cm} 2.35-1.88& \hspace{0.3 cm}2.08& \hspace{0.3 cm} 7& \hspace{0.3 cm} 1.80-1.49& \hspace{0.3 cm}1.62& \hspace{0.3 cm} 52& \hspace{0.3 cm} -& \hspace{0.3 cm} -& \hspace{0.3 cm}- \\
    13-14& \hspace{0.3 cm} 1.88-1.53& \hspace{0.3 cm} 1.67& \hspace{0.3 cm} 16& \hspace{0.3 cm} 1.49-1.24& \hspace{0.3 cm}1.37& \hspace{0.3 cm} 37& \hspace{0.3 cm} 1.75-1.74& \hspace{0.3 cm} 1.75& \hspace{0.3 cm} 1 \\
    14-15& \hspace{0.3 cm} 1.53-1.26& \hspace{0.3 cm} 1.40& \hspace{0.3 cm} 21& \hspace{0.3 cm} 1.24-1.04& \hspace{0.3 cm} 1.14& \hspace{0.3 cm} 53& \hspace{0.3 cm} 1.74-1.73& \hspace{0.3 cm} 1.74& \hspace{0.3 cm} 18\\
    15-16& \hspace{0.3 cm} 1.26-1.06& \hspace{0.3 cm} 1.17& \hspace{0.3 cm} 24& \hspace{0.3 cm} 1.04-0.88& \hspace{0.3 cm} 0.96& \hspace{0.3 cm} 53& \hspace{0.3 cm} 1.73-1.57& \hspace{0.3 cm}1.70& \hspace{0.3 cm} 13\\
    16-17& \hspace{0.3 cm} 1.06-0.90& \hspace{0.3 cm} 0.98& \hspace{0.3 cm} 21& \hspace{0.3 cm} 0.88-0.76& \hspace{0.3 cm} 0.84& \hspace{0.3 cm} 40& \hspace{0.3 cm} 1.57-1.32& \hspace{0.3 cm} 1.44& \hspace{0.3 cm} 57\\
    17-18& \hspace{0.3 cm} 0.90-0.77& \hspace{0.3 cm} 0.84& \hspace{0.3 cm} 9& \hspace{0.3 cm} 0.76-0.67& \hspace{0.3 cm} 0.72& \hspace{0.3 cm} 43& \hspace{0.3 cm} 1.32-1.12& \hspace{0.3 cm} 1.22& \hspace{0.3 cm} 85\\
    18-19& \hspace{0.3 cm} 0.77-0.68& \hspace{0.3 cm} 0.73& \hspace{0.3 cm} 10& \hspace{0.3 cm}-& \hspace{0.3 cm} -& \hspace{0.3 cm} -& \hspace{0.3 cm} 1.12-0.95& \hspace{0.3 cm} 1.03& \hspace{0.3 cm} 58\\
    19-20& \hspace{0.3 cm} 0.68-0.59& \hspace{0.3 cm} 0.65& \hspace{0.3 cm} 3& \hspace{0.3 cm} -& \hspace{0.3 cm} -& \hspace{0.3 cm} -& \hspace{0.3 cm} -& \hspace{0.3 cm} -& \hspace{0.3 cm} -\\ \hline
  \end{tabular}
\end{table*}

\begin{figure}
\includegraphics[width=8.6 cm, height=4.0 cm]{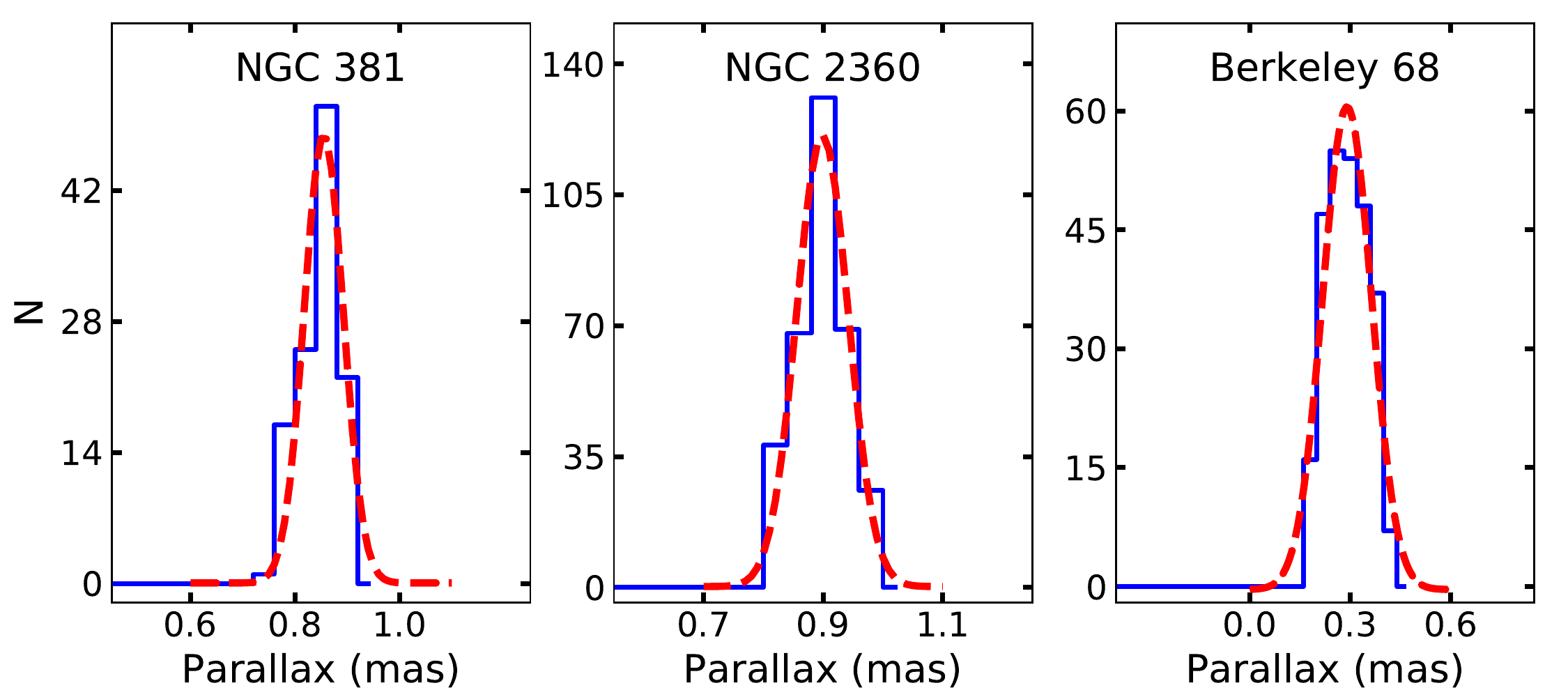}
\caption{Histograms of parallax with bin size 0.04 mas of clusters NGC 381, NGC 2360 and Be 68. The red dashed line represents fitted Gaussian profile for the mean estimation.}
\label{paralx}
\end{figure}
The distance of the stars can also be estimated using astrometric parallax. Since stars associated with the clusters are at the approximately same distance, the mean value of parallax of all member stars can be used to calculate the distance to the clusters. We used parallax from \textit{Gaia} DR2 data for these OCs. As suggested by \citep{2018A&A...616A...9L} we estimated average of individual parallaxes to find average distance to the clusters rather than averaging distances obtained from parallax of individual stars. The mean value of the parallax was derived by fitting Gaussian profile on the histograms of the parallax having bin size of 0.04 mas. The plots for the parallax histograms with Gaussian fit are shown in Figure~\ref{paralx}. We found mean parallax values as 0.855$\pm$0.037, 0.899$\pm$0.043, and 0.290$\pm$0.073 mas. The calculated values of parallax are in agreement with the parallax 0.847, 0.902, and 0.281 obtained by \citet{2018A&A...618A..93C}. 

In some of the recent studies, the evidences of systematic
offsets in \textit{Gaia} DR2  parallaxes were reported and a mean offset of -0.029 mas is generally quoted \citep{2018A&A...616A...2L, 2018A&A...616A...9L, 2018ApJ...862...61S}. After applying this offset to our derived mean parallaxes, we found the distances of 1131$\pm$48, 1078$\pm$50, and 3135$\pm$757 pc for the clusters NGC 381, NGC 2360 and Be 68, respectively. The distances found through \textit{Gaia} DR2 parallaxes are close to the the distances measured using isochrone fitting on CMD of the clusters for the clusters NGC 381 and NGC 2360. In the case of Be 68, the distance estimated through parallax is higher than the value derived by isochrone fitting.

The various physical parameters estimated for the three clusters in the present study are summarized in Table~\ref{clust_par}. 
\section{Dynamical study of the clusters}\label{dynamic}
\subsection{Luminosity function}
%
\begin{figure*}
\centering
\includegraphics[width=17.0 cm, height=4.8 cm]{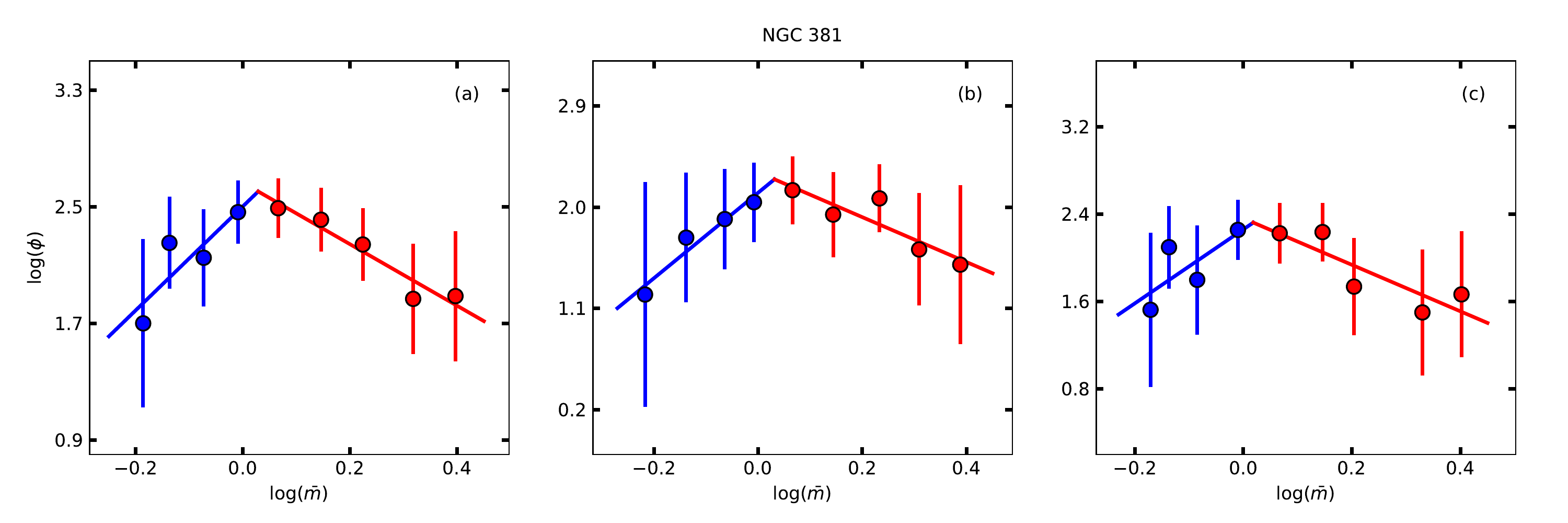}
\includegraphics[width=17.0 cm, height=4.8 cm]{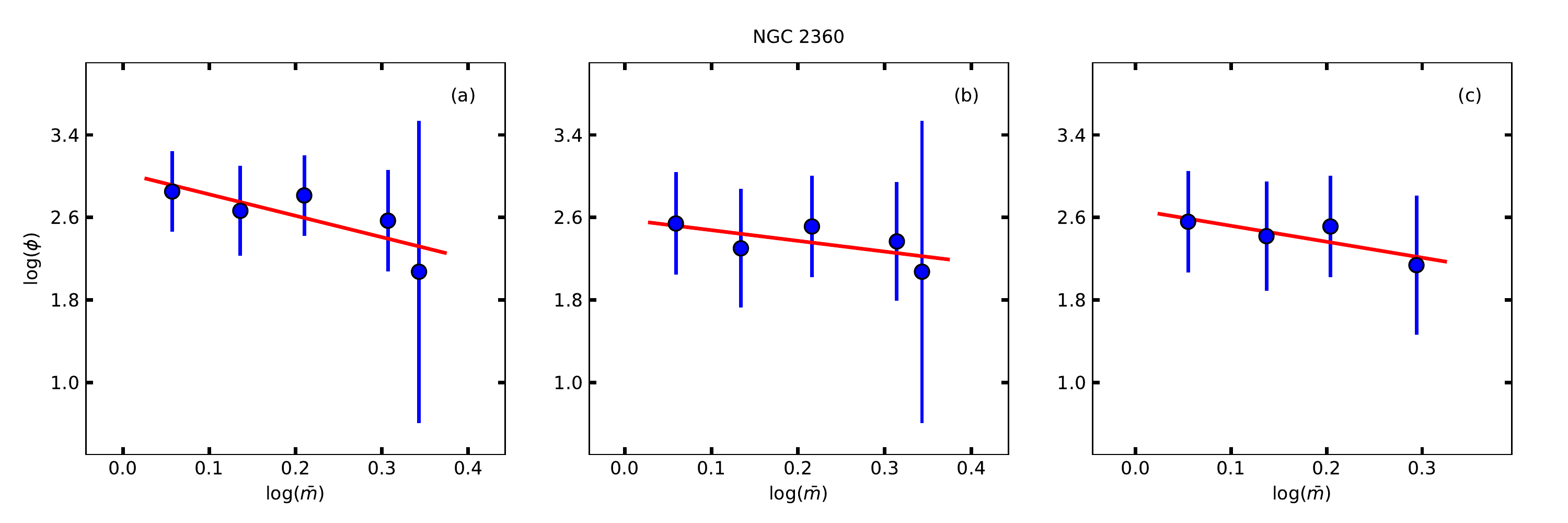}
\includegraphics[width=17.0 cm, height=4.8 cm]{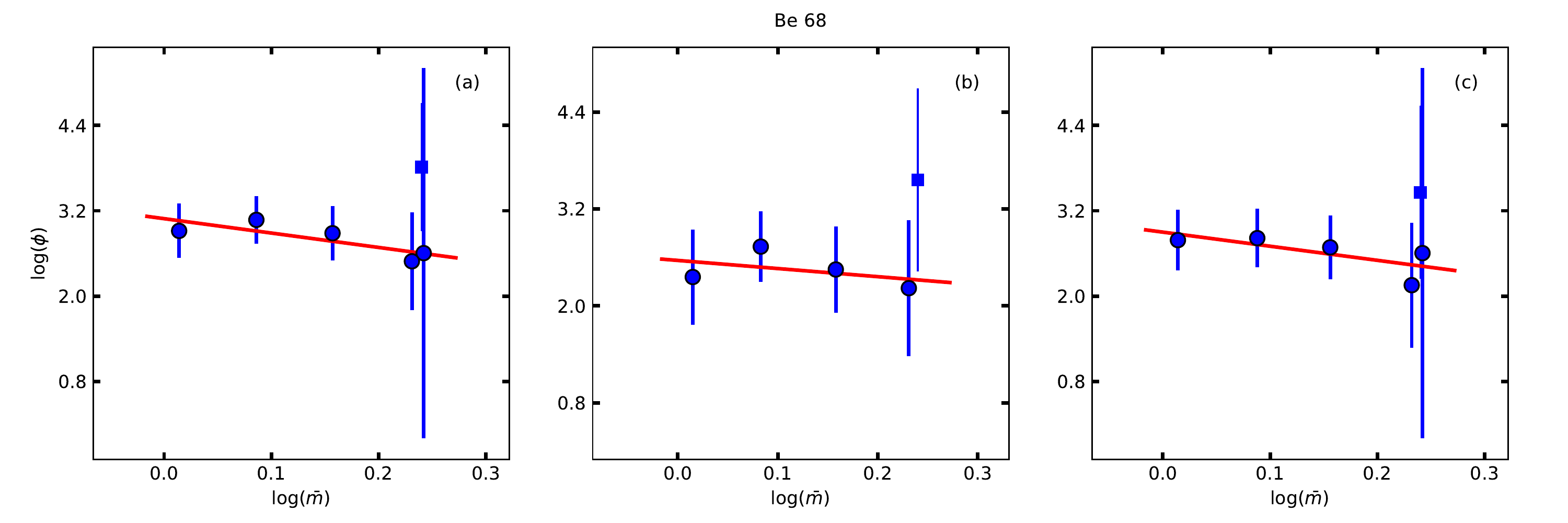}
\caption{Plots of MF slope for NGC 381 (upper panel), NGC 2360 (middle panel), and Be 68 (lower panel) in (a) entire, (b) inner, and (c) outer region. We have shown excluded point corresponding to 14-15 mag bin by squares in plots for Be 68.}
\label{mf1}
\end{figure*}
LF is defined as the total number of cluster stars in different magnitude bins. To correct inaccuracy in the LF caused by the incompleteness of our photometry, we divided numbers of actually detected stars in each magnitude bins by CF of the bins. The estimated LF for these clusters are given in Table~\ref{lf}.

\subsection{Mass function}
The initial mass function (IMF) is defined as the distribution of stellar masses per unit volume in a star formation event. The IMF studies are fundamental in understanding the star formation process and subsequent chemical and dynamical evolution of the star clusters \citep{2002Sci...295...82K}. Thus, the estimation of the correct MF for the stars in the clusters is very important. A direct determination of IMF for the cluster is not possible due to the dynamic evolution of stellar system. Therefore we estimated the present day MF, which is relative number of stars per unit mass and often expressed by power law N (log m) $\propto$ m$^{\Gamma}$, where the slope of MF is given as
$$
\Gamma = \dfrac{d \ log \ N(log \ m)}{d \ log \ m}
$$
where N log(m) is the number of stars per unit logarithmic mass interval. For the mass range $0.4 < M/M_\odot \leq 10.0$, \citet{1955ApJ...121..161S} derived the slope of MF as ${\Gamma}$ = -1.35 in the solar neighbourhood stars which is normally considered as a classical value of MF slope. The mass for each cluster member was determined by comparing observed magnitude with that of the predicted one by fitting isochrone of \citet{2017ApJ...835...77M} on the CMD of the cluster for given age, reddening, distance, and metallicity. We determined mean mass in each magnitude bins of $V$ band. In Table~\ref{lf}, we provide mass range, mean mass and cluster members in different brightness range for the three clusters. The MF slope ($\Gamma$) in the entire region was calculated for all the three clusters using a least square method and resultant slopes are drawn by the solid lines in the leftmost panels of Figure~\ref{mf1}.

Despite there is some scatter seen in the MF slopes, a break in MF slope can be seen at log$(M/M_\odot) \sim $0.0 (or $M \sim 1.06 M_\odot$) for the cluster NGC 381. A similar break was also noticed in some of the other clusters in the past \citep[e.g.,][]{2001A&A...375..863J, 2005A&A...437..483B, 2008MNRAS.384.1675J, 2013MNRAS.434.3236K}. We determined different MF slopes ($\Gamma$) on both sides of the break using a least square fit method. We call stars with mass above $\sim$1 M$_{\odot}$ as high-mass stars while stars having mass below $\sim$1 M$_{\odot}$ as low-mass stars. The MF slopes in the high-mass and low-mass regimes are found to be -2.11$\pm$0.35 and 3.57$\pm$1.54, respectively. We found that a two-stage power law in the MF slope is present in the cluster NGC 381. The two-stage power law in some of the MF slopes of OCs may be due to dynamics of clusters and/or initial conditions in star formation event \citep{2005A&A...437..483B}. The MF slopes for low-mass stars in the clusters NGC 2360 and Be 68 are not estimated separately as we did not find any break in their MF slopes. For the cluster NGC 2360, we derived slope of -2.08$\pm$0.94 for stars in mass range 2.12-1.04 M$_{\odot}$. We derived a single MF slope in the entire region of the cluster Be 68 to be -2.02$\pm$0.82 for stars in mass range 1.73-0.95 M$_{\odot}$. Here, we excluded magnitude bin corresponding to 14-15 mag as the MF value is significantly deviated and contain large uncertainty in this cluster. The slopes derived for high mass stars in the entire region of the cluster Be 68 is comparable, within certain error, with the classical value of \citet{1955ApJ...121..161S}, however, it is steeper for the entire regions of the clusters NGC 381 and NGC 2360. 
\begin{table} 
  \centering
  \caption{The slopes of MF ($\Gamma$) for the clusters. The inner regions have radii equal to 4$^{\prime}$, 5$^{\prime}$, and 5$^{\prime}$ for NGC 381, NGC 2360, and Be 68, respectively.}
  \label{mf_slope}
  \begin{tabular}{c c c c c c c c c}  
  \hline  

 Cluster&mass-range& \multicolumn{3}{c}{$\Gamma$}\\
 \cmidrule(lr){3-5}
 &(M$_{\odot}$)&entire&inner&outer \\ 
 \hline 
 NGC 381:&&&&\\
 & 2.72-1.06& -2.11$\pm$0.35& -2.00$\pm$0.60& -2.13$\pm$0.79\\
 & 1.06-0.59& 3.57$\pm$1.54& 3.80$\pm$0.71& 3.38$\pm$2.23\\
 NGC 2360:&&&&\\
 & 2.12-1.04& -2.08$\pm$0.94& -1.04$\pm$0.68& -1.54$\pm$0.73\\
 Be 68:&&&&\\
 & 1.73-0.95& -2.02$\pm$0.82& -1.00$\pm$1.48& -1.98$\pm$1.12\\

\hline
  \end{tabular}
\end{table}

To study any possible spatial variation in MF slope within clusters, we also estimated MFs for the inner and outer regions separately for all the three clusters  as shown in the middle and rightmost panel of Figure~\ref{mf1}. The inner region is a circular region around the cluster center having radius equal to 4$^{\prime}$, 5$^{\prime}$, and 5$^{\prime}$ for the clusters NGC 381, NGC 2360, and Be 68, respectively. The outer region is taken as the entire region minus the inner region where entire region is considered as the total observed region for the cluster in our observations. The estimated slopes of the MFs for the stars in entire, inner, and outer regions of all the three clusters are given in Table~\ref{mf_slope}. Here, quoted uncertainties come from the linear regression solution in the fit.
 
For the stars having mass above $\sim$1.06 M$_{\odot}$, we estimated MF slope in inner region of NGC 381 to be -2.00$\pm$0.60 which is not very different from the slope of -2.13$\pm$0.74 found in the outer region of this cluster. For the low-mass stars ($\leq$1 M$_{\odot}$) in the cluster NGC 381, the MF slope is estimated as 3.81$\pm$0.71 in the inner region which is found to be steeper in comparison of the slope of 3.38$\pm$2.23 derived in the outer region. This may be caused by lesser number of low-mass stars in the inner region than the outer region which gives some hint of mass segregation in this cluster. We found relatively steeper MF high-mass slopes  in the outer regions of the clusters NGC 2360 and Be 68 which may indicates relatively high concentration of high-mass stars in the inner region possibly due to mass segregation in the clusters \citep{2018AJ....155...44P,2013MNRAS.434.3236K,2020MNRAS.492.3602J}.

\subsection{Mass segregation}\label{segre}
The mass segregation is the phenomenon in which massive stars are more centrally concentrated than low mass stars in clusters and has been reported in many clusters \citep[e.g.,][]{2014MNRAS.437..804J,2016MNRAS.463.3476P,2017AJ....153..122Z,2018MNRAS.473..849D,2020MNRAS.492.3602J}. The mass segregation effect can be triggered due to the dynamical evolution of the cluster or may be imprint of star formation process itself \citep{1988MNRAS.234..831S}. The mass segregation process is believed to be taking place due to equipartition of energy in the cluster members through stellar encounters where massive stars transfer their kinetic energy to low mass stars and accumulate in the central region of the cluster while low-mass stars sink towards outer region \citep[e.g.,][]{1986AJ.....92.1364M}.

\begin{figure}
\centering
\includegraphics[width=8 cm, height=15 cm]{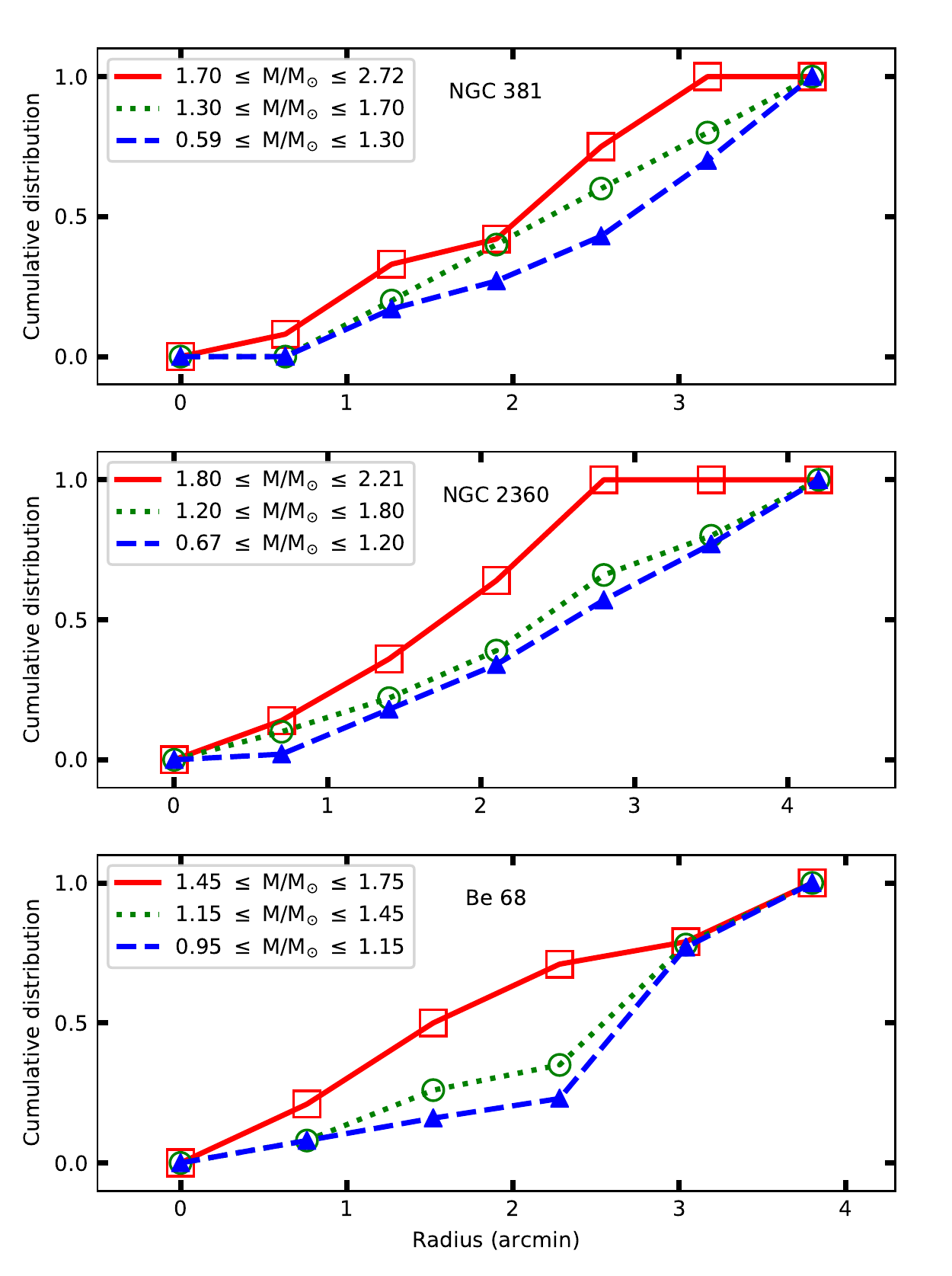}
\caption{The plots of variation of cumulative distribution of stars with radius for various mass ranges for the clusters NGC 381, NGC 2360 and Be 68, respectively. The red solid lines, green dotted lines, and blue dashed lines represent the high, intermediate, and low mass ranges, respectively. The values of mass ranges are shown in upper left of the plots.}
\label{cum_dist}
\end{figure}

The present value of LF and MF in a cluster may be significantly affected by the dynamical evolution and mass segregation. The cumulative distribution of stars with radius for various mass ranges is often used to study the mass segregation in star clusters. To study the effect of dynamical evolution and mass segregation, we determined cumulative radial distributions of member stars for different mass ranges as shown in Figure~\ref{cum_dist}. Here, we considered three different mass ranges, namely, high, intermediate, and low mass ranges which are noted in the upper left part of the plots for each cluster. The estimated cumulative distributions are corrected for the data incompleteness. It can be inferred from the figure that the high mass stars are more concentrated in the central part in comparison of low mass stars. We performed Kolmogorov-Smirnov (K-S) test of the distributions in these mass range to check whether they are different or not. We conclude with $\sim$90$\%$ confidence level for clusters NGC 381 and NGC 2360, and with $\sim$80$\%$   confidence level in Be 68 that the mass segregation effect in these clusters is present.
%
%
\subsection{\textbf{The dynamical relaxation time}}
The dynamical relaxation time T$_{E}$ is defined as time-scale in which velocity distribution of star approaches to Maxwellian velocity distribution in equilibrium through the exchanges of energy among stars in cluster. We calculated dynamic relaxation time to find whether dynamical evolution or star formation process itself is responsible for the mass segregation process in the clusters. The relaxation time, T$_{E}$, as given by \citet{1971ApJ...164..399S}, is numerically defined as
$$
T_E = \frac{8.9 \times 10^5 (N R_h^3/\bar{m})^{1/2}} {\log(0.4N)}
$$
where N is the total number of cluster stars. The R$_{h}$ is half-mass radius of the cluster and $\bar{m}$ is mean mass of the cluster stars. To determine R$_{h}$, first we combined mass of each individual member of a cluster and determined the total mass of each cluster as well as the mean stellar mass. We then estimated the cumulative mass of the stars by increasing the distance from the cluster center and chosen the radius where half of the total cluster mass lies inside the radius. In this way, R$_{h}$ values were estimated as 4$^{\prime}$.5 (1.2 pc), 5$^{\prime}$.0 (1.4 pc), and 5$^{\prime}$.7 (4.2 pc) for the clusters NGC 381, NGC 2360 and Be 68, respectively. We obtained values of T$_{E}$ as 48.5, 78.9, and 87.6 Myr for the clusters NGC 381, NGC 2360 and Be 68, respectively. 

It is important to know whether dynamical evolution or star formation process itself is responsible for the mass segregation effect \citep{1988MNRAS.234..831S}. As in the present cases, the relaxation times of these clusters are found to be smaller than their respective ages which indicates that these clusters are dynamically relaxed due to the mass segregation effect.

\subsection{Tidal radius}
The tidal interactions are very important in understanding the initial structure as well as the dynamic evolution of the clusters \citep{2015MNRAS.449.1811D,2010MNRAS.402.1841C}. The tidal radius is defined as the distance from the center of a cluster where the gravitational acceleration caused by the cluster becomes equal to the tidal acceleration due to the parent Galaxy \citep{1957ApJ...125..451V}. Within the tidal radius of a cluster, stars are generally gravitationally bound to the cluster while stars outside tidal radius are unbound to the cluster. In other words, stars outside the tidal radius are more influenced by the external potential of the Galaxy than the effective potential of the cluster.

To derive tidal radius, R$_{t}$, we followed the relation given by \citet{2000ApJ...545..301K}
$$
R_{t} = \left(\frac{M_{C}}{2 M_{G}}\right)^{1/3}\times R_{G}
$$
Where M$_{C}$ is the total mass of the cluster which we obtained in the earlier section. The total masses for the clusters NGC 381, NGC 2360 and Be 68 were estimated to be 148.56, 379.77, and 325.43 M$_{\odot}$ through their identified cluster members. However, we cannot rule out the exclusion of some of the low mass stars which were faint enough to be detected from our telescope or their proper motions contain too much uncertainty in the \textit{Gaia} DR2 catalogue to be identified through the VPD. Therefore, the present value of total cluster mass may be taken as a lower limit while the mean stellar mass can be taken as an upper limit for the three clusters under observation.

The M$_{G}$ is the Galactic mass within a Galactocentric radius R$_{G}$ and calculated using the relation given by \citet{1987ARA&A..25..377G} as following
$$
M_{G} = 2 \times 10^{8} M_{\odot} \left(\frac{R_{G}}{30pc }\right)^{1.2}
$$
We derived the values of M$_{G}$ to be $\sim$ 1.57$\times$10$^{11}$, 1.79$\times$10$^{11}$, and 1.81$\times$10$^{11}$ M$_{\odot}$ for NGC 381, NGC 2360 and Be 68, respectively and corresponding values of tidal radius are estimated to be 6.49, 9.46, and 9.04 pc, respectively. These estimated tidal radius are larger than the cluster radius values 2.9, 3.4, and 3.5 pc derived using RDP for the cluster NGC 381, NGC 2360, and Be 68, respectively. Thus, any star beyond the derived tidal radius of these clusters would be considered to be gravitationally unbound to the respective cluster. 
%
\begin{table}
\centering
\caption{Parameters estimated from the dynamical study of three clusters.}
\label{dyna}
\begin{tabular}{lccc}
\hline
Cluster parameter                     & NGC 381 & NGC 2360 & Be 68 \\ 
\hline
Number of member stars used           & 116& 327& 264 \\
Mean stellar mass ($\bar{m}/M_\odot$) & 1.28& 1.16&  1.23 \\
Total mass  ($M_\odot$)               & 148.56& 379.77& 325.43 \\
Cluster half radius ($r_h/pc$)        & 1.2& 1.4& 4.2 \\
Tidal radius ($r_t/pc$)               & 6.49& 9.46& 9.04 \\
Relaxation time ($T_E/Myr$)          & 48.5& 78.9& 87.6 \\ 
\hline
\end{tabular}
\end{table}

%

A summary of cluster parameters obtained from the dynamical study of three selected clusters in the present study is given in Table~\ref{dyna}. The number of member stars detected in V band are 327 in the cluster NGC 2360, hence only these 327 stars, instead of total 332 member stars, were used to estimate the dynamic parameters as also mentioned in Table~\ref{dyna}.
\section{Discussion}\label{discussion}
Open star clusters are one of the most vital objects to study stellar and dynamical evolution of stars and their distributions over a large volume of the sky make them ideal candidates to study the dynamical evolution of Galactic disk and probe Galactic structure \citep{1998MNRAS.296.1045C,2003AJ....125.1397C, 2006A&A...445..545P, 2016A&A...593A.116J}. Though more than three thousand open star clusters are already known in the Galaxy \citep{2013A&A...558A..53K, 2018A&A...618A..93C}, not all of them are well studied. In order to investigate some of the most poorly studied OCs in our Galaxy, we carried out in-depth photometric study of three intermediate age clusters NGC 381, NGC 2360 and Be 68. Although few previous studies are already available for two of them but they lack studies of MF and dynamical evolution in those clusters \citep{1994ApJS...90...31P,2002AJ....123..905A,2015NewA...34..195O}. The biggest problem in carrying out any such study is the uncertainty on the true members of the clusters as contamination by field stars makes the estimated cluster parameters less reliable. However, with the availability of recent \textit{Gaia} DR2 catalogue, the informations on the stellar position, their parallaxes and proper motions are precisely available which makes membership determination in the clusters more reliable than the methods used in the past. 

In the present study, we carried out Johnson-Cousins $UBVRcIc$ photometric study of three clusters to derive accurate cluster parameters based on the members identified through proper motion studies from the \textit{Gaia} DR2 catalogues. We identified 116, 332, and 264 members in the clusters NGC 381, NGC 2360, and Be 68, respectively. We estimated reddening in the clusters using $(U-B)/(B-V)$ two-colour diagram which has been further used to estimate the age and distance of these clusters. We found the value of E(B-V) to be 0.36$\pm$0.04, 0.08$\pm$0.03 and 0.52$\pm$0.04 for NGC 381, NGC 2360, and Be 68, respectively. The near-IR photometry has also been widely used in the study of interstellar extinction because they are less effected by blanketing of metallic lines, thus are less sensitive to metallicity than the optical colours. Therefore, we supplemented our optical $UBVRcIc$ data with the 2MASS near-IR $JHK$ data and further derived E(B-V) from the ratio of colour excess $E(J-H)$ and $E(J-K)$. The E(B-V) values derived from the near-IR data are found to be in agreement with those derived through the optical data. We also derived the reddening law for these three clusters and found anomalous reddening law in the directions of the clusters NGC 381 with the $R_V$ values of 3.6. For NGC 2360 and Be 68, $R_V$ was found to be 3.0 and 3.2, respectively which are close to the normal value. The anomalous reddening in the direction of the cluster NGC 381 was further ascertained from the fact that we found a smaller value of reddening vector $\frac{E(B-V)}{E(U-B)}$ = 0.44$\pm$0.03 in comparison of normal value of 0.72. The large value of $R$ and small value of reddening vector in NGC 381 implies that the average dust grain size is larger than the average size in the direction of this cluster.

The prior knowledge of IMF, i.e. the frequency distribution of stellar masses at the time of birth of the star cluster, is of fundamental importance in the studies of the star formation process, chemical and dynamical evolution in the star clusters as clusters are considered to be fossil records of the star formation process. Since it is difficult to estimate the IMF for even young star clusters, one can only drive present day MF in order to understand the dynamical evolution in these clusters. We derived MF slopes for the three clusters. The two-step slopes of MFs in the cluster NGC 381 may be due to dynamics of clusters and initial conditions in star formation event in the cluster \citep[e.g.,][]{2005A&A...437..483B}. We estimated MFs slopes for stars above $\sim$1 M$_{\odot}$ as -2.11$\pm$0.35 and -2.08$\pm$0.94, and -2.02$\pm$0.82 for NGC 381, NGC 2360, and Be 68, respectively. To find any spatial variation in the MF slope within the clusters we estimated MF slopes in the inner and the outer regions of the all these clusters and found that the MF slopes are steeper in the outer region than the inner region of the clusters which is also reported in many OCs in the past \citep[e.g,][]{2006A&A...445..567B, 2008AJ....135.1934S, 2013AJ....145...46L}. The steeper slope of the MF in the outer region may be due to the mass segregation in the clusters. Further evidence of mass segregation in the clusters were found by studying cumulative variations of number density of stars in different mass ranges where we found mass segregation effect in the clusters NGC 381 and NGC 2360 with $\sim$90$\%$ and in Be 68 with $\sim$80$\%$ confidence level. The mass segregation in these clusters may be resulting due to dynamic evolution of the clusters where more massive stars clusters in the core and low-mas stars settle in the outer region of the clusters. In fact it cannot be ruled out that some of low mass stars which have acquired large enough velocity due to equipartition of energy might have actually escaped from these clusters. We calculated dynamical relaxation time, T$_{E}$, as 48.5, 78.9, and 87.6 Myr for the clusters NGC 381, NGC 2360 and Be 68, respectively. The smaller values of T$_{E}$ than the ages of these clusters imply that these clusters are dynamically relaxed.
\section{Summary}\label{summary}
We presented a comprehensive photometric study of three open star clusters NGC 381, NGC 2360, and Be 68 as a part of our ongoing effort to better characterize some poorly studied clusters, particularly young and intermediate-age open clusters in the Galaxy. The $UBVR_cI_c$ photometric data data has been taken with a 1.3-m telescope at Nainital. There have been only a few photometric studies of NGC 381 and NGC 2360 in the past while no previous photometry was available for the cluster Be 68 until now. We combined our observations with the 2MASS JHK near-IR data and recently available \textit{Gaia} DR2 proper motions data. We employed kinematic informations of stars in the clusters to identify their most-likely member stars in order to determine the physical parameters and probe the star formation process and dynamical history of these three clusters. The main points of our study are as following:
\begin{enumerate}

\item We derived the cluster radius from its radial density profile which are found to be 10$^{\prime}$.4$\pm$0$^{\prime}$.1, 12$^{\prime}$.1$\pm$0$^{\prime}$.1, and 4$^{\prime}$.7$\pm$0$^{\prime}$.1 for the clusters NGC 381, NGC 2360, and Be 68, respectively. The angular size of the cluster Be 68 was expected to be small due to its relatively large distance.

\item The recent \textit{Gaia} DR2 catalogue complemented with the deep, ground based photometric catalogue enables us to identify the most likely members in the clusters. Using VPD constructed from proper motions, we found a total of 116, 332, and 264 member stars in the clusters NGC 381, NGC 2360, and Be 68, respectively. All the analysis in the present study has been carried out on the basis of identified cluster members only.

\item We determined the mean proper motion, $\bar{\mu}$, of the clusters NGC 381, NGC 2360, and Be 68 as 2.78$\pm$0.12, 5.61$\pm$0.19, and 2.65$\pm$0.34 mas yr$^{-1}$, respectively.

\item The reddening, E(B-V), values were estimated to be 0.36$\pm$0.04, 0.08$\pm$0.03, and 0.52$\pm$0.04 mag in the direction of NGC 381, NGC 2360, and Be 68, respectively.

\item We found anomalous reddening law in the direction of NGC 381, however, no such behaviour was noticed in NGC 2360 and Be 68. This nature was found to be consistent in both optical and near-IR studies of these clusters.

\item The age were derived to be log(Age/yr) = 8.65$\pm$0.05, 8.95$\pm$0.05, and 9.25$\pm$0.05 for the clusters NGC 381, NGC 2360, and Be 68, respectively and their subsequent distances were estimated as 957$\pm$152, 982$\pm$132, and 2554$\pm$387 pc. The distances derived through \textit{Gaia} DR2 parallaxes were found to be in broad agreement with the distances measured using the isochrone fitting on the CMDs of the clusters.

\item The apparent age spread in the cluster NGC 2360 was found to be $\sim$595 Myr which is higher than those reported in some of the earlier studies.

\item We noticed a two-step slope in the present day MF of the cluster NGC 381 with an apparent break at $\sim$1.06 $M_\odot$. The slopes of the MF ($\Gamma$) in the entire regions of clusters NGC 381, NGC 2360, and Be 68 were estimated as -2.11$\pm$0.35, -2.08$\pm$0.94, and -2.02$\pm$0.82, respectively for the stars having mass above $\sim$1.0 M$_{\odot}$.

\item  The MF of the clusters NGC 381, NGC 2360, and Be 68 was not uniform over the entire region of the cluster extent hence we determined MF slope in inner as well as outer regions separately for these clusters. We noticed flattening of the slopes in the inner regions of these clusters in comparison of their outer regions.

\item We studied dynamical evolution in these clusters and found a deficiency of low-mass stars in core of the clusters. The dominance of massive stars in the core and low-mas stars in the outer regions of the clusters appear to be resulted due to mass segregation process which might be caused by the process of dynamic evolution of the clusters. The estimated relaxation times of these clusters suggest that all the three clusters are dynamically relaxed.

\end{enumerate} 
\section*{Acknowledgements}
This study was done using data from the Two Micron All Sky Survey, which is a joint project of the University of Massachusetts; the Infrared Processing and Analysis Center/California Institute of Technology, funded by the NASA. In the present study data have been used from the European Space Agency (ESA) mission Gaia (https://www.cosmos.esa.int/Gaia), processed by the Gaia Data Processing and Analysis Consortium (DPAC, https://www.cosmos.esa.int/web/GAIA/dpac/consortium).

\bibliographystyle{mnras}

\bibliography{main}

\begin{thebibliography}{}
\makeatletter
\relax
\def\mn@urlcharsother{\let\do\@makeother \do\$\do\&\do\#\do\^\do\_\do\%\do\~}
\def\mn@doi{\begingroup\mn@urlcharsother \@ifnextchar [ {\mn@doi@}
  {\mn@doi@[]}}
\def\mn@doi@[#1]#2{\def\@tempa{#1}\ifx\@tempa\@empty \href
  {http://dx.doi.org/#2} {doi:#2}\else \href {http://dx.doi.org/#2} {#1}\fi
  \endgroup}
\def\mn@eprint#1#2{\mn@eprint@#1:#2::\@nil}
\def\mn@eprint@arXiv#1{\href {http://arxiv.org/abs/#1} {{\tt arXiv:#1}}}
\def\mn@eprint@dblp#1{\href {http://dblp.uni-trier.de/rec/bibtex/#1.xml}
  {dblp:#1}}
\def\mn@eprint@#1:#2:#3:#4\@nil{\def\@tempa {#1}\def\@tempb {#2}\def\@tempc
  {#3}\ifx \@tempc \@empty \let \@tempc \@tempb \let \@tempb \@tempa \fi \ifx
  \@tempb \@empty \def\@tempb {arXiv}\fi \@ifundefined
  {mn@eprint@\@tempb}{\@tempb:\@tempc}{\expandafter \expandafter \csname
  mn@eprint@\@tempb\endcsname \expandafter{\@tempc}}}

\bibitem[\protect\citeauthoryear{{Alonso}, {Arribas}  \&
  {Martinez-Roger}}{{Alonso} et~al.}{1996}]{1996A&A...313..873A}
{Alonso} A.,  {Arribas} S.,   {Martinez-Roger} C.,  1996, \aap, \href
  {https://ui.adsabs.harvard.edu/abs/1996A&A...313..873A} {313, 873}

\bibitem[\protect\citeauthoryear{{Anders} et~al.,}{{Anders}
  et~al.}{2019}]{2019A&A...628A..94A}
{Anders} F.,  et~al., 2019, \mn@doi [\aap] {10.1051/0004-6361/201935765}, \href
  {https://ui.adsabs.harvard.edu/abs/2019A&A...628A..94A} {628, A94}

\bibitem[\protect\citeauthoryear{{Ann} et~al.,}{{Ann}
  et~al.}{2002}]{2002AJ....123..905A}
{Ann} H.~B.,  et~al., 2002, \mn@doi [\aj] {10.1086/338309}, \href
  {https://ui.adsabs.harvard.edu/abs/2002AJ....123..905A} {123, 905}

\bibitem[\protect\citeauthoryear{{Bisht}, {Yadav}, {Ganesh}, {Durgapal},
  {Rangwal}  \& {Fynbo}}{{Bisht} et~al.}{2019}]{2019MNRAS.482.1471B}
{Bisht} D.,  {Yadav} R.~K.~S.,  {Ganesh} S.,  {Durgapal} A.~K.,  {Rangwal} G.,
   {Fynbo} J.~P.~U.,  2019, \mn@doi [\mnras] {10.1093/mnras/sty2781}, \href
  {https://ui.adsabs.harvard.edu/abs/2019MNRAS.482.1471B} {482, 1471}

\bibitem[\protect\citeauthoryear{{Bonatto} \& {Bica}}{{Bonatto} \&
  {Bica}}{2005}]{2005A&A...437..483B}
{Bonatto} C.,  {Bica} E.,  2005, \mn@doi [\aap] {10.1051/0004-6361:20042516},
  \href {https://ui.adsabs.harvard.edu/abs/2005A&A...437..483B} {437, 483}

\bibitem[\protect\citeauthoryear{{Bonatto} \& {Bica}}{{Bonatto} \&
  {Bica}}{2009}]{2009MNRAS.397.1915B}
{Bonatto} C.,  {Bica} E.,  2009, \mn@doi [\mnras]
  {10.1111/j.1365-2966.2009.14877.x}, \href
  {https://ui.adsabs.harvard.edu/abs/2009MNRAS.397.1915B} {397, 1915}

\bibitem[\protect\citeauthoryear{{Bonatto}, {Santos}  \& {Bica}}{{Bonatto}
  et~al.}{2006}]{2006A&A...445..567B}
{Bonatto} C.,  {Santos} J.~F.~C. J.,   {Bica} E.,  2006, \mn@doi [\aap]
  {10.1051/0004-6361:20052793}, \href
  {https://ui.adsabs.harvard.edu/abs/2006A&A...445..567B} {445, 567}

\bibitem[\protect\citeauthoryear{{Bossini} et~al.,}{{Bossini}
  et~al.}{2019}]{2019A&A...623A.108B}
{Bossini} D.,  et~al., 2019, \mn@doi [\aap] {10.1051/0004-6361/201834693},
  \href {https://ui.adsabs.harvard.edu/abs/2019A&A...623A.108B} {623, A108}

\bibitem[\protect\citeauthoryear{{Bostanc{\i}} et~al.,}{{Bostanc{\i}}
  et~al.}{2018}]{2018Ap&SS.363..143B}
{Bostanc{\i}} Z.~F.,  et~al., 2018, \mn@doi [\apss]
  {10.1007/s10509-018-3364-4}, \href
  {https://ui.adsabs.harvard.edu/abs/2018Ap&SS.363..143B} {363, 143}

\bibitem[\protect\citeauthoryear{{Cantat-Gaudin} et~al.,}{{Cantat-Gaudin}
  et~al.}{2018}]{2018A&A...618A..93C}
{Cantat-Gaudin} T.,  et~al., 2018, \mn@doi [\aap]
  {10.1051/0004-6361/201833476}, \href
  {https://ui.adsabs.harvard.edu/abs/2018A&A...618A..93C} {618, A93}

\bibitem[\protect\citeauthoryear{{Cantat-Gaudin} et~al.,}{{Cantat-Gaudin}
  et~al.}{2019}]{2019A&A...624A.126C}
{Cantat-Gaudin} T.,  et~al., 2019, \mn@doi [\aap]
  {10.1051/0004-6361/201834453}, \href
  {https://ui.adsabs.harvard.edu/abs/2019A&A...624A.126C} {624, A126}

\bibitem[\protect\citeauthoryear{{Cardelli}, {Clayton}  \& {Mathis}}{{Cardelli}
  et~al.}{1989}]{1989ApJ...345..245C}
{Cardelli} J.~A.,  {Clayton} G.~C.,   {Mathis} J.~S.,  1989, \mn@doi [\apj]
  {10.1086/167900}, \href
  {https://ui.adsabs.harvard.edu/abs/1989ApJ...345..245C} {345, 245}

\bibitem[\protect\citeauthoryear{{Carney}, {Lee}  \& {Dodson}}{{Carney}
  et~al.}{2005}]{2005AJ....129..656C}
{Carney} B.~W.,  {Lee} J.-W.,   {Dodson} B.,  2005, \mn@doi [\aj]
  {10.1086/426754}, \href
  {https://ui.adsabs.harvard.edu/abs/2005AJ....129..656C} {129, 656}

\bibitem[\protect\citeauthoryear{{Carpenter}}{{Carpenter}}{2001}]{2001AJ....121.2851C}
{Carpenter} J.~M.,  2001, \mn@doi [\aj] {10.1086/320383}, \href
  {https://ui.adsabs.harvard.edu/abs/2001AJ....121.2851C} {121, 2851}

\bibitem[\protect\citeauthoryear{{Carraro}, {Ng}  \& {Portinari}}{{Carraro}
  et~al.}{1998}]{1998MNRAS.296.1045C}
{Carraro} G.,  {Ng} Y.~K.,   {Portinari} L.,  1998, \mn@doi [\mnras]
  {10.1046/j.1365-8711.1998.01460.x}, \href
  {https://ui.adsabs.harvard.edu/abs/1998MNRAS.296.1045C} {296, 1045}

\bibitem[\protect\citeauthoryear{{Carraro}, {Villanova}, {Demarque}, {Moni
  Bidin}  \& {McSwain}}{{Carraro} et~al.}{2008}]{2008MNRAS.386.1625C}
{Carraro} G.,  {Villanova} S.,  {Demarque} P.,  {Moni Bidin} C.,   {McSwain}
  M.~V.,  2008, \mn@doi [\mnras] {10.1111/j.1365-2966.2008.13143.x}, \href
  {https://ui.adsabs.harvard.edu/abs/2008MNRAS.386.1625C} {386, 1625}

\bibitem[\protect\citeauthoryear{{Castro-Ginard}, {Jordi}, {Luri}, {Julbe},
  {Morvan}, {Balaguer-N{\'u}{\~n}ez}  \& {Cantat-Gaudin}}{{Castro-Ginard}
  et~al.}{2018}]{2018A&A...618A..59C}
{Castro-Ginard} A.,  {Jordi} C.,  {Luri} X.,  {Julbe} F.,  {Morvan} M.,
  {Balaguer-N{\'u}{\~n}ez} L.,   {Cantat-Gaudin} T.,  2018, \mn@doi [\aap]
  {10.1051/0004-6361/201833390}, \href
  {https://ui.adsabs.harvard.edu/abs/2018A&A...618A..59C} {618, A59}

\bibitem[\protect\citeauthoryear{{Castro-Ginard}, {Jordi}, {Luri},
  {Cantat-Gaudin}  \& {Balaguer-N{\'u}{\~n}ez}}{{Castro-Ginard}
  et~al.}{2019}]{2019A&A...627A..35C}
{Castro-Ginard} A.,  {Jordi} C.,  {Luri} X.,  {Cantat-Gaudin} T.,
  {Balaguer-N{\'u}{\~n}ez} L.,  2019, \mn@doi [\aap]
  {10.1051/0004-6361/201935531}, \href
  {https://ui.adsabs.harvard.edu/abs/2019A&A...627A..35C} {627, A35}

\bibitem[\protect\citeauthoryear{{Chen}, {Hou}  \& {Wang}}{{Chen}
  et~al.}{2003}]{2003AJ....125.1397C}
{Chen} L.,  {Hou} J.~L.,   {Wang} J.~J.,  2003, \mn@doi [\aj] {10.1086/367911},
  \href {https://ui.adsabs.harvard.edu/abs/2003AJ....125.1397C} {125, 1397}

\bibitem[\protect\citeauthoryear{{Chen} et~al.,}{{Chen}
  et~al.}{2011}]{2011AJ....142...71C}
{Chen} W.~P.,  et~al., 2011, \mn@doi [\aj] {10.1088/0004-6256/142/3/71}, \href
  {https://ui.adsabs.harvard.edu/abs/2011AJ....142...71C} {142, 71}

\bibitem[\protect\citeauthoryear{{Chini} \& {Wargau}}{{Chini} \&
  {Wargau}}{1990}]{1990A&A...227..213C}
{Chini} R.,  {Wargau} W.~F.,  1990, \aap, \href
  {https://ui.adsabs.harvard.edu/abs/1990A&A...227..213C} {227, 213}

\bibitem[\protect\citeauthoryear{{Chumak}, {Platais}, {McLaughlin},
  {Rastorguev}  \& {Chumak}}{{Chumak} et~al.}{2010}]{2010MNRAS.402.1841C}
{Chumak} Y.~O.,  {Platais} I.,  {McLaughlin} D.~E.,  {Rastorguev} A.~S.,
  {Chumak} O.~V.,  2010, \mn@doi [\mnras] {10.1111/j.1365-2966.2009.16021.x},
  \href {https://ui.adsabs.harvard.edu/abs/2010MNRAS.402.1841C} {402, 1841}

\bibitem[\protect\citeauthoryear{{Clari{\'a}}, {Piatti}, {Mermilliod}  \&
  {Palma}}{{Clari{\'a}} et~al.}{2008}]{2008AN....329..609C}
{Clari{\'a}} J.~J.,  {Piatti} A.~E.,  {Mermilliod} J.~C.,   {Palma} T.,  2008,
  \mn@doi [Astronomische Nachrichten] {10.1002/asna.200710970}, \href
  {https://ui.adsabs.harvard.edu/abs/2008AN....329..609C} {329, 609}

\bibitem[\protect\citeauthoryear{{Cordoni}, {Milone}, {Marino}, {Di
  Criscienzo}, {D'Antona}, {Dotter}, {Lagioia}  \& {Tailo}}{{Cordoni}
  et~al.}{2018}]{2018ApJ...869..139C}
{Cordoni} G.,  {Milone} A.~P.,  {Marino} A.~F.,  {Di Criscienzo} M.,
  {D'Antona} F.,  {Dotter} A.,  {Lagioia} E.~P.,   {Tailo} M.,  2018, \mn@doi
  [\apj] {10.3847/1538-4357/aaedc1}, \href
  {https://ui.adsabs.harvard.edu/abs/2018ApJ...869..139C} {869, 139}

\bibitem[\protect\citeauthoryear{{Dalessandro}, {Miocchi}, {Carraro},
  {J{\'\i}lkov{\'a}}  \& {Moitinho}}{{Dalessandro}
  et~al.}{2015}]{2015MNRAS.449.1811D}
{Dalessandro} E.,  {Miocchi} P.,  {Carraro} G.,  {J{\'\i}lkov{\'a}} L.,
  {Moitinho} A.,  2015, \mn@doi [\mnras] {10.1093/mnras/stv395}, \href
  {https://ui.adsabs.harvard.edu/abs/2015MNRAS.449.1811D} {449, 1811}

\bibitem[\protect\citeauthoryear{{Dib}, {Schmeja}  \& {Hony}}{{Dib}
  et~al.}{2017}]{2017MNRAS.464.1738D}
{Dib} S.,  {Schmeja} S.,   {Hony} S.,  2017, \mn@doi [\mnras]
  {10.1093/mnras/stw2465}, \href
  {https://ui.adsabs.harvard.edu/abs/2017MNRAS.464.1738D} {464, 1738}

\bibitem[\protect\citeauthoryear{{Dib}, {Schmeja}  \& {Parker}}{{Dib}
  et~al.}{2018}]{2018MNRAS.473..849D}
{Dib} S.,  {Schmeja} S.,   {Parker} R.~J.,  2018, \mn@doi [\mnras]
  {10.1093/mnras/stx2413}, \href
  {https://ui.adsabs.harvard.edu/abs/2018MNRAS.473..849D} {473, 849}

\bibitem[\protect\citeauthoryear{{Friel} \& {Janes}}{{Friel} \&
  {Janes}}{1993}]{1993A&A...267...75F}
{Friel} E.~D.,  {Janes} K.~A.,  1993, \aap, \href
  {https://ui.adsabs.harvard.edu/abs/1993A&A...267...75F} {267, 75}

\bibitem[\protect\citeauthoryear{{Gaia Collaboration} et~al.,}{{Gaia
  Collaboration} et~al.}{2018}]{2018A&A...616A...1G}
{Gaia Collaboration} et~al., 2018, \mn@doi [\aap]
  {10.1051/0004-6361/201833051}, \href
  {https://ui.adsabs.harvard.edu/abs/2018A&A...616A...1G} {616, A1}

\bibitem[\protect\citeauthoryear{{Gao}}{{Gao}}{2014}]{2014RAA....14..159G}
{Gao} X.-H.,  2014, \mn@doi [Research in Astronomy and Astrophysics]
  {10.1088/1674-4527/14/2/004}, \href
  {https://ui.adsabs.harvard.edu/abs/2014RAA....14..159G} {14, 159}

\bibitem[\protect\citeauthoryear{{Garcia}, {Claria}  \& {Levato}}{{Garcia}
  et~al.}{1988}]{1988Ap&SS.143..317G}
{Garcia} B.,  {Claria} J.~J.,   {Levato} H.,  1988, \mn@doi [\apss]
  {10.1007/BF00637143}, \href
  {https://ui.adsabs.harvard.edu/abs/1988Ap&SS.143..317G} {143, 317}

\bibitem[\protect\citeauthoryear{{Genzel} \& {Townes}}{{Genzel} \&
  {Townes}}{1987}]{1987ARA&A..25..377G}
{Genzel} R.,  {Townes} C.~H.,  1987, \mn@doi [\araa]
  {10.1146/annurev.aa.25.090187.002113}, \href
  {https://ui.adsabs.harvard.edu/abs/1987ARA&A..25..377G} {25, 377}

\bibitem[\protect\citeauthoryear{{Hur}, {Sung}  \& {Bessell}}{{Hur}
  et~al.}{2012}]{2012AJ....143...41H}
{Hur} H.,  {Sung} H.,   {Bessell} M.~S.,  2012, \mn@doi [\aj]
  {10.1088/0004-6256/143/2/41}, \href
  {https://ui.adsabs.harvard.edu/abs/2012AJ....143...41H} {143, 41}

\bibitem[\protect\citeauthoryear{{Jeffries}, {Thurston}  \&
  {Hambly}}{{Jeffries} et~al.}{2001}]{2001A&A...375..863J}
{Jeffries} R.~D.,  {Thurston} M.~R.,   {Hambly} N.~C.,  2001, \mn@doi [\aap]
  {10.1051/0004-6361:20010918}, \href
  {https://ui.adsabs.harvard.edu/abs/2001A&A...375..863J} {375, 863}

\bibitem[\protect\citeauthoryear{{Jose} et~al.,}{{Jose}
  et~al.}{2008}]{2008MNRAS.384.1675J}
{Jose} J.,  et~al., 2008, \mn@doi [\mnras] {10.1111/j.1365-2966.2007.12825.x},
  \href {https://ui.adsabs.harvard.edu/abs/2008MNRAS.384.1675J} {384, 1675}

\bibitem[\protect\citeauthoryear{{Joshi}}{{Joshi}}{2005}]{2005MNRAS.362.1259J}
{Joshi} Y.~C.,  2005, \mn@doi [\mnras] {10.1111/j.1365-2966.2005.09391.x},
  \href {https://ui.adsabs.harvard.edu/abs/2005MNRAS.362.1259J} {362, 1259}

\bibitem[\protect\citeauthoryear{{Joshi}}{{Joshi}}{2007}]{2007MNRAS.378..768J}
{Joshi} Y.~C.,  2007, \mn@doi [\mnras] {10.1111/j.1365-2966.2007.11831.x},
  \href {https://ui.adsabs.harvard.edu/abs/2007MNRAS.378..768J} {378, 768}

\bibitem[\protect\citeauthoryear{{Joshi}, {Joshi}, {Kumar}, {Mondal}  \&
  {Balona}}{{Joshi} et~al.}{2012}]{2012MNRAS.419.2379J}
{Joshi} Y.~C.,  {Joshi} S.,  {Kumar} B.,  {Mondal} S.,   {Balona} L.~A.,  2012,
  \mn@doi [\mnras] {10.1111/j.1365-2966.2011.19890.x}, \href
  {https://ui.adsabs.harvard.edu/abs/2012MNRAS.419.2379J} {419, 2379}

\bibitem[\protect\citeauthoryear{{Joshi}, {Balona}, {Joshi}  \&
  {Kumar}}{{Joshi} et~al.}{2014}]{2014MNRAS.437..804J}
{Joshi} Y.~C.,  {Balona} L.~A.,  {Joshi} S.,   {Kumar} B.,  2014, \mn@doi
  [\mnras] {10.1093/mnras/stt1939}, \href
  {https://ui.adsabs.harvard.edu/abs/2014MNRAS.437..804J} {437, 804}

\bibitem[\protect\citeauthoryear{{Joshi}, {Dambis}, {Pandey}  \&
  {Joshi}}{{Joshi} et~al.}{2016}]{2016A&A...593A.116J}
{Joshi} Y.~C.,  {Dambis} A.~K.,  {Pandey} A.~K.,   {Joshi} S.,  2016, \mn@doi
  [\aap] {10.1051/0004-6361/201628944}, \href
  {https://ui.adsabs.harvard.edu/abs/2016A&A...593A.116J} {593, A116}

\bibitem[\protect\citeauthoryear{{Joshi}, {Maurya}, {John}, {Panchal}, {Joshi}
  \& {Kumar}}{{Joshi} et~al.}{2020}]{2020MNRAS.492.3602J}
{Joshi} Y.~C.,  {Maurya} J.,  {John} A.~A.,  {Panchal} A.,  {Joshi} S.,
  {Kumar} B.,  2020, \mn@doi [\mnras] {10.1093/mnras/staa029}, \href
  {https://ui.adsabs.harvard.edu/abs/2020MNRAS.492.3602J} {492, 3602}

\bibitem[\protect\citeauthoryear{{Kaluzny} \& {Udalski}}{{Kaluzny} \&
  {Udalski}}{1992}]{1992AcA....42...29K}
{Kaluzny} J.,  {Udalski} A.,  1992, \actaa, \href
  {https://ui.adsabs.harvard.edu/abs/1992AcA....42...29K} {42, 29}

\bibitem[\protect\citeauthoryear{{Khalaj} \& {Baumgardt}}{{Khalaj} \&
  {Baumgardt}}{2013}]{2013MNRAS.434.3236K}
{Khalaj} P.,  {Baumgardt} H.,  2013, \mn@doi [\mnras] {10.1093/mnras/stt1239},
  \href {https://ui.adsabs.harvard.edu/abs/2013MNRAS.434.3236K} {434, 3236}

\bibitem[\protect\citeauthoryear{{Kharchenko}, {Piskunov}, {Schilbach},
  {R{\"o}ser}  \& {Scholz}}{{Kharchenko} et~al.}{2013}]{2013A&A...558A..53K}
{Kharchenko} N.~V.,  {Piskunov} A.~E.,  {Schilbach} E.,  {R{\"o}ser} S.,
  {Scholz} R.~D.,  2013, \mn@doi [\aap] {10.1051/0004-6361/201322302}, \href
  {https://ui.adsabs.harvard.edu/abs/2013A&A...558A..53K} {558, A53}

\bibitem[\protect\citeauthoryear{{Kim}, {Figer}, {Lee}  \& {Morris}}{{Kim}
  et~al.}{2000}]{2000ApJ...545..301K}
{Kim} S.~S.,  {Figer} D.~F.,  {Lee} H.~M.,   {Morris} M.,  2000, \mn@doi [\apj]
  {10.1086/317807}, \href
  {https://ui.adsabs.harvard.edu/abs/2000ApJ...545..301K} {545, 301}

\bibitem[\protect\citeauthoryear{{King}}{{King}}{1962}]{1962AJ.....67..471K}
{King} I.,  1962, \mn@doi [\aj] {10.1086/108756}, \href
  {https://ui.adsabs.harvard.edu/abs/1962AJ.....67..471K} {67, 471}

\bibitem[\protect\citeauthoryear{{Kroupa}}{{Kroupa}}{2002}]{2002Sci...295...82K}
{Kroupa} P.,  2002, \mn@doi [Science] {10.1126/science.1067524}, \href
  {https://ui.adsabs.harvard.edu/abs/2002Sci...295...82K} {295, 82}

\bibitem[\protect\citeauthoryear{{Lada} \& {Lada}}{{Lada} \&
  {Lada}}{2003}]{2003ARA&A..41...57L}
{Lada} C.~J.,  {Lada} E.~A.,  2003, \mn@doi [\araa]
  {10.1146/annurev.astro.41.011802.094844}, \href
  {https://ui.adsabs.harvard.edu/abs/2003ARA&A..41...57L} {41, 57}

\bibitem[\protect\citeauthoryear{{Lim}, {Chun}, {Sung}, {Park}, {Lee}, {Sohn},
  {Hur}  \& {Bessell}}{{Lim} et~al.}{2013}]{2013AJ....145...46L}
{Lim} B.,  {Chun} M.-Y.,  {Sung} H.,  {Park} B.-G.,  {Lee} J.-J.,  {Sohn}
  S.~T.,  {Hur} H.,   {Bessell} M.~S.,  2013, \mn@doi [\aj]
  {10.1088/0004-6256/145/2/46}, \href
  {https://ui.adsabs.harvard.edu/abs/2013AJ....145...46L} {145, 46}

\bibitem[\protect\citeauthoryear{{Lindegren} et~al.,}{{Lindegren}
  et~al.}{2018}]{2018A&A...616A...2L}
{Lindegren} L.,  et~al., 2018, \mn@doi [\aap] {10.1051/0004-6361/201832727},
  \href {https://ui.adsabs.harvard.edu/abs/2018A&A...616A...2L} {616, A2}

\bibitem[\protect\citeauthoryear{{Liu} \& {Pang}}{{Liu} \&
  {Pang}}{2019}]{2019ApJS..245...32L}
{Liu} L.,  {Pang} X.,  2019, \mn@doi [\apjs] {10.3847/1538-4365/ab530a}, \href
  {https://ui.adsabs.harvard.edu/abs/2019ApJS..245...32L} {245, 32}

\bibitem[\protect\citeauthoryear{{Luri} et~al.,}{{Luri}
  et~al.}{2018}]{2018A&A...616A...9L}
{Luri} X.,  et~al., 2018, \mn@doi [\aap] {10.1051/0004-6361/201832964}, \href
  {https://ui.adsabs.harvard.edu/abs/2018A&A...616A...9L} {616, A9}

\bibitem[\protect\citeauthoryear{{Mackey} \& {Broby Nielsen}}{{Mackey} \&
  {Broby Nielsen}}{2007}]{2007MNRAS.379..151M}
{Mackey} A.~D.,  {Broby Nielsen} P.,  2007, \mn@doi [\mnras]
  {10.1111/j.1365-2966.2007.11915.x}, \href
  {https://ui.adsabs.harvard.edu/abs/2007MNRAS.379..151M} {379, 151}

\bibitem[\protect\citeauthoryear{{Marigo} et~al.,}{{Marigo}
  et~al.}{2017}]{2017ApJ...835...77M}
{Marigo} P.,  et~al., 2017, \mn@doi [\apj] {10.3847/1538-4357/835/1/77}, \href
  {https://ui.adsabs.harvard.edu/abs/2017ApJ...835...77M} {835, 77}

\bibitem[\protect\citeauthoryear{{Marino}, {Milone}, {Casagrande}, {Przybilla},
  {Balaguer-N{\'u}{\~n}ez}, {Di Criscienzo}, {Serenelli}  \&
  {Vilardell}}{{Marino} et~al.}{2018}]{2018ApJ...863L..33M}
{Marino} A.~F.,  {Milone} A.~P.,  {Casagrande} L.,  {Przybilla} N.,
  {Balaguer-N{\'u}{\~n}ez} L.,  {Di Criscienzo} M.,  {Serenelli} A.,
  {Vilardell} F.,  2018, \mn@doi [\apjl] {10.3847/2041-8213/aad868}, \href
  {https://ui.adsabs.harvard.edu/abs/2018ApJ...863L..33M} {863, L33}

\bibitem[\protect\citeauthoryear{{Mathieu} \& {Latham}}{{Mathieu} \&
  {Latham}}{1986}]{1986AJ.....92.1364M}
{Mathieu} R.~D.,  {Latham} D.~W.,  1986, \mn@doi [\aj] {10.1086/114269}, \href
  {https://ui.adsabs.harvard.edu/abs/1986AJ.....92.1364M} {92, 1364}

\bibitem[\protect\citeauthoryear{{Mathis}}{{Mathis}}{1990}]{1990ARA&A..28...37M}
{Mathis} J.~S.,  1990, \mn@doi [\araa] {10.1146/annurev.aa.28.090190.000345},
  \href {https://ui.adsabs.harvard.edu/abs/1990ARA&A..28...37M} {28, 37}

\bibitem[\protect\citeauthoryear{{Medhi} \& {Tamura}}{{Medhi} \&
  {Tamura}}{2013}]{2013MNRAS.430.1334M}
{Medhi} B.~J.,  {Tamura} M.,  2013, \mn@doi [\mnras] {10.1093/mnras/sts714},
  \href {https://ui.adsabs.harvard.edu/abs/2013MNRAS.430.1334M} {430, 1334}

\bibitem[\protect\citeauthoryear{{Michalska}}{{Michalska}}{2019}]{2019MNRAS.487.3505M}
{Michalska} G.,  2019, \mn@doi [\mnras] {10.1093/mnras/stz1500}, \href
  {https://ui.adsabs.harvard.edu/abs/2019MNRAS.487.3505M} {487, 3505}

\bibitem[\protect\citeauthoryear{{Neckel} \& {Chini}}{{Neckel} \&
  {Chini}}{1981}]{1981A&AS...45..451N}
{Neckel} T.,  {Chini} R.,  1981, \aaps, \href
  {https://ui.adsabs.harvard.edu/abs/1981A&AS...45..451N} {45, 451}

\bibitem[\protect\citeauthoryear{{Nilakshi}, {Sagar}, {Pandey}  \&
  {Mohan}}{{Nilakshi} et~al.}{2002}]{2002A&A...383..153N}
{Nilakshi} {Sagar} R.,  {Pandey} A.~K.,   {Mohan} V.,  2002, \mn@doi [\aap]
  {10.1051/0004-6361:20011719}, \href
  {https://ui.adsabs.harvard.edu/abs/2002A&A...383..153N} {383, 153}

\bibitem[\protect\citeauthoryear{{Oralhan}, Karata{\c s}, {Schuster}, {Michel}
  \& {Chavarr{\'\i}a}}{{Oralhan} et~al.}{2015}]{2015NewA...34..195O}
{Oralhan} {\.I}.~A.,  Karata{\c s} Y. Y.,  {Schuster} W.~J.,  {Michel} R.,
  {Chavarr{\'\i}a} C.,  2015, \mn@doi [\na] {10.1016/j.newast.2014.06.011},
  \href {https://ui.adsabs.harvard.edu/abs/2015NewA...34..195O} {34, 195}

\bibitem[\protect\citeauthoryear{{Pandey} et~al.,}{{Pandey}
  et~al.}{2013}]{2013ApJ...764..172P}
{Pandey} A.~K.,  et~al., 2013, \mn@doi [\apj] {10.1088/0004-637X/764/2/172},
  \href {https://ui.adsabs.harvard.edu/abs/2013ApJ...764..172P} {764, 172}

\bibitem[\protect\citeauthoryear{{Panwar}, {Pandey}, {Samal}, {Battinelli},
  {Ogura}, {Ojha}, {Chen}  \& {Singh}}{{Panwar}
  et~al.}{2018}]{2018AJ....155...44P}
{Panwar} N.,  {Pandey} A.~K.,  {Samal} M.~R.,  {Battinelli} P.,  {Ogura} K.,
  {Ojha} D.~K.,  {Chen} W.~P.,   {Singh} H.~P.,  2018, \mn@doi [\aj]
  {10.3847/1538-3881/aa9f1b}, \href
  {https://ui.adsabs.harvard.edu/abs/2018AJ....155...44P} {155, 44}

\bibitem[\protect\citeauthoryear{{Parker}, {Goodwin}, {Wright}, {Meyer}  \&
  {Quanz}}{{Parker} et~al.}{2016}]{2016MNRAS.459L.119P}
{Parker} R.~J.,  {Goodwin} S.~P.,  {Wright} N.~J.,  {Meyer} M.~R.,   {Quanz}
  S.~P.,  2016, \mn@doi [\mnras] {10.1093/mnrasl/slw061}, \href
  {https://ui.adsabs.harvard.edu/abs/2016MNRAS.459L.119P} {459, L119}

\bibitem[\protect\citeauthoryear{{Phelps} \& {Janes}}{{Phelps} \&
  {Janes}}{1994}]{1994ApJS...90...31P}
{Phelps} R.~L.,  {Janes} K.~A.,  1994, \mn@doi [\apjs] {10.1086/191857}, \href
  {https://ui.adsabs.harvard.edu/abs/1994ApJS...90...31P} {90, 31}

\bibitem[\protect\citeauthoryear{{Piatti}}{{Piatti}}{2016}]{2016MNRAS.463.3476P}
{Piatti} A.~E.,  2016, \mn@doi [\mnras] {10.1093/mnras/stw2248}, \href
  {https://ui.adsabs.harvard.edu/abs/2016MNRAS.463.3476P} {463, 3476}

\bibitem[\protect\citeauthoryear{{Piatti} \& {Bonatto}}{{Piatti} \&
  {Bonatto}}{2019}]{2019MNRAS.490.2414P}
{Piatti} A.~E.,  {Bonatto} C.,  2019, \mn@doi [\mnras] {10.1093/mnras/stz2798},
  \href {https://ui.adsabs.harvard.edu/abs/2019MNRAS.490.2414P} {490, 2414}

\bibitem[\protect\citeauthoryear{{Piskunov}, {Kharchenko}, {R{\"o}ser},
  {Schilbach}  \& {Scholz}}{{Piskunov} et~al.}{2006}]{2006A&A...445..545P}
{Piskunov} A.~E.,  {Kharchenko} N.~V.,  {R{\"o}ser} S.,  {Schilbach} E.,
  {Scholz} R.~D.,  2006, \mn@doi [\aap] {10.1051/0004-6361:20053764}, \href
  {https://ui.adsabs.harvard.edu/abs/2006A&A...445..545P} {445, 545}

\bibitem[\protect\citeauthoryear{{Rui}, {Hosek}, {Lu}, {Clarkson}, {Anderson},
  {Morris}  \& {Ghez}}{{Rui} et~al.}{2019}]{2019ApJ...877...37R}
{Rui} N.~Z.,  {Hosek} Matthew~W. J.,  {Lu} J.~R.,  {Clarkson} W.~I.,
  {Anderson} J.,  {Morris} M.~R.,   {Ghez} A.~M.,  2019, \mn@doi [\apj]
  {10.3847/1538-4357/ab17e0}, \href
  {https://ui.adsabs.harvard.edu/abs/2019ApJ...877...37R} {877, 37}

\bibitem[\protect\citeauthoryear{{Sagar}, {Miakutin}, {Piskunov}  \&
  {Dluzhnevskaia}}{{Sagar} et~al.}{1988}]{1988MNRAS.234..831S}
{Sagar} R.,  {Miakutin} V.~I.,  {Piskunov} A.~E.,   {Dluzhnevskaia} O.~B.,
  1988, \mn@doi [\mnras] {10.1093/mnras/234.4.831}, \href
  {https://ui.adsabs.harvard.edu/abs/1988MNRAS.234..831S} {234, 831}

\bibitem[\protect\citeauthoryear{{Salpeter}}{{Salpeter}}{1955}]{1955ApJ...121..161S}
{Salpeter} E.~E.,  1955, \mn@doi [\apj] {10.1086/145971}, \href
  {https://ui.adsabs.harvard.edu/abs/1955ApJ...121..161S} {121, 161}

\bibitem[\protect\citeauthoryear{{Sanders}}{{Sanders}}{1971}]{1971A&A....14..226S}
{Sanders} W.~L.,  1971, \aap, \href
  {https://ui.adsabs.harvard.edu/abs/1971A&A....14..226S} {14, 226}

\bibitem[\protect\citeauthoryear{{Sariya} \& {Yadav}}{{Sariya} \&
  {Yadav}}{2015}]{2015A&A...584A..59S}
{Sariya} D.~P.,  {Yadav} R.~K.~S.,  2015, \mn@doi [\aap]
  {10.1051/0004-6361/201526688}, \href
  {https://ui.adsabs.harvard.edu/abs/2015A&A...584A..59S} {584, A59}

\bibitem[\protect\citeauthoryear{{Schmidt - Kaler}}{{Schmidt -
  Kaler}}{1982}]{Schmidt-Kaler}
{Schmidt - Kaler} T.,  1982, \mn@doi [New series]
  {10.1088/0004-6256/135/5/1934}, \href
  {https://ui.adsabs.harvard.edu/abs/2008AJ....135.1934S} {Group VI, vol. 2b.
  springer - verlag, p. 14}

\bibitem[\protect\citeauthoryear{{Sharma}, {Pandey}, {Ogura}, {Aoki}, {Pandey},
  {Sandhu}  \& {Sagar}}{{Sharma} et~al.}{2008}]{2008AJ....135.1934S}
{Sharma} S.,  {Pandey} A.~K.,  {Ogura} K.,  {Aoki} T.,  {Pandey} K.,  {Sandhu}
  T.~S.,   {Sagar} R.,  2008, \mn@doi [\aj] {10.1088/0004-6256/135/5/1934},
  \href {https://ui.adsabs.harvard.edu/abs/2008AJ....135.1934S} {135, 1934}

\bibitem[\protect\citeauthoryear{{Siegel}, {LaPorte}, {Porterfield}, {Hagen}
  \& {Gronwall}}{{Siegel} et~al.}{2019}]{2019AJ....158...35S}
{Siegel} M.~H.,  {LaPorte} S.~J.,  {Porterfield} B.~L.,  {Hagen} L. M.~Z.,
  {Gronwall} C.~A.,  2019, \mn@doi [\aj] {10.3847/1538-3881/ab21e1}, \href
  {https://ui.adsabs.harvard.edu/abs/2019AJ....158...35S} {158, 35}

\bibitem[\protect\citeauthoryear{{Skrutskie} et~al.,}{{Skrutskie}
  et~al.}{2006}]{2006AJ....131.1163S}
{Skrutskie} M.~F.,  et~al., 2006, \mn@doi [\aj] {10.1086/498708}, \href
  {https://ui.adsabs.harvard.edu/abs/2006AJ....131.1163S} {131, 1163}

\bibitem[\protect\citeauthoryear{{Sneden}, {Gehrz}, {Hackwell}, {York}  \&
  {Snow}}{{Sneden} et~al.}{1978}]{1978ApJ...223..168S}
{Sneden} C.,  {Gehrz} R.~D.,  {Hackwell} J.~A.,  {York} D.~G.,   {Snow} T.~P.,
  1978, \mn@doi [\apj] {10.1086/156247}, \href
  {https://ui.adsabs.harvard.edu/abs/1978ApJ...223..168S} {223, 168}

\bibitem[\protect\citeauthoryear{{Spitzer} \& {Hart}}{{Spitzer} \&
  {Hart}}{1971}]{1971ApJ...164..399S}
{Spitzer} Lyman J.,  {Hart} M.~H.,  1971, \mn@doi [\apj] {10.1086/150855},
  \href {https://ui.adsabs.harvard.edu/abs/1971ApJ...164..399S} {164, 399}

\bibitem[\protect\citeauthoryear{{Stassun} \& {Torres}}{{Stassun} \&
  {Torres}}{2018}]{2018ApJ...862...61S}
{Stassun} K.~G.,  {Torres} G.,  2018, \mn@doi [\apj]
  {10.3847/1538-4357/aacafc}, \href
  {https://ui.adsabs.harvard.edu/abs/2018ApJ...862...61S} {862, 61}

\bibitem[\protect\citeauthoryear{{Stetson}}{{Stetson}}{1987}]{1987PASP...99..191S}
{Stetson} P.~B.,  1987, \mn@doi [\pasp] {10.1086/131977}, \href
  {https://ui.adsabs.harvard.edu/abs/1987PASP...99..191S} {99, 191}

\bibitem[\protect\citeauthoryear{{Stetson}}{{Stetson}}{1992}]{1992ASPC...25..297S}
{Stetson} P.~B.,  1992, in {Worrall} D.~M.,  {Biemesderfer} C.,   {Barnes} J.,
  eds,  Astronomical Society of the Pacific Conference Series Vol. 25,
  Astronomical Data Analysis Software and Systems I. p.~297

\bibitem[\protect\citeauthoryear{{Strai{\v{z}}ys} et~al.,}{{Strai{\v{z}}ys}
  et~al.}{2019}]{2019A&A...623A..22S}
{Strai{\v{z}}ys} V.,  et~al., 2019, \mn@doi [\aap]
  {10.1051/0004-6361/201833987}, \href
  {https://ui.adsabs.harvard.edu/abs/2019A&A...623A..22S} {623, A22}

\bibitem[\protect\citeauthoryear{{Tapia}, {Roth}, {Costero}  \&
  {Navarro}}{{Tapia} et~al.}{1984}]{1984IAUS..105..353T}
{Tapia} M.,  {Roth} M.,  {Costero} R.,   {Navarro} S.,  1984, in {Maeder} A.,
  {Renzini} A.,  eds,  IAU Symposium Vol. 105, Observational Tests of the
  Stellar Evolution Theory. p.~353

\bibitem[\protect\citeauthoryear{{Tapia}, {Roth}, {Marraco}  \& {Ruiz}}{{Tapia}
  et~al.}{1988}]{1988MNRAS.232..661T}
{Tapia} M.,  {Roth} M.,  {Marraco} H.,   {Ruiz} M.~T.,  1988, \mn@doi [\mnras]
  {10.1093/mnras/232.3.661}, \href
  {https://ui.adsabs.harvard.edu/abs/1988MNRAS.232..661T} {232, 661}

\bibitem[\protect\citeauthoryear{{Turner}}{{Turner}}{1989}]{1989AJ.....98.2300T}
{Turner} D.~G.,  1989, \mn@doi [\aj] {10.1086/115300}, \href
  {https://ui.adsabs.harvard.edu/abs/1989AJ.....98.2300T} {98, 2300}

\bibitem[\protect\citeauthoryear{{Webb} \& {Vesperini}}{{Webb} \&
  {Vesperini}}{2018}]{2018MNRAS.479.3708W}
{Webb} J.~J.,  {Vesperini} E.,  2018, \mn@doi [\mnras] {10.1093/mnras/sty1723},
  \href {https://ui.adsabs.harvard.edu/abs/2018MNRAS.479.3708W} {479, 3708}

\bibitem[\protect\citeauthoryear{{Yadav}, {Sariya}  \& {Sagar}}{{Yadav}
  et~al.}{2013}]{2013MNRAS.430.3350Y}
{Yadav} R.~K.~S.,  {Sariya} D.~P.,   {Sagar} R.,  2013, \mn@doi [\mnras]
  {10.1093/mnras/stt136}, \href
  {https://ui.adsabs.harvard.edu/abs/2013MNRAS.430.3350Y} {430, 3350}

\bibitem[\protect\citeauthoryear{{Yu} et~al.,}{{Yu}
  et~al.}{2018}]{2018AJ....155...91Y}
{Yu} P.-C.,  et~al., 2018, \mn@doi [\aj] {10.3847/1538-3881/aaa45b}, \href
  {https://ui.adsabs.harvard.edu/abs/2018AJ....155...91Y} {155, 91}

\bibitem[\protect\citeauthoryear{{Zeidler}, {Nota}, {Grebel}, {Sabbi},
  {Pasquali}, {Tosi}  \& {Christian}}{{Zeidler}
  et~al.}{2017}]{2017AJ....153..122Z}
{Zeidler} P.,  {Nota} A.,  {Grebel} E.~K.,  {Sabbi} E.,  {Pasquali} A.,  {Tosi}
  M.,   {Christian} C.,  2017, \mn@doi [\aj] {10.3847/1538-3881/153/3/122},
  \href {https://ui.adsabs.harvard.edu/abs/2017AJ....153..122Z} {153, 122}

\bibitem[\protect\citeauthoryear{{de Juan Ovelar} et~al.,}{{de Juan Ovelar}
  et~al.}{2020}]{2020MNRAS.491.2129D}
{de Juan Ovelar} M.,  et~al., 2020, \mn@doi [\mnras] {10.1093/mnras/stz3128},
  \href {https://ui.adsabs.harvard.edu/abs/2020MNRAS.491.2129D} {491, 2129}

\bibitem[\protect\citeauthoryear{{von Hoerner}}{{von
  Hoerner}}{1957}]{1957ApJ...125..451V}
{von Hoerner} S.,  1957, \mn@doi [\apj] {10.1086/146321}, \href
  {https://ui.adsabs.harvard.edu/abs/1957ApJ...125..451V} {125, 451}

\makeatother
\end{thebibliography}
\end{document}